\documentclass[11pt,a4paper]{article}
\pdfoutput=1
\usepackage{jcappub}
\usepackage{amsmath}
\usepackage{graphicx}
\usepackage{epstopdf}
\usepackage{latexsym}
\usepackage{amssymb}

\usepackage{hyperref}
\hypersetup{
    colorlinks=true,
    linkcolor=blue,
    citecolor=blue,
}
\usepackage{bm}
\usepackage{textpos}
\usepackage{caption}
\usepackage{subcaption}
\usepackage{appendix}
\allowdisplaybreaks
\usepackage{wrapfig}
\usepackage{sidecap}
\usepackage{lscape}
\usepackage{array}

\begin{document}

\title{Generalized dark energy interactions with multiple fluids}
\author[a]{Carsten van de Bruck}
\author[a]{,~Jurgen Mifsud}
\author[b]{, Jos\'{e} P. Mimoso}
\author[b]{and Nelson J. Nunes}
\affiliation[a]{Consortium for Fundamental Physics, School of Mathematics and Statistics, University of Sheffield, Hounsfield Road, Sheffield S3 7RH, UK}
\affiliation[b]{Instituto de Astrof\'isica e Ci\^encias do Espa\c{c}o, Faculdade de Ci\^encias da Universidade de Lisboa, Campo Grande, PT1749-016 Lisboa, Portugal} 
\emailAdd{c.vandebruck@sheffield.ac.uk}
\emailAdd{jmifsud1@sheffield.ac.uk}
\emailAdd{jpmimoso@fc.ul.pt}
\emailAdd{njnunes@fc.ul.pt}


\abstract{
In the search for an explanation for the current acceleration of the Universe, scalar fields are the most simple and useful tools to build models of dark energy. This field, however, must in principle couple with the rest of the world and not necessarily in the same way to different particles or fluids.
We provide the most complete dynamical system analysis to date, consisting of a canonical scalar field conformally and disformally coupled to both dust and radiation. We perform a detailed study of the existence and stability conditions of the systems and comment on constraints imposed on the disformal coupling from Big-Bang Nucleosynthesis and given current limits on the variation of the fine--structure constant.
}

\maketitle

\section{Introduction}
The origin of the accelerated expansion belongs to the great unresolved puzzles in cosmology. There are a plethora of different models proposed to explain this phenomenon, such as the simple cosmological constant, minimally coupled slowly rolling scalar fields, scalar--tensor extensions of the Einstein--Hilbert action, $f(R)$ gravity, Gauss--Bonnet extensions of General Relativity, massive gravity, etc. Each of these models make predictions for cosmology and possible local (on earth and the solar system) experiments. See Ref. \cite{Copeland:2006wr,Sotiriou:2008rp,Clifton:2011jh} for recent reviews. 

Scalar field models of dark energy, in which the present accelerated expansion is due to a slowly rolling scalar field, are among the most studied models \cite{Wetterich:1987fm,Peebles:1987ek,Wetterich:1994bg,Caldwell:1997ii,Barreiro:1999zs}. This is on the one hand due to their simplicity and on the other hand these models can be phenomenologically quite rich and make verifiable predictions. In the simplest models, the scalar field is not coupled to any other matter form. In the simplest extension, the field couples non--minimally to dark matter (coupled quintessence) \cite{Wetterich:1994bg,Amendola:1999er,Holden:1999hm,Copeland:2003cv}. From the field--theoretical side, there is no reason to believe that the scalar field is decoupled from the rest of the world, unless there is a symmetry which forbids coupling to the standard model fields, for example \cite{Carroll:1998zi}. 

To couple the scalar field in a non--trivial way to matter, it is usually assumed that matter feels a different metric than the one in the gravitational sector. For example, dark matter particles could propagate such that the geodesics are with respect to a metric ${\tilde g}_{\mu\nu}$ which is related to the gravitational metric via a conformal transformation (we will provide a more formal definition in the next section). Recently, extensions have been proposed in which the metric ${\tilde g}_{\mu\nu}$ is related to the gravitational metric via a {\it disformal} transformation (again, we will provide a formal definition in the next section) \cite{Bekenstein,Zuma2,Zuma5,Zumalacarregui:2012us,Jack,Sakstein:2014aca,Sakstein:2014isa}. As a result, these models can in principle have a very different phenomenology compared to the simplest possible cases. 

In this paper we provide a comprehensive dynamical system analysis of models which allow for disformal couplings of a scalar field to matter. In particular, we allow the scalar to be coupled differently to two different fluids, such as dust and radiation or two dust components. We find the fixed points for the different cases, and evaluate the conditions for their existence and stability. We do recover all results from the literature, but extend those where necessary (for example in the disformally coupled single fluid case, a thorough stability analysis had not yet been done) and discuss the two--fluid case in the presence of disformal couplings for the first time.

The paper is organised as follows: In the next Section, we present the model and the equations of motion in full generality. In Section \ref{sec:general_system} we present the dynamical system equations, and solve for a single fluid case with arbitrary equation of state (EOS) in Section \ref{sec:single}. In Section \ref{sec:two_fluids} we present the different cases with two fluids. The cosmological consequences are discussed in Section \ref{sec:consequences}, in which we also apply our solutions to the variation of the fine--structure constant. A summary and conclusions can be found in Section \ref{sec:conclusions}. In an Appendix we present a dynamical system which can be studied with any functional form of the scalar field potential and scalar--matter couplings. We also collect in the Appendix the expressions for the eigenvalues. 

\section{Cosmology}
We consider a scalar--tensor theory in the Einstein frame with action
\begin{equation}\label{action}
\mathcal{S} = \int d^4 x \sqrt{-g} \left[ \frac{M_{\rm Pl}^2}{2} R - \frac{1}{2} g^{\mu\nu}\partial_\mu \phi \partial_\nu \phi - V(\phi) \right] + \sum_i \mathcal{S}_i \left[\tilde g_{\mu\nu}^i, \chi_i\right],
\end{equation}
where the fields $\chi_i$ propagate on geodesics defined by the metrics
\begin{equation}\label{disformal_relation}
\tilde g_{\mu\nu}^i = C_i(\phi) g_{\mu\nu} + D_i(\phi) \partial_\mu\phi \partial_\nu \phi\;, 
\end{equation}
with $C_i(\phi),\;D_i(\phi)$ being the conformal and disformal coupling functions respectively. In the most general case, the functions $C_i$ and $D_i$ can depend on the kinetic term $X = -\frac{1}{2} g^{\mu\nu}\partial_\mu \phi \partial_\nu \phi$ as well, but in this paper they will be a function of the scalar field $\phi$ only. The introduction of disformal couplings is the simplest extension of the models discussed in the literature which are based on conformal couplings only. But we should mention that the simple look of the action above is deceiving: if we were to study the theory in the frame in which ordinary matter is decoupled from the scalar, the theory in this frame is a Horndeski--theory in which the scalar field is in general coupled disformally to all other matter forms \cite{Bettoni:2013diz,Zuma1}. Furthermore, the action above can also be motivated from a higher--dimensional setup, in which dark matter is confined on a slow--moving brane moving in a higher-dimensional space \cite{Koivisto}. In such a model, the scalar field describes the position of the brane and dark matter is coupled disformally. Working in the Einstein frame will simplify our calculations enormously. 

Variation of the action (\ref{action}) with respect to the metric $g_{\mu\nu}$ leads to the field equations in the Einstein frame
\begin{equation}\label{EFE}
G_{\mu\nu}\equiv R_{\mu\nu}\,-\,\frac{1}{2}g_{\mu\nu}R = \kappa^2\left(T^\phi_{\mu\nu}+\sum_i T^i_{\mu\nu}\right)\;,
\end{equation}
where the energy--momentum tensors for the scalar field, $\phi$, and the other fields, $\chi_i$, are defined by 
\begin{eqnarray}
T^\phi_{\mu\nu}&=&\partial_\mu \phi \partial_\nu \phi\,-\,g_{\mu\nu}\left(\frac{1}{2}g^{\rho\sigma}\partial_\rho\phi\partial_\sigma\phi\;\;+\;\;V(\phi)\right)~,\nonumber\\
T^i_{\mu\nu}&=&-\frac{2}{\sqrt{-g}} \frac{\delta\left(\sqrt{-\tilde{g}^i}\tilde{\mathcal{L}}_i\right)}{\delta g^{\mu\nu}}\;,  \nonumber
\end{eqnarray}
respectively. We further define $\kappa^2\equiv M_\text{Pl}^{-2}\equiv 8\pi G$ such that $M_\text{Pl}=2.4\times 10^{18}$ GeV is the reduced Planck mass. The equation of motion of the scalar field simplifies to the following equation
\begin{equation}
\Box\phi=V_{,\phi}-\sum_i Q_i\;,
\end{equation}
where
\begin{equation}
Q_i=\frac{C_{i,\phi}}{2C_i}T_i+\frac{D_{i,\phi}}{2C_i}T_i^{\mu\nu}\nabla_\mu\phi\nabla_\nu\phi-\nabla_\mu\left[\frac{D_i}{C_i}T^{\mu\nu}_i\nabla_\nu\phi\right]\;,
\end{equation}
where $T_i$ is the trace of $T_i^{\mu\nu}$. The Einstein tensor $G_{\mu\nu}$ is divergenceless, but in our theory this does not imply that all $(i+1)$ energy--momentum tensors on the right hand side of Eq.~(\ref{EFE}) are independently conserved. Indeed, we find the following conservation equation for each $i$--component
\begin{equation}\label{cons}
\nabla^\mu T^i_{\mu\nu}=Q_i\nabla_\nu\phi\;.
\end{equation}
On specifying a perfect fluid energy--momentum tensor for each $i$--component, 
\begin{equation}
T^{\mu\nu}_i=(\rho_i+p_i)u^\mu u^\nu+p_i g^{\mu\nu}\;,
\end{equation}
where $\rho_i$ and $p_i$ are the Einstein frame $i^\text{th}$ fluid energy density and pressure respectively, we find the following modified conservation equation
\begin{equation}
u^{\mu}\nabla_\mu\rho_i+(\rho_i+p_i)\nabla_\mu u^\mu=-Q_i u^\mu\nabla_\mu\phi\;,
\end{equation}
after projecting Eq.~(\ref{cons}) along the 4--velocity $u^\mu$. From now on we will consider the standard flat Friedmann-Robertson-Walker (FRW) metric, such that $ds^2 = -dt^2 + a^2(t) \delta_{ij} dx^i dx^j$, as our Einstein frame metric. Furthermore, for background cosmology, a time dependent scalar field is considered, and we denote a coordinate time derivative by a dot. In this setting, the modified Klein-Gordon equation, fluid conservation equation, and Friedmann equations simplify as follows  
\begin{eqnarray}
\ddot \phi + 3 H \dot \phi + V_{,\phi} &=& \sum_i Q_i\;, \\
\dot \rho_i + 3H\rho_i(1+w_i) &=& -Q_i \dot \phi\;, \\
H^2 &=& \frac{\kappa^2}{3}\left(\rho_\phi + \sum_i\rho_i\right), \\
\dot H &=& -\frac{\kappa^2}{2}\left(\rho_\phi(1+w_\phi) + \sum_i\rho_i(1+w_i)\right),
\end{eqnarray}
where we define the field's energy density as $\rho_\phi=\frac{1}{2}\dot{\phi}^2+V(\phi)$, its pressure as $p_\phi=\frac{1}{2}\dot{\phi}^2-V(\phi)$, and the equation of state parameters for the field and fluids are $w_\phi=p_\phi/\rho_\phi$, and $w_i=p_i/\rho_i$, respectively. 
Since we will be interested in the Einstein frame dynamics of the scalar field in the presence of radiation and matter, we remark that for an FRW cosmology with two perfect fluids coupled to the scalar field, it can be found that \cite{Carsten}
\begin{eqnarray}
Q_1 &=& \frac{{\cal A}_2}{{\cal A}_1{\cal A}_2-D_1 D_2 \rho_1 \rho_2}\left( {\cal B}_1 - D_1 \rho_1 \frac{{\cal B}_2}{{\cal A}_2}\right) \;, \\
Q_2 &=& \frac{{\cal A}_1}{{\cal A}_1{\cal A}_2-D_1 D_2 \rho_1 \rho_2}\left( {\cal B}_2 - D_2 \rho_2 \frac{{\cal B}_1}{{\cal A}_1}\right) \;, 
\end{eqnarray}
where 
\begin{eqnarray}
{\cal A}_i &=& C_i + D_i \left(\rho_i-\dot \phi^2\right)\;, \\
{\cal B}_i &=& \left[ \frac{1}{2} {C_i}_{,\phi} (-1 + 3 w_i) -  \frac{1}{2} {D_i}_{,\phi} \dot \phi^2
+ D_i \left(3 (1+w_i) H\dot \phi + V_{,\phi} + \frac{{C_i}_{,\phi}}{C_i} \dot\phi^2 \right)  \right] \rho_i\;.
\end{eqnarray}
%

\section{Dynamical System Analysis}
\label{sec:general_system}
%
We are now going to reduce the above system of equations to a set of first order autonomous differential equations. We first introduce the variables
\begin{eqnarray}\label{dynamical_variables1}
x^2 &\equiv& \frac{\kappa^2 \phi^{\prime^2}}{6}, 
\hspace{1cm}  y^2 \equiv \frac{\kappa^2 V}{3H^2}, 
 \hspace{1cm}  z_i^2 \equiv \frac{\kappa^2 \rho_i}{3H^2},
 \hspace{1cm}  \lambda_V \equiv - \frac{1}{\kappa}\frac{V_{,\phi}}{V}, \\
 \hspace{1cm}  \lambda_C^i &\equiv& - \frac{1}{\kappa}\frac{{C_i}_{,\phi}}{C_i}, 
 \hspace{1cm}  \lambda_D^i \equiv - \frac{1}{\kappa}\frac{{D_i}_{,\phi}}{D_i}, 
  \hspace{1cm}  \sigma_i \equiv  \frac{D_i H^2}{\kappa^2 C_i},\label{dynamical_variables2} 
\end{eqnarray}
where we use the number of e-folds, $N\equiv\ln a(t)$, instead of the Einstein frame coordinate time, $t$, as the time coordinate, and denote derivatives with respect to $N$ by a prime. In these new variables the Friedmann--scalar field system of equations can be written as follows
\begin{eqnarray} 
x^\prime &=& -\left(3 + \frac{H'}{H}\right) x + \sqrt{\frac{3}{2}} \left( \lambda_V y^2 +  \frac{\kappa Q_1}{3H^2} +     \frac{\kappa Q_2}{3 H^2} \right) \;,\label{start} \\
y^\prime &=& - \sqrt{\frac{3}{2}} \left( \lambda_V x + \sqrt{\frac{2}{3}} \frac{H'}{H} \right) y\;, \\
z_i^\prime &=& -\frac{3}{2} \left( 1+w_i + \frac{2}{3} \frac{H'}{H} + \frac{1}{3} \sqrt{\frac{2}{3}} \frac{\kappa Q_i}{H^2} \frac{x}{z_i^2} \right) z_i\;, \\
\sigma_i^\prime &=& \left( \sqrt{6} (\lambda_C^i - \lambda_D^i) x + 2 \frac{H'}{H} \right) \sigma_i\;,  
\end{eqnarray}
where
\begin{equation}
\frac{H'}{H} = -\frac{3}{2} \left( 2 x^2 + \sum_{i=1}^2 (1+w_i) z_i^2\right)\;,
\end{equation}
subject to the Friedmann equation constraint
\begin{equation}\label{Friedmann}
x^2 + y^2 + \sum_{i=1}^2 z_i^2  = 1\;.
\end{equation}
For two species coupled to the scalar field, we can write 
\begin{eqnarray}
\frac{Q_1}{H^2} &=& \frac{3 A_2}{A_1 A_2 - 9 \sigma_1 \sigma_2 z_1^2 z_2^2} \left( \frac{\nu_1}{3 z_1^2} - \frac{\nu_2 \sigma_1}{A_2} \right) z_1^2\;, \\
 \frac{Q_2}{H^2} &=& \frac{3 A_1}{A_1 A_2 - 9 \sigma_1 \sigma_2 z_1^2 z_2^2} \left( \frac{\nu_2}{3 z_2^2} - \frac{\nu_1 \sigma_2}{A_1} \right) z_2^2\;,
 \end{eqnarray}
where 
\begin{align}
 A_i &\equiv \frac{{\cal A}_i}{C_i} = 1+ 3\sigma_i \left( z_i^2 - 2 x^2\right)\;, \\
\nu_i &\equiv \frac{{\cal B}_i}{C_i H^2} =  3 \left[ \frac{1}{2} \lambda_C^i (1-3w_i) + 3 \sigma_i \left( (\lambda_D^i-2\lambda_C^i) x^2+ \sqrt{6} (1+w_i)x -\lambda_V y^2\right) \right] \frac{z_i^2}{\kappa}.\label{end}
\end{align}
In what follows, we will be considering exponential forms for $C_i,\;D_i$ and V. This ensures that the autonomous system of equations is closed. Indeed, if we choose a different functional form for either one of them, we require a set of evolution equations for the functions $\lambda_{V,C,D}$ (see for example Ref. \cite{Steinhardt:1999nw,Ng:2001hs,Barrow:2013uza}), or alternatively include the equation $\phi' = \sqrt{6}x/\kappa$ \cite{Nunes:2000yc}. We discuss this issue in Appendix \ref{appendix:0}. Other useful quantities are the following 
\begin{align}
\Omega_\phi &= x^2+y^2\;, \label{omega_phi}\\
w_\phi &= \frac{x^2-y^2}{x^2+y^2}\;, \label{w_phi}\\
Z_i &= \sqrt{1-6\sigma_i x^2}\;, \label{Zi}\\
w_i &= \tilde{w}_i\left(1-6\sigma_i x^2\right)\;,
\end{align}
where $\tilde{w}_i$ is the equation of state parameter of fluid $i=1,2$ in the frame defined by the metric $\tilde{g}_{\mu\nu}^i$. For example, in the case of dust and radiation, $\tilde{w}_i=0$ and $\tilde{w}_i=1/3$, respectively. We define $Z_i$ in (\ref{Zi}) as follows 
\begin{equation}
\sqrt{\frac{-\tilde{g}^i}{-g}}=C_i^2\sqrt{1+\frac{D_i}{C_i}g^{\mu\nu}\phi_{,\mu}\phi_{,\nu}}=C_i^2 Z_i\;.
\end{equation}
A potential problem for the theory is when $Z_i=0$ due to a metric singularity. This has been discussed in Ref. \cite{Sakstein:2014aca}. We also define an effective equation of state parameter \cite{Copeland:2006wr,Amendola:2014kwa}, $w_{\text{eff}}$, such that
\begin{equation}\label{acceleration}
\frac{H^{\prime}}{H}\equiv-\frac{3}{2}\left(1+w_{\text{eff}}\right)\;,
\end{equation}
which implies that
\begin{equation}
w_{\text{eff}}=x^2-y^2+w_1z_1^2+w_2\left(1-x^2-y^2-z_1^2\right)\;.
\end{equation}
We require $w_{\text{eff}}<-1/3$ in order to obtain an accelerated expansion of the Universe. At a fixed point, $\left(x^c,\;y^c,\;z_i^c,\;\sigma_i^c\right)$, the dynamical system is at rest, and furthermore the acceleration equation (\ref{acceleration}) implies a power law solution of the scale factor, i.e.
\begin{equation}
a\propto (t-t_0)^{\frac{2}{3\left(1+w_{\text{eff}}^c\right)}}\;,
\end{equation}
where $w_{\text{eff}}^c=w_{\text{eff}}\left(x^c,\;y^c,\;z_i^c,\;\sigma_i^c\right)$ and $t_0$ is a constant of integration. When $w_{\text{eff}}^c=-1$, the Universe is undergoing eternal de Sitter exponential expansion with a constant Hubble parameter. 
%
\section{Single Fluid--Arbitrary EOS}
\label{sec:single}
%
For a single fluid with an Einstein frame equation of state parameter, $w$, the relevant equations reduce to the following

\begin{align}
\begin{split} 
x^\prime &= -3x + \frac{3}{2}\left(1+\left(\frac{1-w}{1+w}\right)x^2-y^2\right)(1+w) x + \sqrt{\frac{3}{2}}\lambda_V y^2 \\
&+  \sqrt{\frac{3}{2}}\frac{1-x^2-y^2}{1+3\sigma(1-3x^2-y^2)}\left(\frac{1}{2}\lambda_C (1-3w) + 3\sigma\left(\sqrt{6}x(1+w)-\lambda_V y^2 + (\lambda_D-2\lambda_C)x^2\right) \right)\;, \label{single1}\\
\end{split}\\
y^\prime &= - \sqrt{\frac{3}{2}} \lambda_V xy + \frac{3}{2}\left(1+\left(\frac{1-w}{1+w}\right)x^2-y^2\right)(1+w) y\;, \label{single2}\\
\sigma^\prime &= \sqrt{6} (\lambda_C - \lambda_D) x\sigma -3\left(1+\left(\frac{1-w}{1+w}\right)x^2-y^2\right)(1+w)\sigma\;, \label{single3}  
\end{align}
with the Friedmann equation constraint 
\begin{equation}
x^2+y^2+z^2=1\;,
\end{equation}
where the other variables have the same definition as in the general two fluid case discussed in Section \ref{sec:general_system}. We remark that for the special case of a pressureless fluid, i.e. $w=0$, with the following couplings and scalar field potential
\begin{equation}\label{couplings}
C(\phi) = e^{2\alpha\kappa\phi},\hspace{1cm}
D(\phi) = \frac{e^{2(\alpha+\beta)\kappa\phi}}{M^4}, \hspace{1cm}
V(\phi) = V_0^4 e^{-\lambda\kappa\phi}, 
\end{equation}
where $\alpha$, $\beta$, $\lambda$, $M$, and $V_0$ are all considered to be constant, we recover the dynamical system studied in Ref. \cite{Sakstein:2014aca}. The latter parameter $V_0$ is a mass scale associated with the scalar potential. We will be considering the couplings and scalar field potential as defined in Eq.~(\ref{couplings}), and furthermore, we will re-parametrise our single fluid equation of state parameter to $\gamma\equiv \tilde{w}+1$ such that $0\leq\gamma\leq2$. We remark that the above system coincides with the conformally coupled case given in Ref. \cite{Amendola:1999er,Gumjudpai:2005ry} in the limit $\beta\rightarrow-\infty$ and furthermore, the uncoupled system presented in Ref. \cite{Copeland:1997et} is recovered when $\beta\rightarrow-\infty$ and $\alpha=0$. 

\subsection{Fixed Points}
The fixed points for a single fluid with an arbitrary constant equation of state parameter, $\gamma$, are found by setting equations (\ref{single1})--(\ref{single3}) equal to zero. The fixed points for this system, labelled (1)-(8), are tabulated in Table \ref{table1}. We list the cosmological parameters, $\Omega_\phi$ and $w_\phi$, together with $Z$ in Table \ref{table2}. In Table \ref{table3} we explicitly write down the equation of state parameter dependent fixed points for the particular cases of dust and radiation, since the remaining will be identical to the generic case found in Table \ref{table1}. We use the same numbering system for radiation and dust fixed points and label the radiation fixed points by a subscript $(r)$ and the dust fixed points by a subscript $(d)$. For simplicity, we do not rename (1), (2) and (7) for the radiation and dust cases, although we relabel (3) and (4) for the specific cases of dust and radiation, as described above, in order to use them in the two fluid cases discussed in Section \ref{sec:two_fluids}. In Table \ref{table3a} we give the effective equation of state together with the required parameter values for accelerated expansion for all dust fixed points. Fixed point (5) is obtained when considering $y=0,\;\sigma\neq0$ in equations (\ref{single1}) and (\ref{single3}) giving, in the generic case, fixed points (3), (4), and (5). For the specific case of dust, only fixed points $(3)_{(d)}$ and $(4)_{(d)}$ are obtained, ending up with seven fixed points which coincide with the fixed points found in Ref. \cite{Sakstein:2014aca}. As expected, for radiation, i.e., $\gamma=4/3$, we obtain the full set of eight fixed points.  
{\setlength\extrarowheight{9pt}
\begin{table}
\begin{center}
\begin{tabular}{ c c c c c} 
 \hline
\hline
 Name &  $x$ & $y$ & $\sigma$  \\ 
\hline
(1) & -1 & 0 & 0    \\ 
(2) & 1 & 0 & 0    \\ 
(3) & $\frac{\sqrt{2} \beta -\sqrt{2 \beta ^2-3}}{\sqrt{3}}$ & 0 & $\frac{1}{18} \left(2 \beta  \left(2\beta+\sqrt{4 \beta ^2-6} \right)-3\right)$   \\ 
(4) & $\frac{\sqrt{2} \beta + \sqrt{2 \beta ^2-3} }{\sqrt{3}}$ & 0 & $\frac{1}{18} \left(2 \beta \left(2\beta - \sqrt{4 \beta ^2-6}\right)-3\right)$    \\ 
(5) & $\frac{\sqrt{\frac{3}{2}} \gamma }{\alpha  (4-3 \gamma )+2 \beta }$ & 0 & $\frac{(\alpha  (4-3 \gamma )+2 \beta )^2 \left(2 \alpha ^2 (4-3 \gamma )^2+4
   \alpha  \beta  (4-3 \gamma )-3 (\gamma -2) \gamma \right)}{9 (\gamma -1)
   \gamma  \left(3 \left(6 \alpha ^2-1\right) \gamma ^2-24 \alpha  \gamma  (2
   \alpha +\beta )+8 (2 \alpha +\beta )^2\right)}$    \\ 
(6) & $\frac{\sqrt{\frac{2}{3}} \alpha  (4-3 \gamma )}{\gamma -2}$ & 0 & 0    \\ 
(7) & $\frac{\lambda }{\sqrt{6}}$ & $\sqrt{1-\frac{\lambda ^2}{6}}$ & 0    \\ 
(8) & $\frac{\sqrt{\frac{3}{2}} \gamma }{(4-3\gamma)\alpha +\lambda }$ & $\frac{\sqrt{2 \alpha ^2 (4-3 \gamma )^2+\alpha  (8-6 \gamma ) \lambda -3
   (\gamma -2) \gamma }}{\sqrt{2 (\alpha  (4-3 \gamma )+\lambda )^2}}$ & $0$    \\ 
 \hline
\hline
\end{tabular}
\end{center}
\caption{\label{table1} Fixed points of the system (\ref{single1})--(\ref{single3}) for the single fluid case.}
\end{table}}
{\setlength\extrarowheight{9pt}
\setlength{\tabcolsep}{1.6pt}
\begin{table}
\begin{center}
\begin{tabular}{ c c c c c} 
 \hline
\hline
 Name &  $\Omega_\phi$ & $w_\phi$ & $Z$ \\ 
\hline
(1) & 1 & 1 & 1    \\ 
(2) & 1 & 1 & 1    \\ 
(3) & $\frac{1}{3} \left(\sqrt{2 \beta ^2-3}-\sqrt{2} \beta \right)^2$ & 1 & 0    \\ 
(4) & $\frac{1}{3} \left(\sqrt{2 \beta ^2-3}+\sqrt{2} \beta \right)^2$ & 1 & 0    \\ 
(5) & $\frac{3 \gamma ^2}{2 (\alpha  (4-3 \gamma )+2 \beta )^2}$ & 1 & $\sqrt{\frac{\gamma  \left(-2 \alpha ^2 (4-3 \gamma )^2+4 \alpha  \beta  (3 \gamma
   -4)+3 (\gamma -2) \gamma \right)}{(\gamma -1) \left(3 \left(6 \alpha
   ^2-1\right) \gamma ^2-24 \alpha  \gamma  (2 \alpha +\beta )+8 (2 \alpha +\beta
   )^2\right)}+1}$    \\ 
(6) & $\frac{2 \alpha ^2 (4-3 \gamma )^2}{3 (\gamma -2)^2}$ & 1 & 1    \\ 
(7) & $1$ & $\frac{1}{3} \left(\lambda ^2-3\right)$ & 1    \\ 
(8) & $\frac{\alpha ^2 (4-3 \gamma )^2+\alpha  (4-3 \gamma ) \lambda +3 \gamma }{(\alpha
    (4-3 \gamma )+\lambda )^2}$ & $\frac{3 \gamma ^2}{\alpha ^2 (4-3 \gamma )^2+\alpha  (4-3 \gamma ) \lambda +3
   \gamma }-1$ & 1   \\ 
 \hline
\hline
\end{tabular}
\end{center}
\caption{\label{table2}The cosmological parameters $\Omega_\phi$ and $w_\phi$, together with the quantity $Z$ as defined in equations (\ref{omega_phi}), (\ref{w_phi}), and (\ref{Zi}), for the system (\ref{single1})--(\ref{single3}) corresponding to the single fluid case.}
\end{table}}
%
\subsubsection{Existence Conditions}
\label{sec:single_fluid_existence}
For this analysis we will be using the fact that $0\leq\Omega_\phi\leq1$, such that the fluid energy density is non-negative, $\rho\geq0$. As already mentioned, we will be considering $0\leq\gamma\leq2$ and that $x,\;y,\;\sigma,\;Z\in\mathbb{R}$.
%
\subsubsection*{\textnormal{\underline{Arbitrary EOS}}}
We will now make some remarks on the existence of the fixed points (1)-(8) found in Table \ref{table1}.
\begin{itemize}
\item \textit{Kination}: Fixed points (1) and (2) always exist as they are independent from the introduced parameters. These scalar field kinetic dominated solutions 
 are characterised by a stiff equation of state, $w_\phi=1$, and as expected there is no metric singularity as $Z=1$. 
\item \textit{Disformal}: For the disformal fixed points (3) and (4), we find that $\beta\geq\sqrt{3/2}$ for (3) and $\beta\leq-\sqrt{3/2}$ for (4). 
Both points give a stiff fluid with a metric singularity as $Z=0$, and they are found to be independent from the fluid equation of state parameter, although, as we will see, their stability does depend on $\gamma$.
\item \textit{Mixed}: As already remarked in the beginning of this section, (5) is not defined for dust and, furthermore, this fixed point does not exist for the choice $\gamma=0$ as well. The allowed parameter values of $\alpha,\;\beta$ and $\gamma$ must satisfy the inequality $3\gamma^2<2\left(\alpha(4-3\gamma)+2\beta\right)^2$ together with the condition $Z\in\mathbb{R}$.
This disformal fixed point is also characterized by a stiff fluid $(w_\phi=1)$, although in this case we can avoid the metric singularity if we choose the right parameter values, such that $Z\in\mathbb{R}\setminus\{0\}$. 
\item \textit{Conformal kinetic}: In order to define a finite $x$-coordinate of (6), we restrict the range of $\gamma$ to $0\leq\gamma<2$. For $\gamma=4/3$, all parameter values are allowed, although we end up with an indeterminate value of $w_\phi$. For $\gamma\neq4/3$, the solution is characterised by a stiff fluid. The existence of this fixed point is as follows
\begin{equation*}
\begin{split}
&\;\;\;\;\;\;\;\; \text{For} \;\; \gamma\in[0,4/3)\cup(4/3,2)\;,\;\hspace{1cm} \alpha^2<\frac{3(\gamma-2)^2}{2(4-3\gamma)^2}\;,\\
&\;\;\;\;\;\;\;\; \text{For} \;\; \gamma=\frac{4}{3}\;, \; \hspace{1cm} \forall\alpha\;. 
\end{split}
\end{equation*}
\item \textit{Scalar field dominated}: For fixed point (7) we find that this is defined if $\lambda^2<6$. This is a scalar field dominated solution $(\Omega_\phi=1)$ with a scalar field equation of state parameter $w_\phi=-1+\lambda^2/3$. 
\item \textit{Conformal scaling}: For the last fixed point, (8), we require the following inequalities to be satisfied in order to be defined
\begin{eqnarray*}
\alpha^2(4-3\gamma)^2+\alpha(4-3\gamma)\lambda &>& -\frac{3}{2}\gamma(2-\gamma)\;,\;\\
\alpha(4-3\gamma)\lambda &>& 3\gamma-\lambda^2\;.
\end{eqnarray*}
In the absence of the conformal coupling, $\alpha=0$, (8) is a cosmological scaling solution \cite{Gumjudpai:2005ry,Wetterich:1987fm,Copeland:1997et}, such that $w_\phi=\gamma-1$.  
\end{itemize}
We shall now consider the existence of the dust $(\gamma=1)$ fixed points, denoted by the index $(d)$, and radiation $(\gamma=4/3)$ fixed points, denoted by the index $(r)$;
\subsubsection*{\textnormal{\underline{Dust $(\gamma=1)$}}}
The existence arguments for fixed points (1)-(4) found in Table \ref{table1} also hold for $(1),\;(2),\;(3)_{(d)},\linebreak\;(4)_{(d)}$, respectively. Furthermore, the existence of $(7)$ is equivalent to the general case. Regarding the conformal kinetic dominated fixed point $(6)_{(d)}$, we require that $\alpha^2<3/2$. For the last fixed point, $(8)_{(d)}$, we require that $\alpha(\alpha+\lambda)>-3/2$ and $\lambda(\alpha+\lambda)>3$. We note that for non-negative values of $\alpha,\;\lambda$ and $\beta$, this analysis coincides with that of Ref. \cite{Sakstein:2014aca}. 
%
\subsubsection*{\textnormal{\underline{Radiation $(\gamma=4/3)$}}}
Similar to the dust case, the first four radiation fixed points and $(7)$ are respectively equivalent to (1)-(4) and (7) found in Table \ref{table1}, hence the existence of these fixed points follows from the general fluid discussion. For fixed point $(5)_{(r)}$, we find that $\beta^2>2$ in order to satisfy the condition $\Omega_\phi<1$, and that the coupling between the two disformally related metrics, $Z$, is made sure to be real valued. This solution is characterized by a stiff fluid equation of state, and for $\beta^2>2$, the radiation and Einstein frame metrics are both well-defined without a singularity. The radiation fluid dominated solution, $(6)_{(r)}$, always exists, irrespective of the parameter values. It is characterised by an indeterminate scalar field equation of state. The last radiation fixed point, $(8)_{(r)}$, is a scaling solution which exists when $\lambda^2>4$.   
{\setlength\extrarowheight{9pt}
\begin{table}
\begin{center}
\begin{tabular}{ c c c c c c c c} 
 \hline
\hline
 Name &  $x$ & $y$ & $\sigma$ & $\Omega_\phi$ & $w_\phi$ & $Z$ \\ 
\hline
$(5)_{(r)}$ & $\frac{\sqrt{\frac{2}{3}}}{\beta }$ & 0 & $\frac{\beta ^2}{3 \beta ^2-2}$ & $\frac{2}{3\beta^2}$ & 1 & $\sqrt{1+\frac{4}{2-3\beta^2}}$ \\ 
$(6)_{(d)}$ & $-\sqrt{\frac{2}{3}} \alpha$ & 0 & 0 & $\frac{2\alpha^2}{3}$ & 1 & 1 \\ 
$(6)_{(r)}$ & 0 & 0 & 0  & 0 & - & 1   \\
$(8)_{(d)}$ & $\sqrt{\frac{3}{2}}\frac{1}{\alpha +\lambda }$ & $\frac{\sqrt{3+2\alpha(\alpha+\lambda)}}{\sqrt{2 (\alpha+\lambda)^2}}$ & 0 & $\frac{3+\alpha(\alpha+\lambda)}{(\alpha+\lambda)^2}$ & $-1+\frac{3}{3+\alpha(\alpha+\lambda)}$ & 1   \\
$(8)_{(r)}$ &  $2 \sqrt{\frac{2}{3}}\frac{1}{\lambda}$ & $\frac{2 }{\sqrt{3\lambda^2}}$ & 0  & $\frac{4}{\lambda^2}$ & $\frac{1}{3}$ & 1 \\  
 \hline
\hline
\end{tabular}
\end{center}
\caption{\label{table3} The equation of state parameter dependent fixed points of the system (\ref{single1})--(\ref{single3}), the single fluid case,  specified for dust and radiation, together with the respective cosmological parameters.}
\end{table}}
{\setlength\extrarowheight{9pt}
\setlength{\tabcolsep}{4.5pt}
\begin{table}
\begin{center}
\begin{tabular}{ c c c c c c c p{6.1cm}} 
 \hline
\hline
Quantity & $(1)$ & $(2)$ & $(3)_{(d)}$ & $(4)_{(d)}$ & $(6)_{(d)}$ & $(7)$ & $(8)_{(d)}$\\ 
\hline
$w_\text{eff}$ & 1 & 1 & $\Omega_\phi$ & $\Omega_\phi$ & $\frac{2\alpha^2}{3}$ & $\frac{\lambda^2}{3}-1$ & $\frac{-\alpha}{\alpha+\lambda}$   \\ 
Acceleration & No & No & No & No & No & $\lambda^2<2$ & $\lambda\leq-\sqrt{2},\;\alpha<\lambda/2\;,\;\;\;\;\;\;\;\;$ $\text{or},\;-\sqrt{2}<\lambda<0,\;\alpha<\left(3-\lambda^2\right)/\lambda\;,$ or, $0<\lambda<\sqrt{2},\;\alpha>\left(3-\lambda^2\right)/\lambda\;,$ $\text{or},\;\lambda\geq\sqrt{2},\;\alpha>\lambda/2\;\;\;\;\;\;\;\;\;$ \\ 
 \hline
\hline
\end{tabular}
\end{center}
\caption{\label{table3a} The effective equation of state, $w_\text{eff}$, together with the required parameter values for an accelerated expansion, for all dust fixed points $(1)-(8)_{(d)}$. We take into consideration the existence of the fixed point when determining the required parameters for acceleration.}
\end{table}}
\subsubsection{Stability Conditions}
We now study the stability of the fixed points by analysing the eigenvalues of the matrix $\mathcal{M}$, which is constructed after considering a small perturbation around each fixed point. In what follows, no zero eigenvalues are obtained, and hence the Hartman--Grobman theorem guarantees that the stability around a fixed point can be studied by the linear approximation (see for example Ref. \cite{Sakstein:2014aca,Barrow:2013uza} for stability analysis with zero eigenvalue). We give the matrix elements of $\mathcal{M}$ in Appendix \ref{appendix:a} together with the corresponding eigenvalues $e_{1,2,3}$ in Appendix \ref{appendix:b}. Since our system is three-dimensional and not two-dimensional, as in the purely conformal case, the stability analysis will be different from the lower dimensional system. We will restrict our stability analysis to the dust and radiation cases only, as in the general fluid case there is freedom in four parameters. This is due to the fact that even if the fixed points are independent from the disformal coupling, the eigenvalues can still contain $\beta$. 
%
\subsubsection*{\textnormal{\underline{Dust $(\gamma=1)$}}}
\begin{figure}
\centering
\begin{subfigure}[b]{0.48\textwidth}
  \includegraphics[width=1\textwidth]{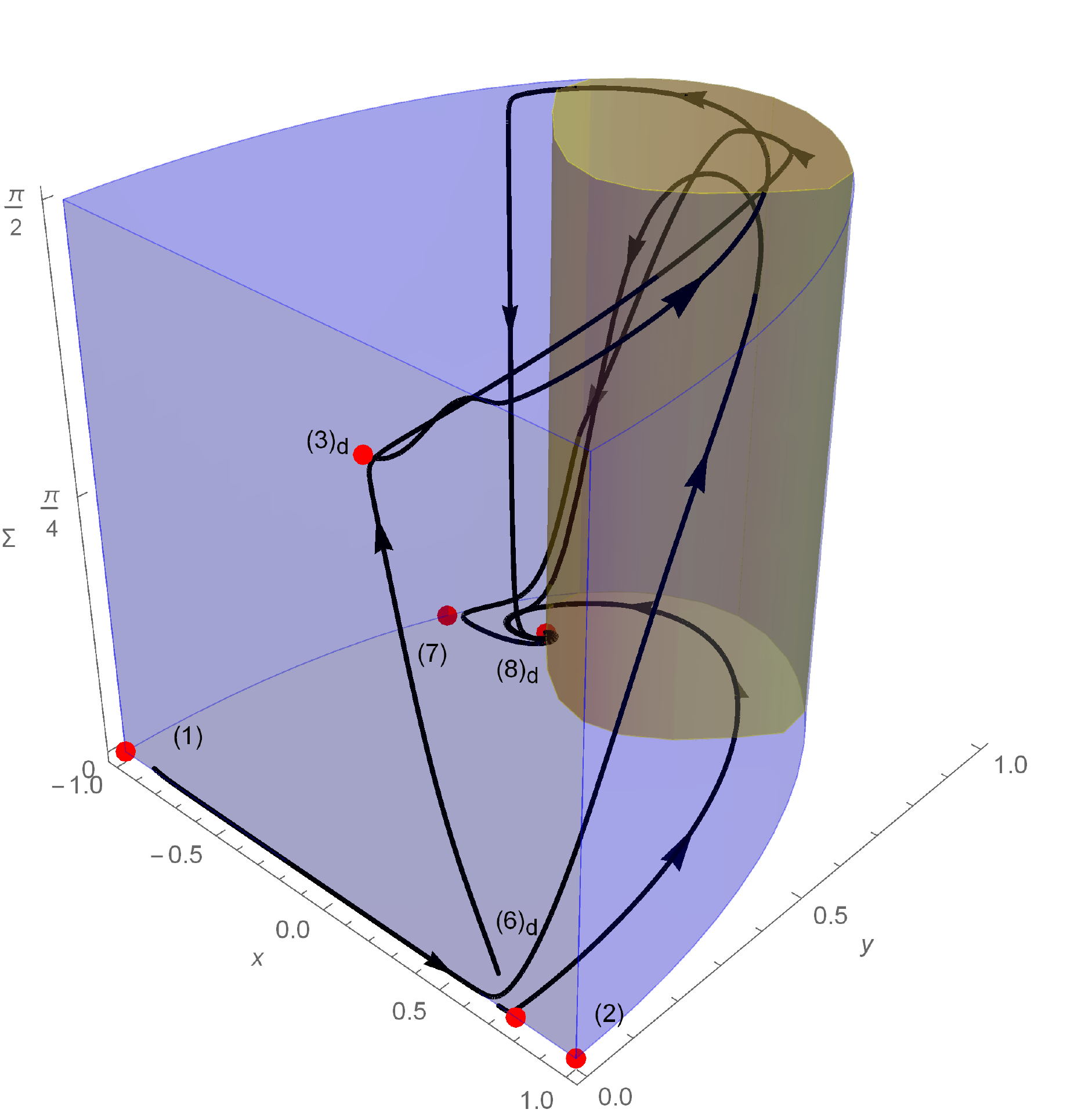}
 \end{subfigure} 
\begin{subfigure}[b]{0.48\textwidth}
  \includegraphics[width=1\textwidth]{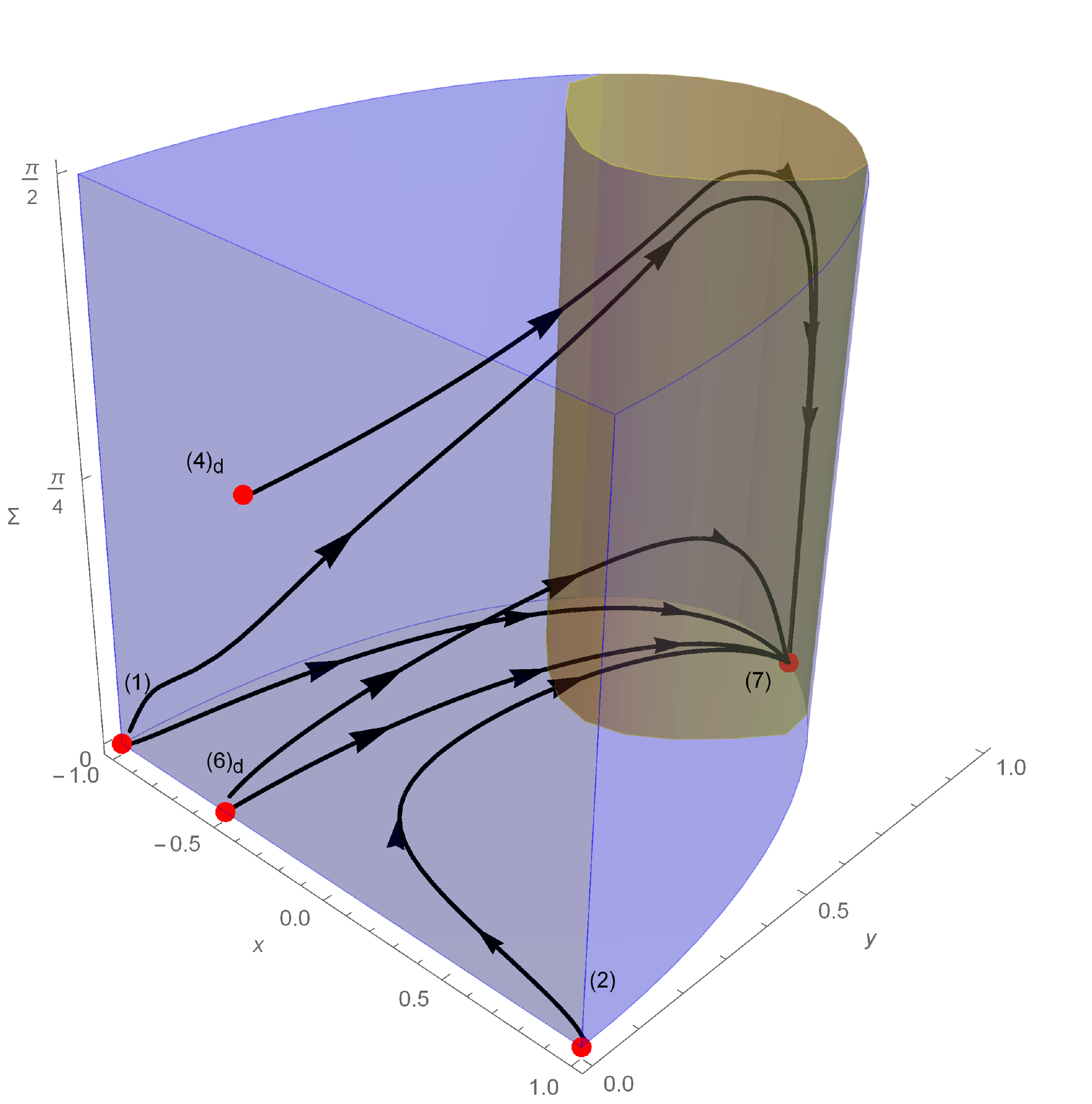}
 \end{subfigure} 
\caption{In this figure we show the phase space for the single dust case with different attractors. The different solution trajectories correspond to different initial conditions. The blue region is the allowed region, whereas the yellow region is where the Universe undergoes an accelerated expansion. On the left the attractor is $(8)_{(d)}$ with $\alpha=-0.94,\;\beta=3,\;\lambda=-1.88$, and on the right the attractor is $(7)$ with $\alpha=0.6,\;\beta=-2,\;\lambda=0.7$.} 
\label{fig:3dplots_dust}
\end{figure}

\begin{itemize}
\item $(1)$: this can either be a stable node, an unstable node or a saddle point depending on the chosen values of $\alpha,\;\beta$ and $\lambda$. It is a stable node if $\beta>-\sqrt{3/2},\;\lambda<-\sqrt{6}$ and $\alpha>\sqrt{3/2}$. Consequently, this fixed point can become stable when a disformal coupling is introduced. Also, in the following fixed points we find that the introduction  of a disformal coupling widens up the region of parameter space that renders a fixed point stable.

\item $(2)$: similarly, the other scalar field kinetic dominated fixed point can either be a stable node, an unstable node or a saddle point according to the chosen parameter values. It is a stable node if $\beta<\sqrt{3/2},\;\lambda>\sqrt{6}$ and $\alpha<-\sqrt{3/2}$.
\item $(3)_{(d)}$: it is either a stable node or a saddle point. Indeed, we find that this is a stable point if $\beta>\sqrt{3/2},\;\alpha<-\beta+\sqrt{(-3+2\beta^2)/2}$ and $\lambda>2\beta$.
\item $(4)_{(d)}$: the remaining disformal fixed point can either be a stable node or a saddle point. It is stable if $\beta<-\sqrt{3/2},\;\alpha>-\beta-\sqrt{(-3+2\beta^2)/2}$ and $\lambda<2\beta$. 
\item $(6)_{(d)}$: the conformal kinetic dust solution can either be a stable node or a saddle point. It cannot be an unstable fixed point as $e_1<0$ when $-\sqrt{3/2}<\alpha<\sqrt{3/2}$. It is found to be stable in the following regions
\begin{eqnarray*}
-\sqrt{\frac{3}{2}}&<&\alpha<0\;,\hspace{1cm} \lambda>\frac{-3-2\alpha^2}{2\alpha}\;, \hspace{1cm}\beta<\frac{-3-2\alpha^2}{4\alpha}\;,\nonumber \\
\text{or}, \hspace{1cm} 0&<&\alpha<\sqrt{\frac{3}{2}}\;,  \hspace{1cm} \lambda<\frac{-3-2\alpha^2}{2\alpha}\;,\hspace{1cm} \beta>\frac{-3-2\alpha^2}{4\alpha}\;.
\end{eqnarray*}

\item $(7)$: for parameter values satisfying either one of the following inequalities
%
\begin{eqnarray*}
-\sqrt{6}&<&\lambda<0 \,,\hspace{1cm} \beta>\frac{\lambda}{2}\;, \hspace{1cm} \alpha>\frac{3-\lambda^2}{\lambda}\;,\\
\text{or}, \hspace{1cm} 0 &<&\lambda<\sqrt{6}\;,\hspace{1cm} \beta<\frac{\lambda}{2}\;, \hspace{1cm} \alpha<\frac{3-\lambda^2}{\lambda}\;,
\end{eqnarray*}

we have a stable node, otherwise it is a saddle point. This dust fixed point cannot be unstable, since $e_2<0$ for $-\sqrt{6}<\lambda<\sqrt{6}$. 
\item $(8)_{(d)}$: the conformal scaling fixed point is found to be either a saddle point, a stable spiral, a stable node or a spiral saddle. We show all four distinct natures of this fixed point when $\beta=-0.9,\;0.5,\;5$ in Fig \ref{fig:3}.

\begin{figure}[h!]
\begin{center}
  \includegraphics[width=0.4\textwidth]{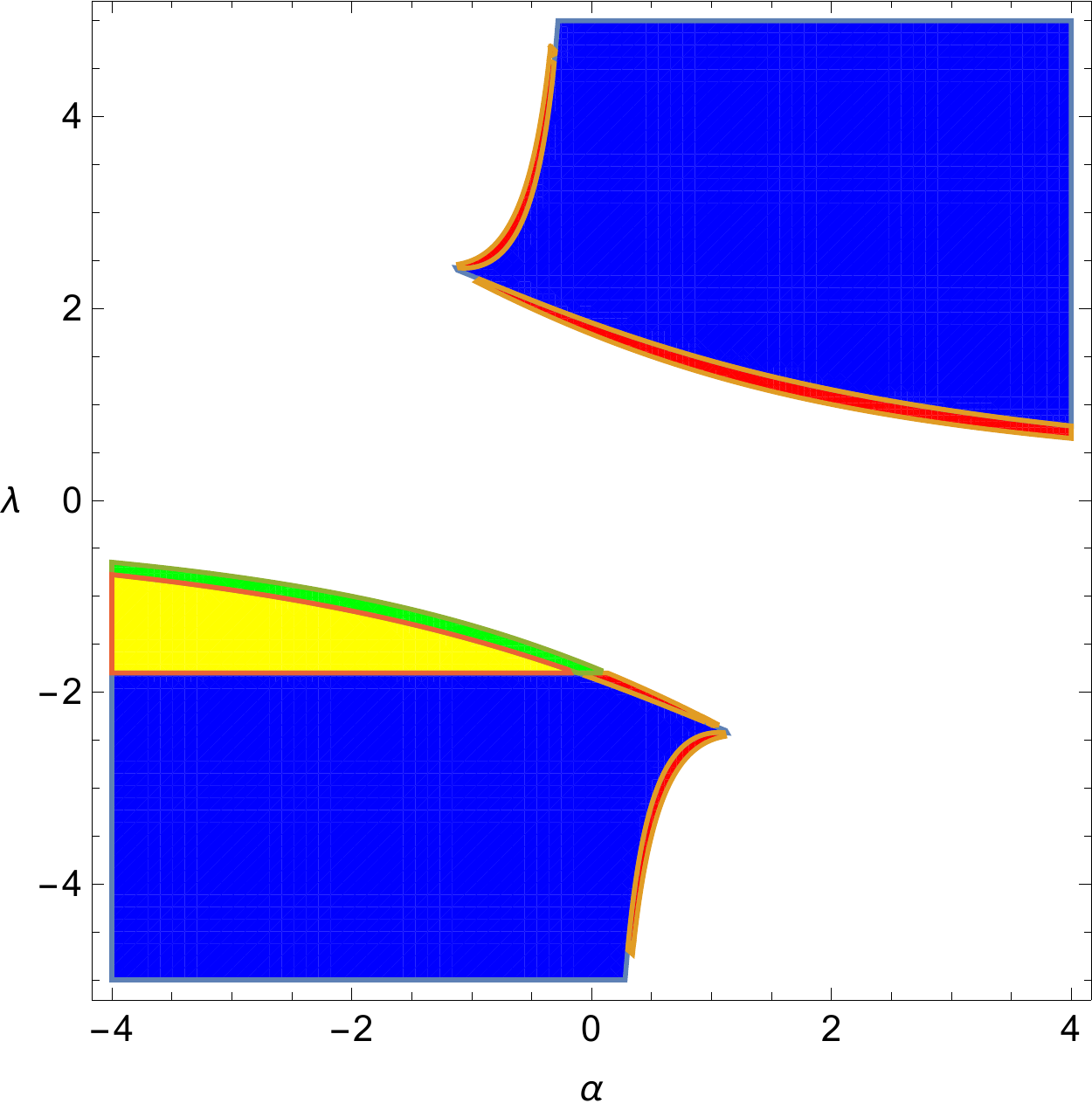}
  \includegraphics[width=0.4\textwidth]{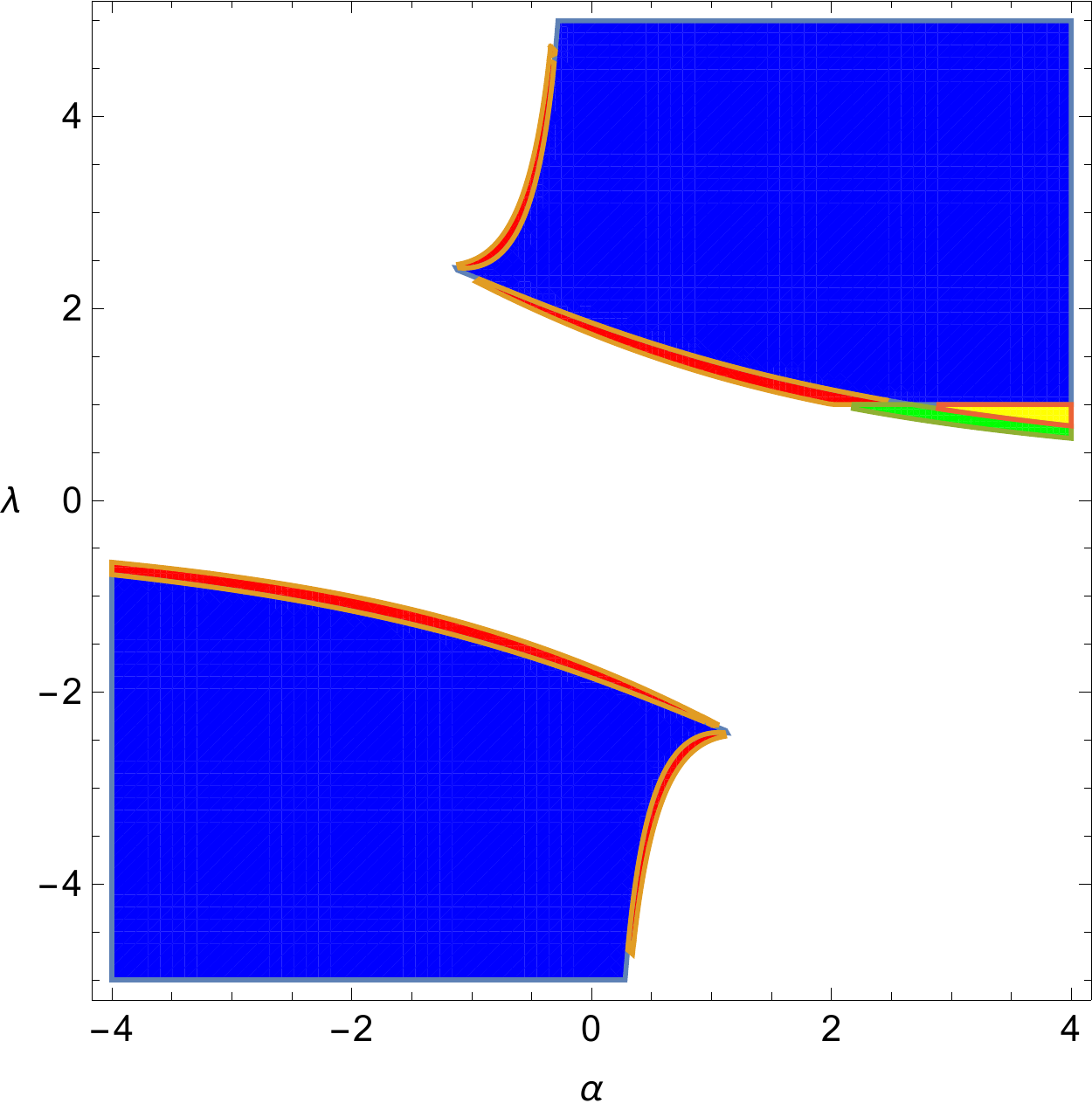}
\end{center}
\begin{center}
\includegraphics[width=0.54\textwidth]{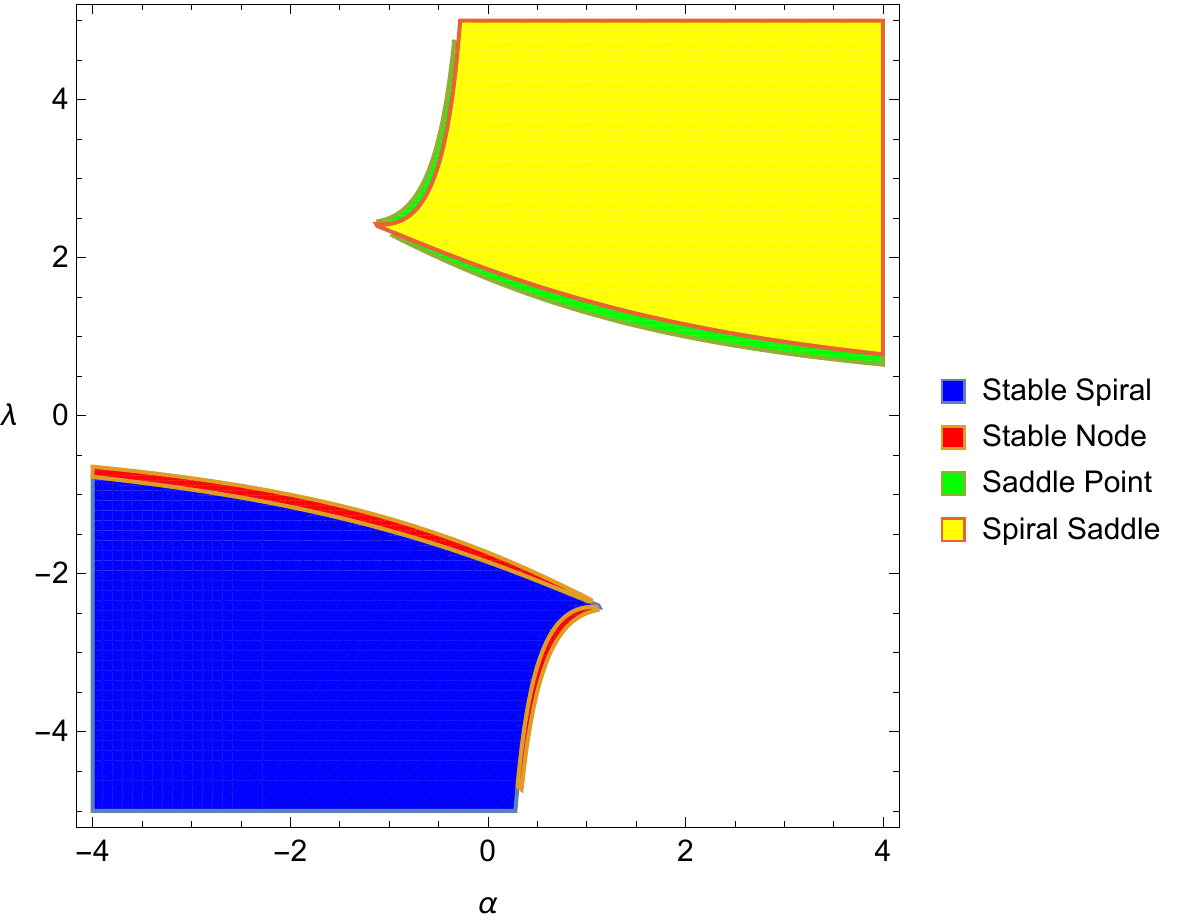}
\caption{The regions show an illustration of the parameter values of $\alpha$ and $\lambda$ for the dust fixed point $(8)_{(d)}$ when $\beta=-0.9$ (left), $\beta=0.5$ (right), and $\beta=5$ (bottom), with each region corresponding to a distinct nature of the fixed point.} 
\label{fig:3}
\end{center}
\end{figure}
\end{itemize}
The three-dimensional single fluid system is invariant under $y\rightarrow-y$ and furthermore, the $(x,\;y,\;\sigma)$-phase space is non-compact, since $-1\leq x\leq1,\;0\leq y\leq\sqrt{1-x^2},\;0\leq\sigma<\infty$. We restrict the range of $\sigma$ to non-negative values due to stability problems \cite{Sakstein:2014aca,Jack}. We compactify this phase space by introducing the variable $\Sigma=\arctan\sigma$. The phase space is now compact, with $x,\;y,\;\Sigma$ lying in the range $-1\leq x\leq 1,\;0\leq y\leq\sqrt{1-x^2},\;0\leq\Sigma<\pi/2$. We are also aware of fixed points at $\Sigma=\pi/2$ \cite{Sakstein:2014aca}, although this is beyond the scope of our study. The compactified phase space is described by a semi-circular prism of length $\pi/2$. Furthermore, we can at most have six fixed points for any parameter choice. This is due to the fact that for a particular choice of $\beta$, either $(3)_{(d)}$ or $(4)_{(d)}$ exists, but not both at the same time. Two illustrations containing some solution trajectories with different attractors are shown in Fig \ref{fig:3dplots_dust}. The $x-y$ plane in these three-dimensional phase spaces, depict the purely conformal case.
%
\subsubsection*{\textnormal{\underline{Radiation $(\gamma=4/3)$}}}
\begin{figure}
\centering
\begin{subfigure}[b]{0.48\textwidth}
  \includegraphics[width=1\textwidth]{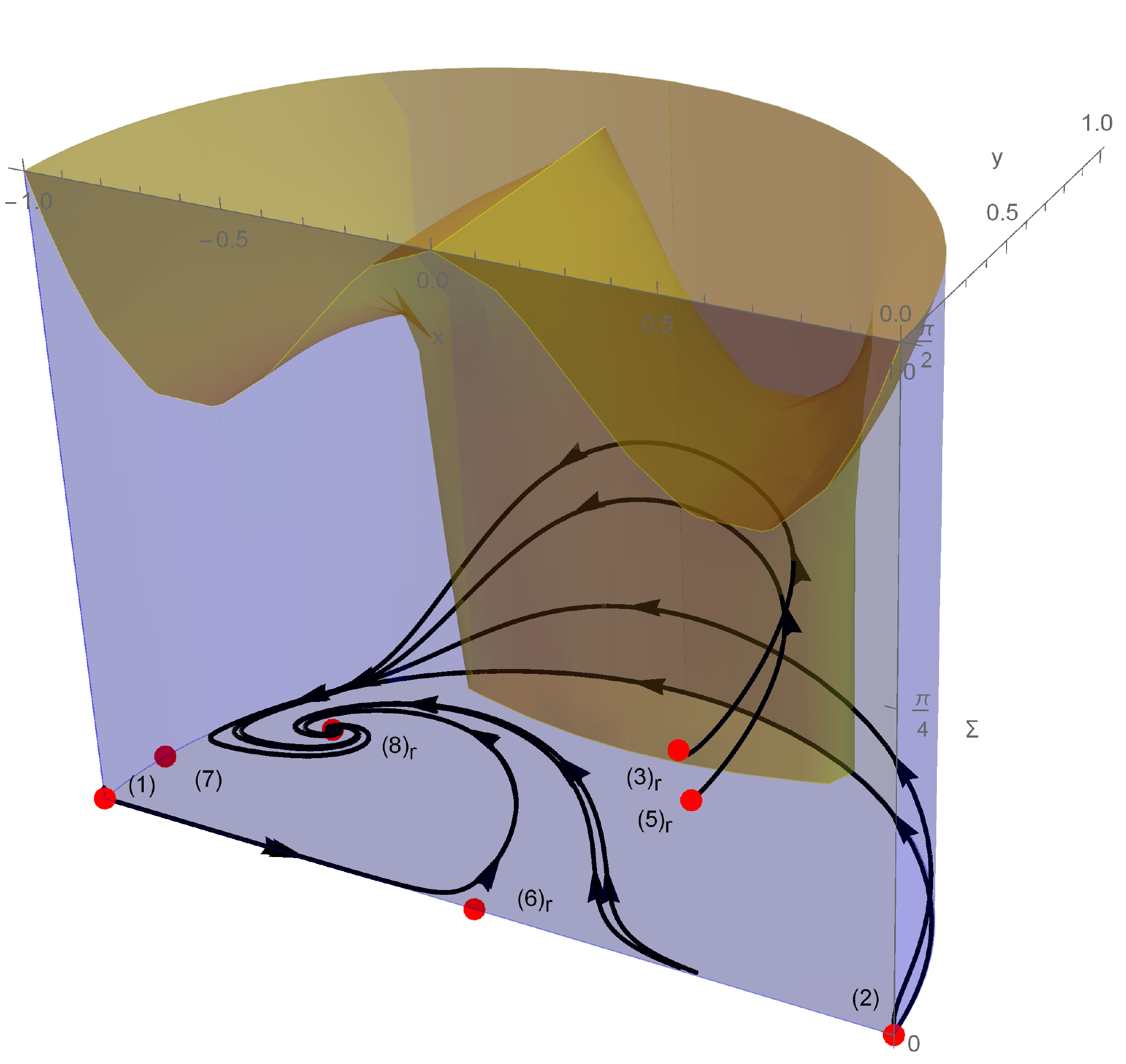}
 \end{subfigure} 
\begin{subfigure}[b]{0.48\textwidth}
  \includegraphics[width=1\textwidth]{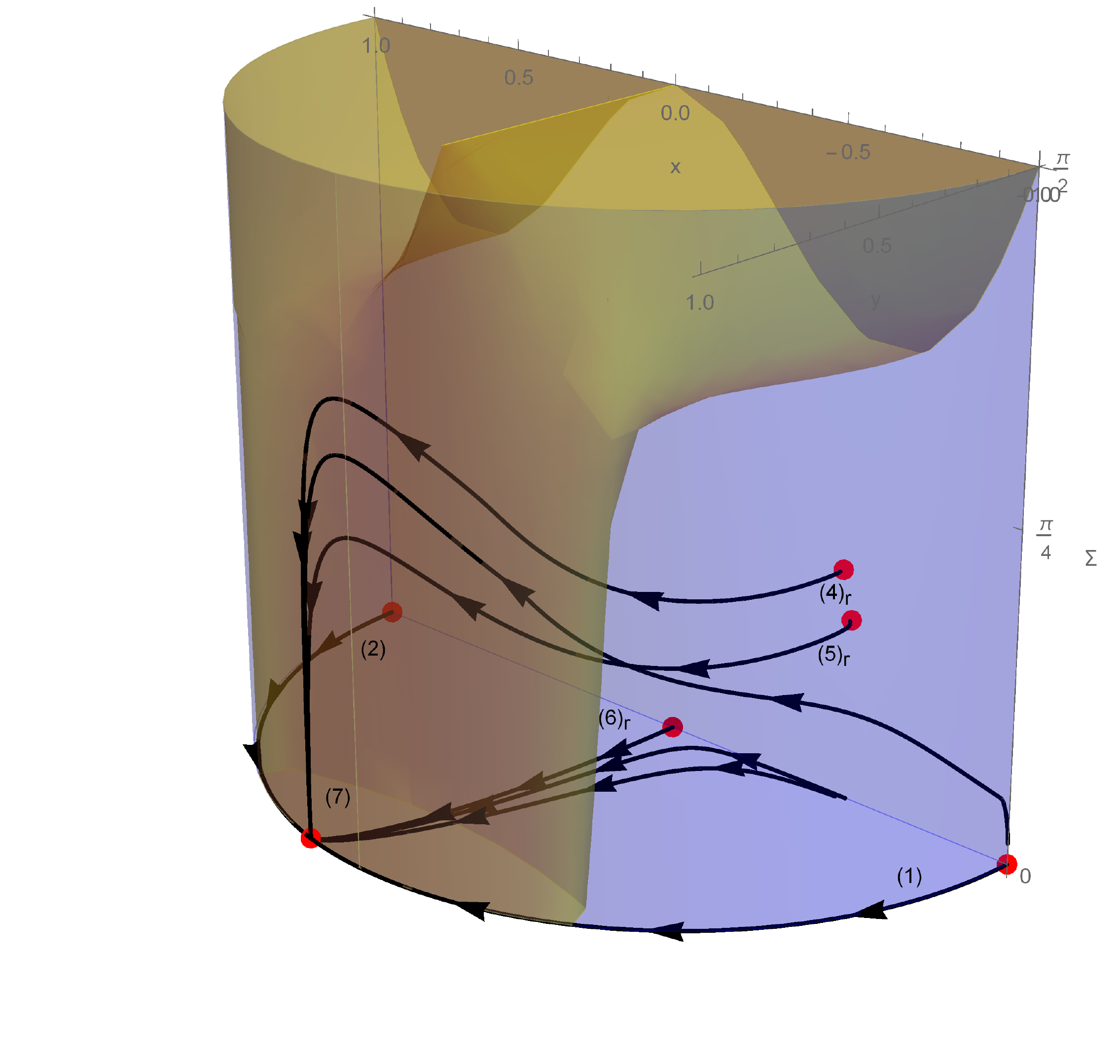}
 \end{subfigure} 
\caption{In this figure we show the phase space for the radiation single fluid case with different attractors. The different solution trajectories correspond to different initial conditions. The blue region is the allowed region, whereas the yellow region is where the Universe undergoes an accelerated expansion. On the left the attractor is $(8)_{(r)}$ with $\beta=1.5,\;\lambda=-2.395$, and on the right the attractor is $(7)$ with $\beta=-1.5,\;\lambda=0.448$.} 
\label{fig:3dplots_radiation}
\end{figure}

\begin{itemize}
\item $(1)$: since $e_2=2$, then this kination fixed point cannot be stable. Indeed, it can either be an unstable node or a saddle point. It is found to be an unstable node if $\lambda>-\sqrt{6}$ and $\beta<-\sqrt{3/2}$, and it is a saddle point if either $\lambda<-\sqrt{6}$ and $\beta\neq-\sqrt{3/2}$, or if $\lambda>-\sqrt{6}$ and $\beta>-\sqrt{3/2}$. 
\item $(2)$: the other kination fixed point is found to be an unstable node if $\lambda<\sqrt{6}$ and $\beta>\sqrt{3/2}$. It can also be a saddle point if $\lambda<\sqrt{6}$ and $\beta<\sqrt{3/2}$, or if $\lambda>\sqrt{6}$ and $\beta\neq\sqrt{3/2}$.
\item $(3)_{(r)}$: in this case, we can have a stable node for $\beta>\sqrt{2}$ and $\lambda>2\beta$. It can also be a saddle point for $\beta>\sqrt{2}$ and $\lambda<2\beta$, or if $\sqrt{3/2}<\beta<\sqrt{2}$, then it is a saddle point when $\lambda\neq2\beta$.
\item $(4)_{(r)}$: this disformal fixed point is a stable node when $\beta<-\sqrt{2}$ and $\lambda<2\beta$. For $\beta<-\sqrt{2}$ and $\lambda>2\beta$, together with the other choice of $-\sqrt{2}<\beta<-\sqrt{3/2}$ and $\lambda\neq2\beta$, we find that $(4)_{(r)}$ is a saddle point.  
\item $(5)_{(r)}$: the mixed fixed point, which is missing in the case of dust, is found to be only a saddle point, since $e_1$ and $e_2$ have opposite signs in the available range of $\beta$. Indeed, this is true for $\beta<-\sqrt{2}$ such that $\lambda\neq2\beta$, and for the choice $\beta>\sqrt{2}$ and $\lambda\neq2\beta$.
\item $(6)_{(r)}$: this radiation dominated fixed point is a saddle point, as its eigenvalues are $e_1=-1,\;e_2=-4$ and $e_3=2$. 
\item $(7)$: this scalar field dominated fixed point is either a stable node or a saddle point, since $e_2<0$ in the fixed point existence range of $-\sqrt{6}<\lambda<\sqrt{6}$. It is found to be stable for $-2<\lambda<0$ and $\beta>\lambda/2$, and also when $0<\lambda<2$ such that $\beta<\lambda/2$.
\item $(8)_{(r)}$: the conformal scaling fixed point cannot be unstable as $\mathfrak{R}(e_2)<0$ and $\mathfrak{R}(e_3)<0$ $\forall\lambda\in\mathbb{R}\setminus[-2,2]$. We find that for $2<\lambda\leq8/\sqrt{15}$ and $-8/\sqrt{15}\leq\lambda<-2$, it is a stable node for $\beta<\lambda/2$ and $\beta>\lambda/2$ respectively, and it is a saddle point if $\beta>\lambda/2$ and $\beta<\lambda/2$ respectively. For $\lambda>8/\sqrt{15}$, it is found to be a stable spiral when $\beta<\lambda/2$, and a spiral saddle if $\beta>\lambda/2$. Similarly, for $\lambda<-8/\sqrt{15}$, $(8)_{(r)}$ is a stable spiral when $\beta>\lambda/2$, and a spiral saddle if $\beta<\lambda/2$. 
\end{itemize}
Similar to the single fluid dust case, we show the phase space together with some solution trajectories in Fig \ref{fig:3dplots_radiation}.
In this case, for any particular choice of the parameters, we can at most have seven fixed points, where two of them are disformal fixed points. We should remark that when either one of the disformal fixed points is the global attractor of the system, it is found that the solution converges very slowly, in agreement with the results found in Ref. \cite{Sakstein:2014aca}.
%
%
\section{Two Fluids}
\label{sec:two_fluids}
We will now investigate the dynamical system presented in Section \ref{sec:general_system} for a two fluid scenario. Three different particular cases are studied in the sections that follow. In each case, we will be considering at least one conformally--disformally coupled fluid. When the obtained fixed point can be generated from a single fluid system, we will use an identical label to that corresponding to a single fluid fixed point. Despite the fact that the majority of the two fluid fixed points reduce to the single fluid fixed points, our aim is to generalise the conformally coupled fluid system \cite{Amendola:1999er,Brookfield:2007au,Amendola:2014kwa} to a conformally--disformally coupled fluid system. Because a generic treatment is very cumbersome, we will consider in what follows, dust and radiation fluids only.  
\subsection{Two Fluids--Conformal-disformal dust and conformal-disformal radiation}
\label{sec:cdcd_dust_rad}
In this section, we study the full solution of the two fluid conformal--disformal system presented in Section \ref{sec:general_system} for the particular case of dust and radiation perfect fluids, i.e. $\gamma_1=1,\;\gamma_2=4/3$. This system will be five-dimensional, in which we choose our dynamical variables to be $x,\;y,\;z_1,\;\sigma_1$ and $\sigma_2$. In the absence of disformal couplings, this system reduces to the three-dimensional conformally coupled case presented in Ref. \cite{Amendola:1999er}. We choose our couplings and scalar field potential to be of exponential forms 
\begin{equation}\label{couplings_potential}
C_i(\phi)=e^{2\alpha_i\kappa\phi},\hspace{1cm} D_i(\phi)=\frac{e^{2(\alpha_i+\beta_i)\kappa\phi}}{M_i^4},\hspace{1cm} V(\phi)=V_0^4 e^{-\lambda\kappa\phi}.
\end{equation}
 For completeness, we list all fixed points in Table \ref{table8} together with the corresponding cosmological parameters in Table \ref{table9}. We also include $z_2$  in Table \ref{table8} in order to link the single fluid cases studied in Section \ref{sec:single} with this scenario. The new fixed points for this two fluid system which cannot be obtained from a single fluid system are $(a)$ and $(b)$. We shall refer to fixed point $(a)$ as the \textit{conformal dust radiation} fixed point, which was obtained in Ref. \cite{Amendola:1999er}, and refer to fixed point $(b)$ as the \textit{disformal dust radiation} fixed point.
{\setlength\extrarowheight{9pt}
\setlength{\tabcolsep}{5.5pt}
\begin{table}
\begin{center}
\begin{tabular}{ c c c c c c c c} 
 \hline
\hline
 Name   &  $x$ & $y$ & $z_1$ & $z_2$ & $\sigma_1$ & $\sigma_2$  \\ 
\hline
(1) & -1 & 0 & 0 & 0 & 0 & 0   \\ 
$(6)_{(r)}$ & 0 & 0 & 0 & 1 & 0 & 0 \\ 
(2) & 1 & 0 & 0 & 0 & 0 & 0  \\ 
$(a)$ & $-\frac{1}{\sqrt{6}\alpha_1}$ & 0 & $\frac{1}{\sqrt{3\alpha_1^2}}$ & $\sqrt{1-\frac{1}{2\alpha_1^2}}$ & 0 & 0   \\ 
$(6)_{(d)}$ & $-\sqrt{\frac{2}{3}}\alpha_1$ & 0 & $\sqrt{1-\frac{2\alpha_1^2}{3}}$ & 0 & 0 & 0    \\ 
$(b)$ & $\frac{\sqrt{\frac{2}{3}}}{\beta_1 }$ & 0 & $\frac{2}{\sqrt{3\beta_1^2}}$ & $\sqrt{1-\frac{2}{\beta_1^2}}$ & $\frac{\beta_1^2}{4}$ & 0   \\ 
$(3)_{(d)}$ & $\frac{\sqrt{2} \beta_1 -\sqrt{2 \beta_1^2-3}}{\sqrt{3}}$ & 0 & 
$\epsilon_1^d$ & 
0 & $\epsilon_2^d$ & 0   \\ 
$(4)_{(d)}$ & $\frac{\sqrt{2} \beta_1 + \sqrt{2 \beta_1^2-3} }{\sqrt{3}}$ & 0 & 
$\epsilon_3^d$ & 
0 & $\epsilon_4^d$ & 0   \\
$(5)_{(r)}$ & $\frac{\sqrt{\frac{2}{3}}}{\beta_2 }$ & 0 & 0 & $\sqrt{1-\frac{2}{3\beta_2^2}}$ & 0 & $\frac{\beta_2^2}{3\beta_2^2-2}$   \\ 
$(3)_{(r)}$ & $\frac{\sqrt{2} \beta_2 -\sqrt{2 \beta_2^2-3}}{\sqrt{3}}$ & 0 & 0 & $\epsilon^r_1$ & 0 & $\epsilon^r_2$ \\ 
$(4)_{(r)}$ &  $\frac{\sqrt{2} \beta_2 + \sqrt{2 \beta_2^2-3} }{\sqrt{3}}$ & 0 & 0 & $\epsilon^r_3$ & 0 & $\epsilon^r_4$ \\
$(8)_{(r)}$ & $\frac{2\sqrt{\frac{2}{3}}}{\lambda}$ & $\frac{2}{\sqrt{3\lambda^2}}$ & 0 & $\sqrt{1-\frac{4}{\lambda^2}}$ & 0 & 0 \\ 
(7) & $\frac{\lambda }{\sqrt{6}}$ & $\sqrt{1-\frac{\lambda ^2}{6}}$ & 0 & 0 & 0 & 0 \\ 
$(8)_{(d)}$ &  $\sqrt{\frac{3}{2}}\frac{1}{\alpha_1+\lambda}$ & $\sqrt{\frac{3+2\alpha_1(\alpha_1+\lambda)}{2(\alpha_1+\lambda)^2}}$ & $\sqrt{\frac{\lambda(\alpha_1+\lambda)-3}{(\alpha_1+\lambda)^2}}$ & 0 & 0 & 0  \\  
\hline
\end{tabular}
\end{center}
\caption{\label{table8} The fixed points for conformally--disformally coupled dust and conformally--disformally coupled radiation. The $\epsilon^{d,r}$ terms are as follows: $\epsilon_1^d\equiv\sqrt{\frac{2}{3}} \sqrt{\beta _1 \left(\sqrt{4 \beta _1^2-6}-2 \beta _1\right)+3}$, $\epsilon_2^d\equiv\frac{1}{18} \left(2 \beta_1  \left(2\beta_1+\sqrt{4 \beta_1^2-6} \right)-3\right)$,    $\epsilon_3^d\equiv\sqrt{6-2 \beta _1 \left(2 \beta _1+\sqrt{4 \beta _1^2-6}\right)}/\sqrt{3}$ ,  $\epsilon_4^d\equiv\frac{1}{18} \left(2 \beta_1 \left(2\beta_1 - \sqrt{4 \beta_1^2-6}\right)-3\right)$, $\epsilon^r_1\equiv\sqrt{\frac{2}{3}} \sqrt{\beta _2 \left(\sqrt{4 \beta _2^2-6}-2 \beta _2\right)+3}$,  $\epsilon^r_2\equiv\frac{1}{18} \left(2 \beta_2  \left(2\beta_2+\sqrt{4 \beta_2^2-6} \right)-3\right)$,  $\epsilon_3^r\equiv\sqrt{6-2 \beta _2 \left(2 \beta _2+\sqrt{4 \beta _2^2-6}\right)}/\sqrt{3}$, and  $\epsilon^r_4\equiv\frac{1}{18} \left(2 \beta_2 \left(2\beta_2 - \sqrt{4 \beta_2^2-6}\right)-3\right)$.}
\end{table}}
{\setlength\extrarowheight{9pt}
\setlength{\tabcolsep}{1.4pt}
\begin{table}
\begin{center}
\begin{tabular}{ c c c c c c} 
 \hline
\hline
 Name   &  $\Omega_\phi$ & $w_\phi$ & $Z_1$ & $Z_2$ & $w_\text{eff}$  \\ 
\hline
(1) & 1 & 1 & 1 & 1 & 1   \\ 
$(6)_{(r)}$ & 0 & - & 1 & 1 & $\frac{1}{3}$  \\ 
(2) & 1 & 1 & 1 & 1 & 1    \\ 
$(a)$ & $\frac{1}{6\alpha_1^2}$ & 1 & 1 & 1 & $\frac{1}{3}$    \\ 
$(6)_{(d)}$ & $\frac{2}{3}\alpha_1^2$ & 1 & 1 & 1 & $\frac{2\alpha_1^2}{3}$    \\ 
$(b)$ & $\frac{2}{3\beta_1^2}$ & 1 & 0 & 1 & $\frac{1}{3}$   \\ 
$(3)_{(d)}$ & $\frac{1}{3}\left(-\sqrt{2} \beta_1 +\sqrt{2 \beta_1^2-3}\right)^2$ & 1 & 0 & 1 & $\frac{1}{3}\left(2\beta_1\left(2\beta_1-\sqrt{4\beta_1^2-6}\right)-3\right)$   \\ 
$(4)_{(d)}$ & $\frac{1}{3}\left(\sqrt{2} \beta_1 +\sqrt{2 \beta_1^2-3}\right)^2$ & 1 & 0 & 1 & $\frac{1}{3}\left(2\beta_1\left(2\beta_1+\sqrt{4\beta_1^2-6}\right)-3\right)$   \\ 
$(5)_{(r)}$ & $\frac{2}{3\beta_2^2}$ & 1 & 1 & $\sqrt{1+\frac{4}{2-3\beta_2^2}}$ & $\frac{1}{3}$   \\ 
$(3)_{(r)}$ & $\frac{1}{3}\left(-\sqrt{2} \beta_2 +\sqrt{2 \beta_2^2-3}\right)^2$ & 1 & 1 & 0 & $\frac{1}{3}\left(2\beta_2\left(2\beta_2-\sqrt{4\beta_2^2-6}\right)-3\right)$   \\ 
$(4)_{(r)}$ & $\frac{1}{3}\left(\sqrt{2} \beta_2 +\sqrt{2 \beta_2^2-3}\right)^2$ & 1 & 1 & 0 & $\frac{1}{3}\left(2\beta_2\left(2\beta_2+\sqrt{4\beta_2^2-6}\right)-3\right)$   \\ 
$(8)_{(r)}$ & $\frac{4}{\lambda^2}$ & $\frac{1}{3}$ & 1 & 1 & $\frac{1}{3}$   \\ 
(7) & 1 & $\frac{\lambda^2}{3}-1$ & 1 & 1 & $\frac{\lambda^2}{3}-1$ \\ 
$(8)_{(d)}$ &  $\frac{3+\alpha_1(\alpha_1+\lambda)}{(\alpha_1+\lambda)^2}$ & $\frac{3}{3+\alpha_1(\alpha_1+\lambda)}-1$ & 1 & 1 & $-\frac{\alpha_1}{\alpha_1+\lambda}$   \\  
 \hline
\hline
\end{tabular}
\end{center}
\caption{\label{table9} Listed are, respectively, the cosmological parameters $\Omega_\phi$ and $w_\phi$, together with $Z_1$, $Z_2$, and the effective equation of state parameter $w_\text{eff}$, for conformally--disformally coupled dust and conformally--disformally coupled radiation fixed points.}
\end{table}}
%
\subsubsection{Existence Conditions}
\label{sec:cdcd_dust_rad_existence}

In this section we will only comment on the existence of the conformal dust radiation fixed point and the disformal dust radiation fixed point, since the existence of the other fixed points follows from Section \ref{sec:single_fluid_existence}. Indeed, we find that fixed point $(a)$ exists when $\alpha_1^2>1/2$ and fixed point $(b)$ exists whenever  $\beta_1^2>2$.
\par
For the disformal fixed points $(b)$, $(3)_{(d)},\;(4)_{(d)},\;(3)_{(r)},\;(4)_{(r)}$, we can clearly observe that when fluid $i$ has a non-zero value of $\sigma_i$, then the disformal metric transformation $Z_i=0$. The mixed disformal fixed point $(5)_{(r)}$, however, avoids the singularity for both metrics.

\subsubsection{Stability Conditions}
\begin{figure}
\centering
\begin{subfigure}[b]{0.475\textwidth}
  \includegraphics[width=1\textwidth]{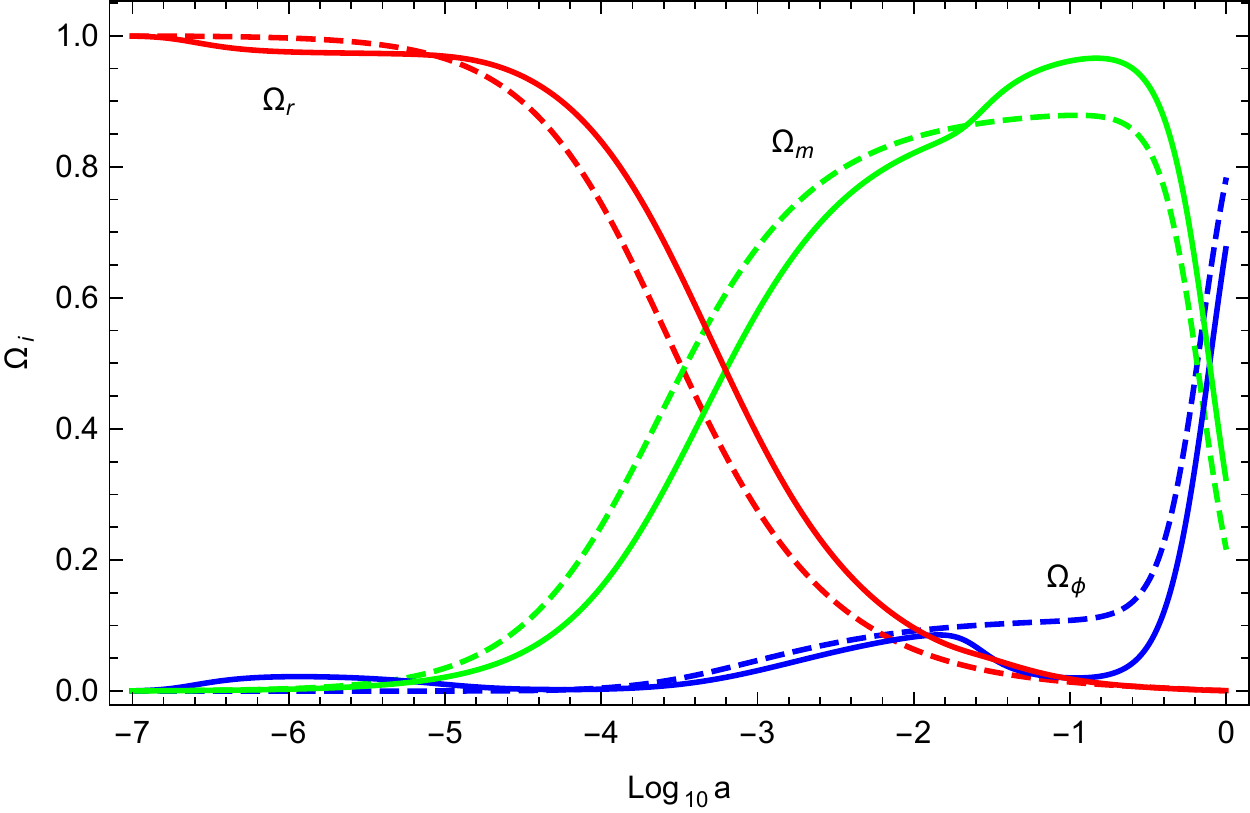} 
\end{subfigure}
\begin{subfigure}[b]{0.495\textwidth}
  \includegraphics[width=1\textwidth]{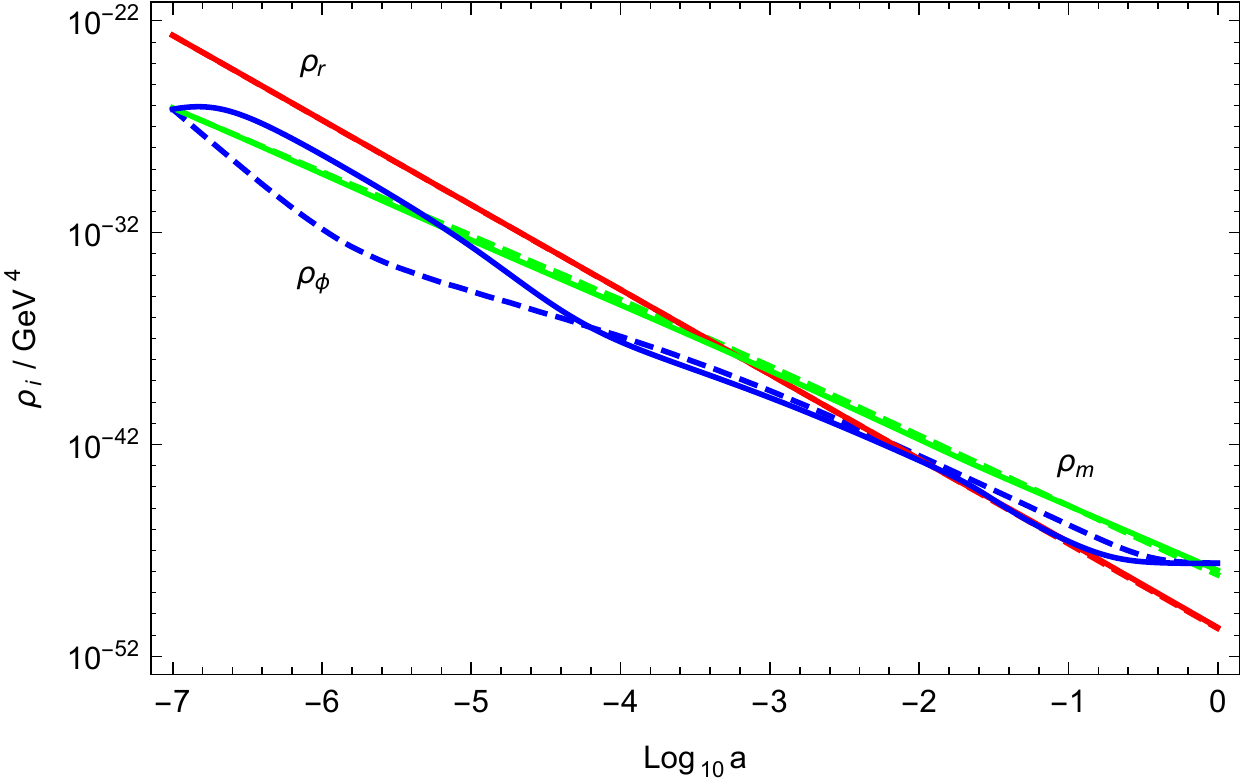} 
\end{subfigure}
\begin{subfigure}[b]{0.495\textwidth}
  \includegraphics[width=1\textwidth]{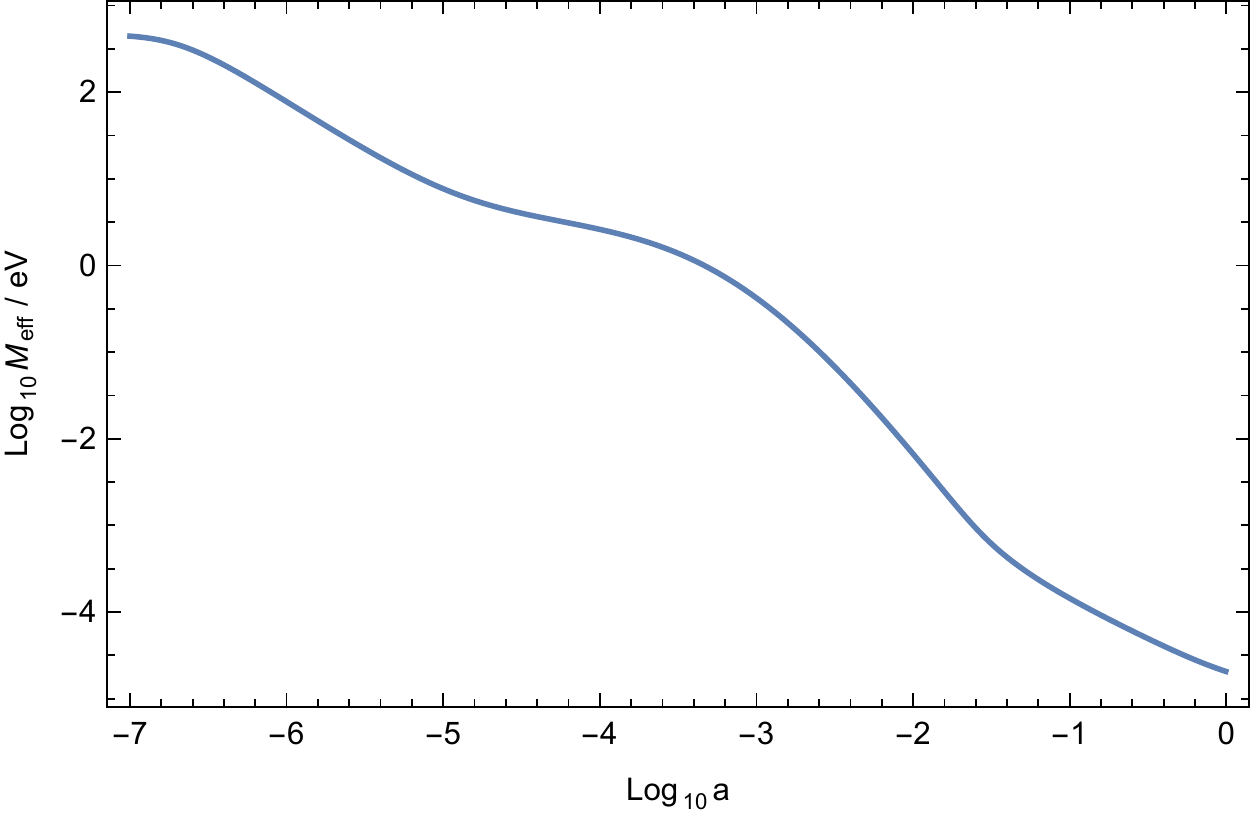} 
\end{subfigure}
\caption{The above plots show the evolution of $\Omega_{r,m,\phi}$ (left) and $\rho_{r,m,\phi}$ (right) when the future attractor is  $(7)$, in the conformal--disformal dust and conformal--disformal radiation case. In each plot we use the same initial conditions and compare the conformally coupled dust case (dashed line) with the conformally--disformally coupled dust-radiation case (solid line) as discussed in Section \ref{sec:cdcd_dust_rad}. The bottom plot shows the evolution of the effective mass scale defined in Eq.~(\ref{meff}). The parameters are fixed to $\alpha_1=-0.41,\;\beta_1=5.81,\;\beta_2=5.11,\;\lambda=-0.08$.} 
\label{fig:Omega7}
\end{figure}

We will now discuss the region in which the above fixed points are found to be stable by using the eigenvalues $e_{1,2,3,4,5}$ \footnote{These eigenvalues are not all included in the text due to the length of the algebraic expressions. We only write down the eigenvalues of the new fixed points $(a)$ and $(b)$ in Appendix \ref{appendix:c1}.}. We only discuss the regions in which the fixed point is found to be stable. 
\begin{itemize}
\item The kinetic dominated fixed points (1) and (2) cannot be stable since $e_1=2$ for both of them. Also, the radiation dominated fixed point, $(6)_{(r)}$, is found to be a saddle point since $e_1,\;e_2,\;e_4<0$ and $e_3,\;e_5>0$. 
\item The conformal dust radiation fixed point, $(a)$, can either be a stable node or a stable spiral. Indeed, it is found to be a stable node when either of the following conditions are satisfied
%
\begin{eqnarray*}
-\sqrt{\frac{2}{3}}&\leq&\alpha_1<-\frac{1}{\sqrt{2}}\;, \hspace{1cm} \beta_1<-2\alpha_1\;,\hspace{1cm} \beta_2<-2\alpha_1\;,\hspace{1cm}\lambda>-4\alpha_1\;,\\
\text{or},\hspace{1cm} \frac{1}{\sqrt{2}}&<&\alpha_1\leq \sqrt{\frac{2}{3}}\;,\hspace{1cm} \beta_1>-2\alpha_1\;, \hspace{1cm}
\beta_2>-2\alpha_1\;,\hspace{1cm}\lambda<-4\alpha_1\;,
\end{eqnarray*}
and is a stable spiral when either of the following holds
%
\begin{eqnarray*}
\alpha_1&<&-\sqrt{\frac{2}{3}}\;,\hspace{1cm} \beta_1<-2\alpha_1\;, \hspace{1cm} \beta_2<-2\alpha_1\;,\hspace{1cm} \lambda>-4\alpha_1\;,\\
\text{or}, \hspace{1cm} \alpha_1&>&\sqrt{\frac{2}{3}}\;,\hspace{1cm} \beta_1>-2\alpha_1\;,\hspace{1cm}\beta_2>-2\alpha_1\;,\hspace{1cm}\lambda<-4\alpha_1\;.
\end{eqnarray*}

\begin{figure}
\centering
 \begin{subfigure}[b]{0.4\textwidth}
  \includegraphics[width=1\textwidth]{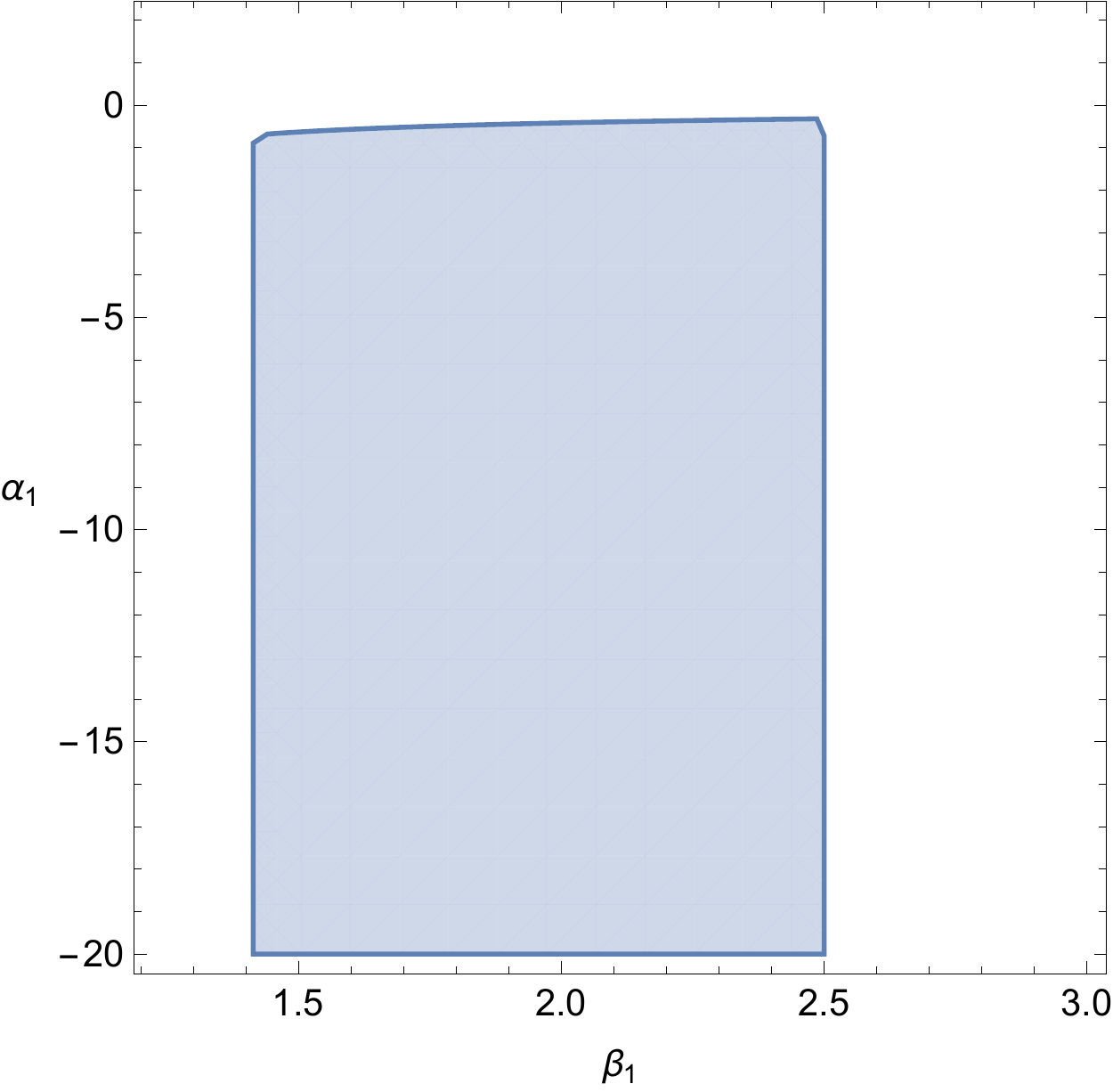}
 \end{subfigure}
  \begin{subfigure}[b]{0.4\textwidth}
  \includegraphics[width=0.9755\textwidth]{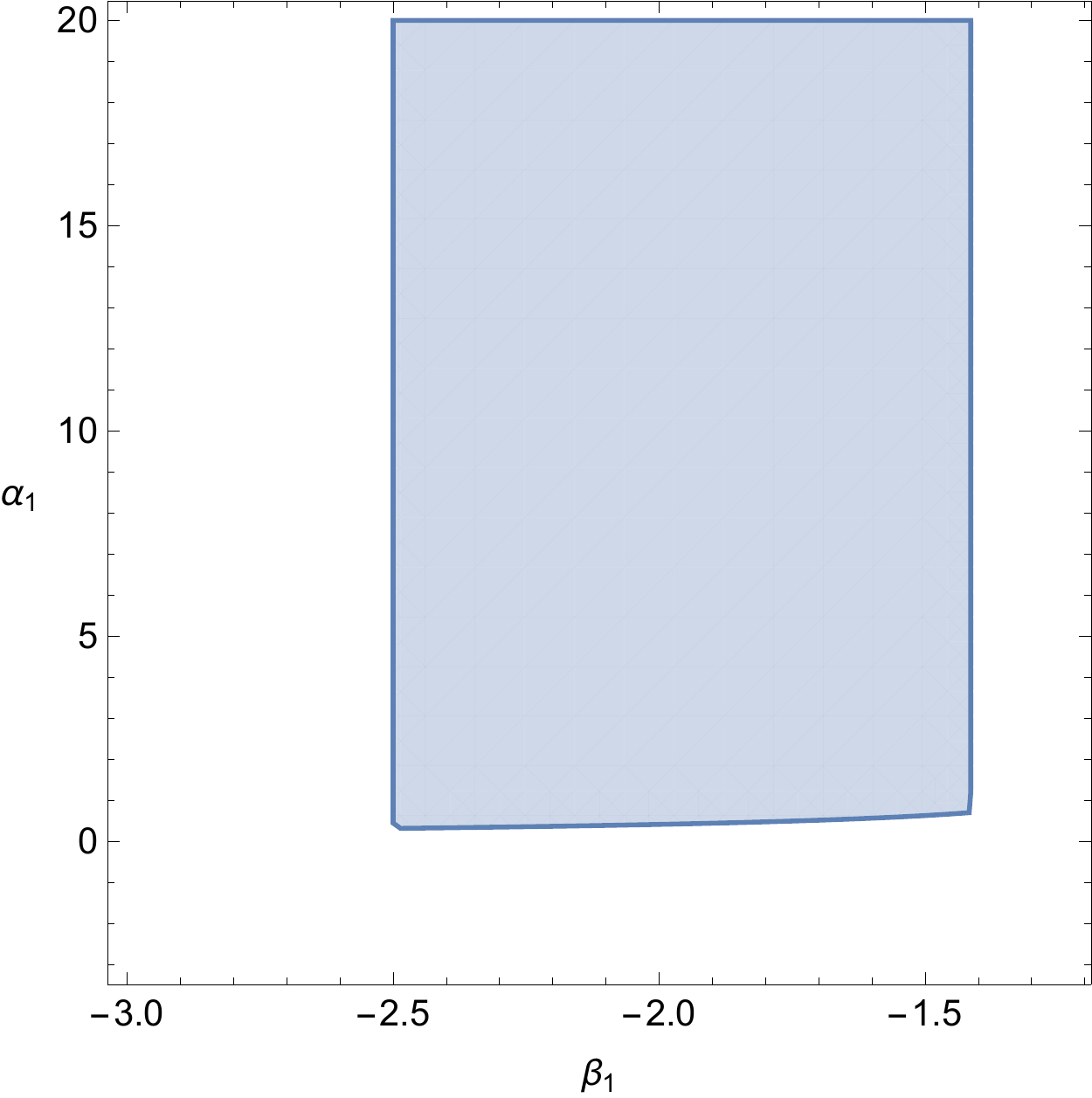}
 \end{subfigure} 
 \caption{The stable node regions for the conformal--disformal dust and conformal--disformal radiation case, illustrate the parameter values of $\alpha_1$ and $\beta_1$ for fixed points $(3)_{(d)}$ (left), and $(4)_{(d)}$ (right). The other free parameters for $(3)_{(d)}$ are chosen to be $(\beta_2,\lambda)=(0.5,5)$, and we set the parameters for $(4)_{(d)}$ to  $(\beta_2,\lambda)=(0.5,-5)$, so that all eigenvalues have a negative real part.}
 
\label{fig:dust_rad}
\end{figure}
\item The conformal dust kinetic fixed point, $(6)_{(d)}$, is a stable node when either of the following conditions are satisfied
\begin{eqnarray*}
-\frac{1}{\sqrt{2}}&<&\alpha_1<0\,, \hspace{0.5cm} \lambda>\frac{-3-2\alpha_1^2}{2\alpha_1}\;, \hspace{0.5cm} \beta_1<\frac{-3-2\alpha_1^2}{4\alpha_1}\;, \hspace{0.5cm}\beta_2<\frac{-3-2\alpha_1^2}{4\alpha_1}\;,\\
\text{or}, \hspace{0.5cm} 0&<&\alpha_1<\frac{1}{\sqrt{2}}\,,\hspace{0.5cm}\lambda<\frac{-3-2\alpha_1^2}{2\alpha_1}\;,\hspace{0.5cm} \beta_1>\frac{-3-2\alpha_1^2}{4\alpha_1}\;,\hspace{0.5cm} \beta_2>\frac{-3-2\alpha_1^2}{4\alpha_1}\;.
\end{eqnarray*}

\item For the disformal dust radiation fixed point, $(b)$, we know that this exists if $\beta_1^2>2$, and furthermore this two fluid disformal fixed point is found to be a saddle point. 
\item The existence of fixed point $(3)_{(d)}$ requires $\beta_1\geq\sqrt{3/2}$. By imposing that $e_{1,2}<0$ we get that $\beta_2<\beta_1,\;\text{and}\;\lambda>2\beta_1$. The constraint on $\alpha_1$ in terms of $\beta_1$ is obtained from $e_{3,4,5}$. An illustration of some allowed parameter values is shown in Fig \ref{fig:dust_rad}.
\begin{figure}[h!]
\begin{center}
  \includegraphics[width=0.4\textwidth]{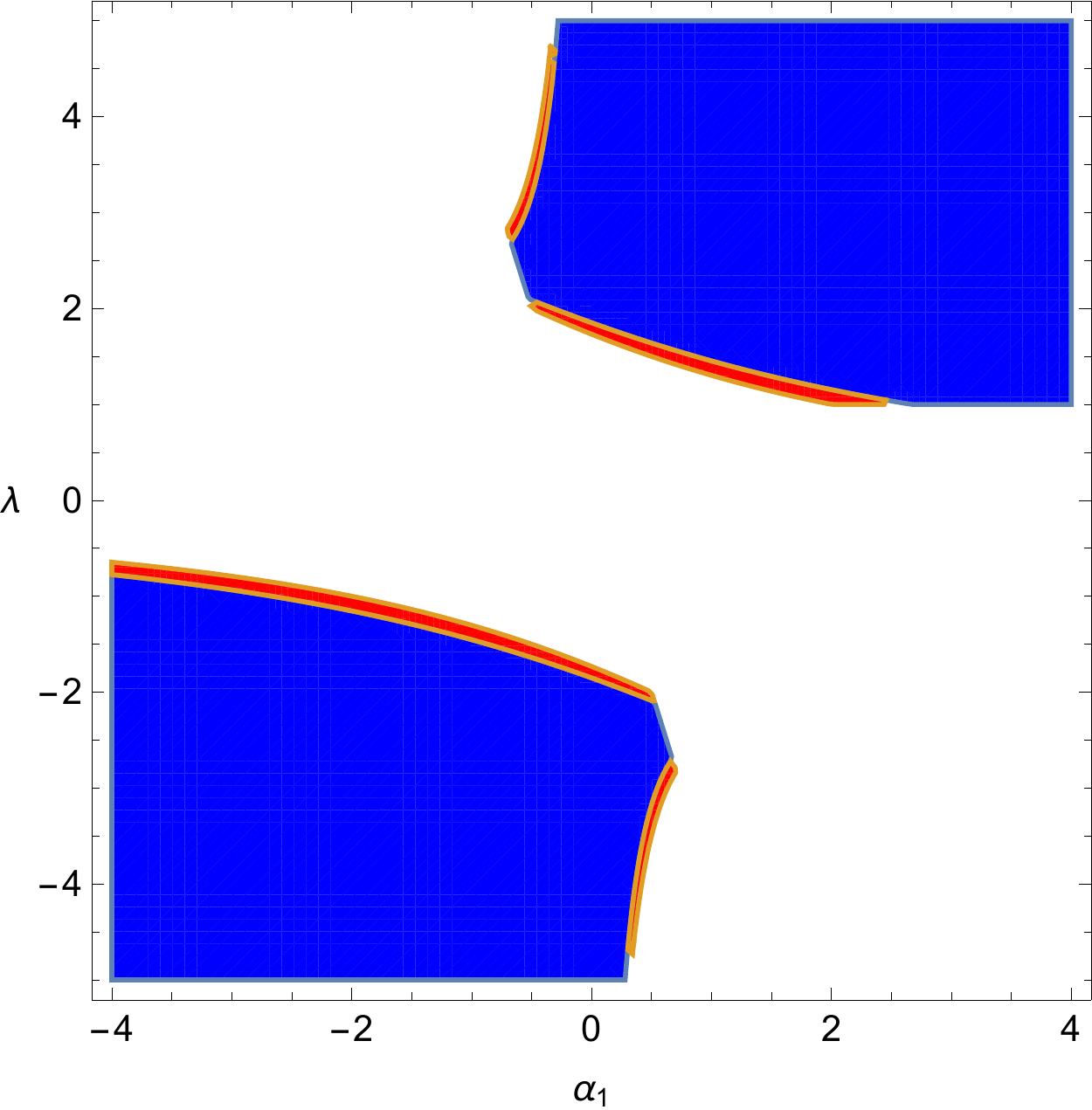}
  \includegraphics[width=0.4\textwidth]{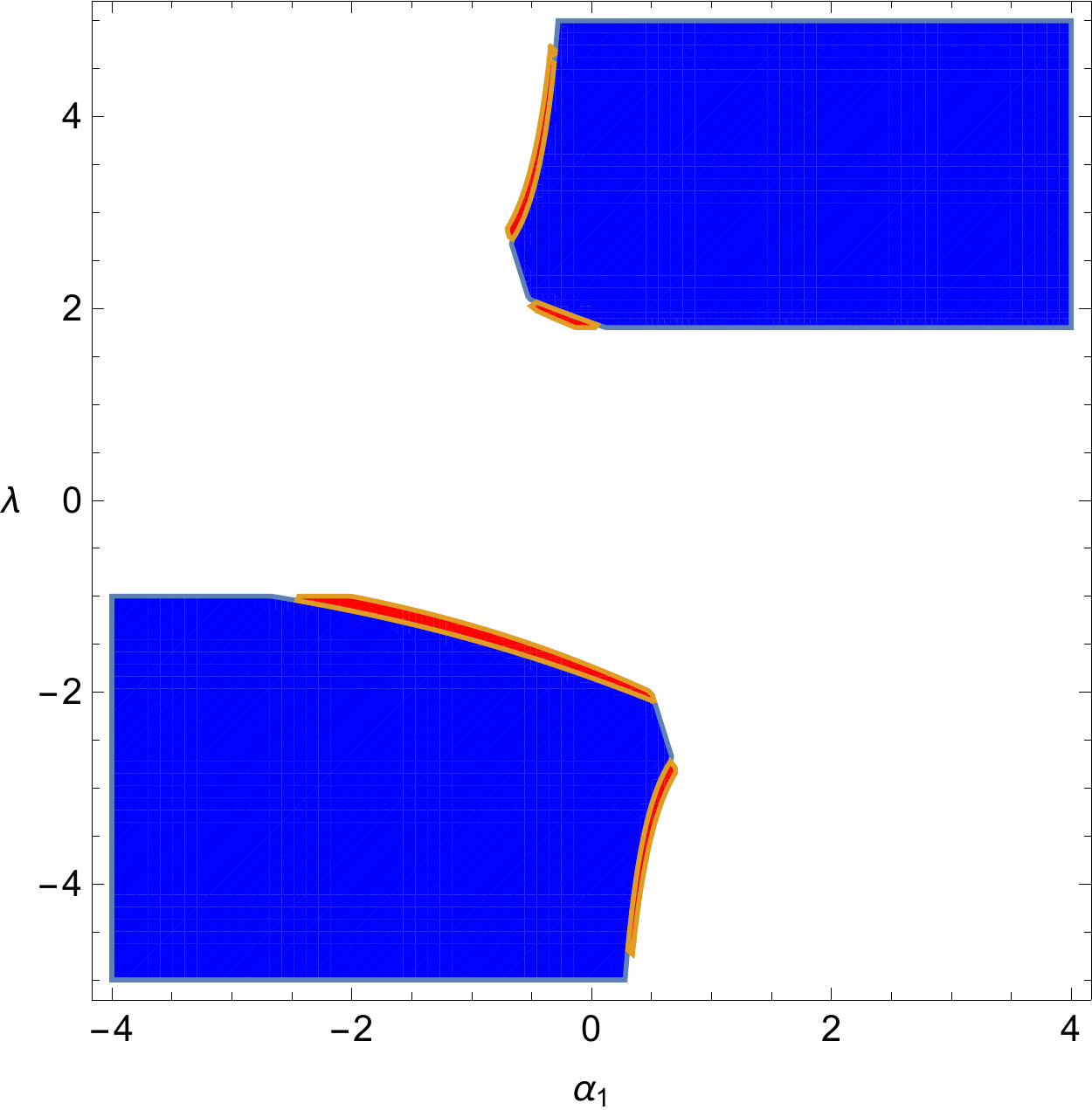}
\end{center}
\begin{center}
\includegraphics[width=0.54\textwidth]{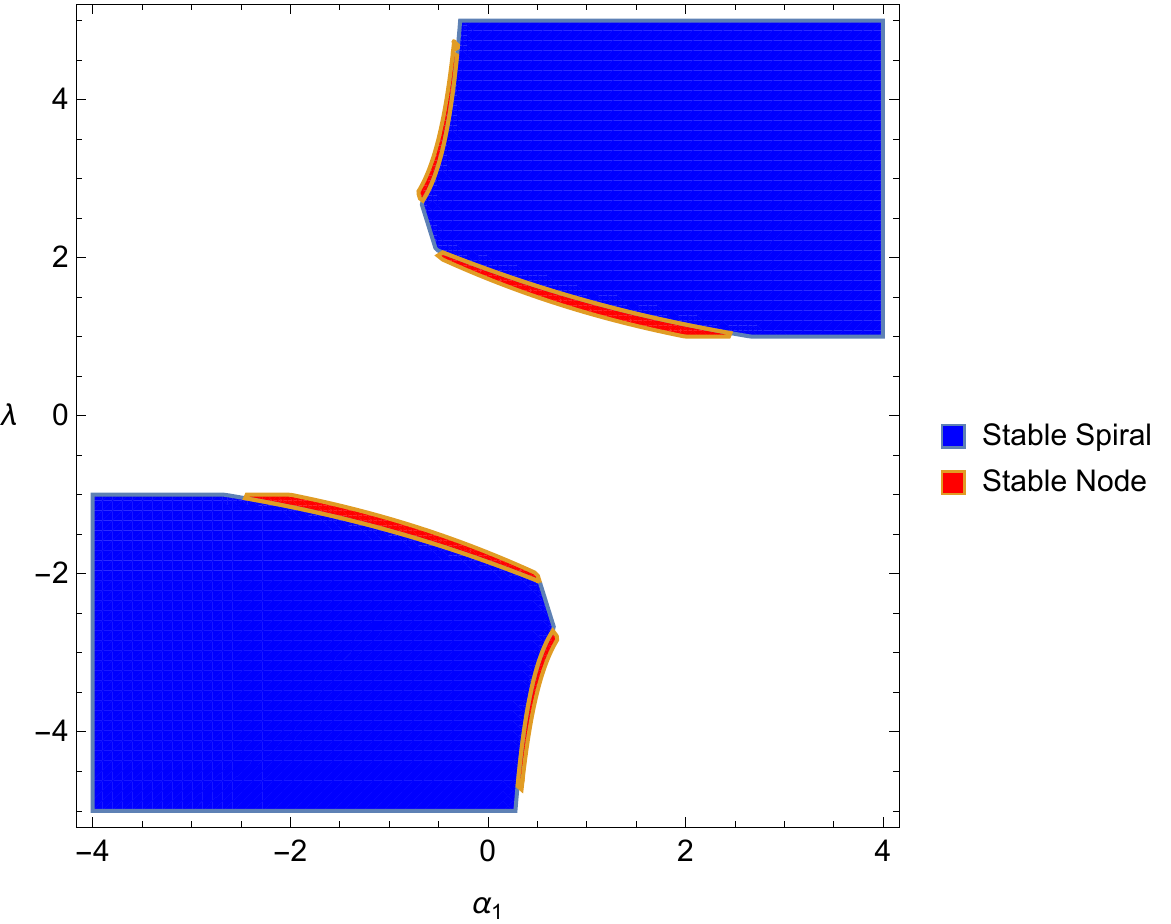}
\caption{The stable regions for the conformal--disformal dust and conformal--disformal radiation scenario, illustrate the parameter values of $\alpha_1$ and $\lambda$ for fixed point $(8)_{(d)}$ when $(\beta_1,\beta_2)=(0.1,0.5)$ (left), $(\beta_1,\beta_2)=(-0.5,0.9)$ (right), and $(\beta_1,\beta_2)=(-0.5,0.5)$ (bottom).} 
\label{fig:full_sys}
\end{center}
\end{figure}
\item The existence of the other disformal fixed point, $(4)_{(d)}$, implies that $\beta_1\leq-\sqrt{3/2}$. Furthermore, $e_1,\;e_2<0$ give $\beta_2>\beta_1,\;\text{and}\;\lambda<2\beta_1$. From $e_{3,4,5}$ we get constraints on the choice of $\alpha_1$ in terms of $\beta_1$. An illustration of some values is given in Fig \ref{fig:dust_rad}.
\item The only non-singular disformal fixed point, $(5)_{(r)}$, is a saddle point when $\beta_2^2>2$. This is due to the opposite signs of $e_3$ and $e_4$. 
\item  For the next disformal fixed point, $(3)_{(r)}$, we find that this is a stable node when the following inequalities are satisfied
\begin{equation*}
\beta_2>\sqrt{2}\;, \hspace{1cm}\;\beta_1<\beta_2\;,\hspace{1cm} \alpha_1<-\beta_2+\sqrt{\frac{2\beta_2^2-3}{2}}\;,\hspace{1cm} \lambda>2\beta_2\;.
\end{equation*}
\item Similarly, fixed point $(4)_{(r)}$ is found to be a stable node when the chosen parameters satisfy the following inequalities
\begin{equation*}
\beta_2<-\sqrt{2}\;,\;\hspace{1cm} \beta_1>\beta_2\;,\;\hspace{1cm} \alpha_1>-\beta_2-\sqrt{\frac{2\beta_2^2-3}{2}}\;,\hspace{1cm}\lambda<2\beta_2\;.
\end{equation*}
\item The conformal radiation scaling fixed point $(8)_{(r)}$ is found to be a stable node when either of the following holds
\begin{equation*}
\begin{split}
-\frac{8}{\sqrt{15}}&\leq\lambda<-2\,,\hspace{1cm} \beta_1>\frac{\lambda}{2}\;,\;\hspace{1cm} \beta_2>\frac{\lambda}{2}\;,\hspace{1cm}\;\alpha_1>-\frac{\lambda}{4}\;,\\
\text{or}, \hspace{1cm} 2&<\lambda\leq\frac{8}{\sqrt{15}}\;,\;\hspace{1cm} \beta_1<\frac{\lambda}{2}\;,\;\hspace{1cm} \beta_2<\frac{\lambda}{2}\;,\;\hspace{1cm} \alpha_1<-\frac{\lambda}{4}\;,
\end{split}
\end{equation*}
and is a stable spiral when either of the following is satisfied
\begin{equation*}
\begin{split}
\lambda&<-\frac{8}{\sqrt{15}}\;,\;\hspace{1cm}\beta_1>\frac{\lambda}{2}\;,\;\hspace{1cm}\beta_2>\frac{\lambda}{2}\;,\;\hspace{1cm}\alpha_1>-\frac{\lambda}{4}\;,\\
\text{or},\;\hspace{1cm} \lambda&>\frac{8}{\sqrt{15}}\;,\;\hspace{1cm} \beta_1<\frac{\lambda}{2}\;,\;\hspace{1cm}\beta_2<\frac{\lambda}{2}\;,\;\hspace{1cm}\alpha_1<-\frac{\lambda}{4}\;.
\end{split}
\end{equation*}
\item The scalar field dominated fixed point (7) is a stable node in the following regions 
\begin{equation*}
\begin{split}
-2&<\lambda<0\;,\;\hspace{1cm} \beta_1>\frac{\lambda}{2}\;,\;\hspace{1cm} \beta_2>\frac{\lambda}{2}\;,\hspace{1cm}\;\alpha_1>\frac{3-\lambda^2}{\lambda}\;,\\
\text{or},\hspace{1cm}0&<\lambda<2\;,\;\hspace{1cm}\beta_1<\frac{\lambda}{2}\;,\;\hspace{1cm}\beta_2<\frac{\lambda}{2}\;,\hspace{1cm} \;\alpha_1<\frac{3-\lambda^2}{\lambda}\;.
\end{split}
\end{equation*}
\item For the conformal dust scaling fixed point, $(8)_{(d)}$, an illustration of the possible values of $\alpha_1$ and $\lambda$, for some fixed values of the other parameters that render this point stable, is given in Fig \ref{fig:full_sys}. Indeed, we find that this point can either be a stable node or a stable spiral.    
\end{itemize}
We will be interested in cosmologically acceptable trajectories, by which we mean that the trajectory should start in the radiation era, then evolve to a matter dominated era, and finally reproduce our present day accelerating Universe. In our examples, we will use $\Omega_{0,m}\simeq0.308$, $\Omega_{0,\phi}\simeq0.692$, $H_0\simeq67.8\,\text{km}\,\text{s}^{-1}\,\text{Mpc}^{-1}$, and $w_{0,\phi}\simeq-1$ as our present day cosmological parameters \cite{P2}. These imply that the present values of the dynamical system variables should be $x_0\simeq 0$, $y_0\simeq 0.832$, $z_{0,1}\simeq 0.555$. Furthermore, since the scalar field plays an important role in the late time Universe, the trajectories in the radiation dominated epoch should start near $x=y=0$. We give an example in Fig \ref{fig:Omega7}, showing the evolution of $\Omega_i$ and $\rho_i$ such that the future attractor is the scalar field dominated fixed point (7). In this plot we compare the conformally coupled scenario with the disformally coupled case by evolving the equations from the same initial conditions. We also define a time-dependent effective mass scale, 
\begin{equation}
\label{meff}
M_\text{eff}=\left(|D_r-D_m|\right)^{-1/4}\;.
\end{equation}
This effective mass scale is also plotted in Fig \ref{fig:Omega7}.
\par 
By assuming standard  Big-Bang Nucleosynthesis (BBN), we can use a conservative constraint of $\Omega_\phi(\text{MeV})<0.2$ \cite{Ferreira:1997hj,Cyburt:2004yc} (a tighter constraint of $\Omega_\phi(\text{MeV})<0.045$ was obtained in Ref. \cite{PhysRevD.64.103508}) to limit the range of the parameters $\alpha_1,\;\beta_2,$ and $\lambda$. Indeed, if we further assume that the non-singular fixed points $(a),\;(5)_{(r)},$ and $(8)_{(r)}$ are reached by BBN, we get, from $\Omega_\phi <0.2$, that $\alpha_1^2>5/6,\;\beta_2^2>10/3,$ and $\lambda^2>20$ respectively.
\subsection{Two Fluids--Conformal dust and conformal-disformal radiation}
\label{sec:ccd_dust_rad}
%
In this section we consider another particular case of  the general system discussed in Section \ref{sec:general_system}. Indeed, we will be interested in a conformally coupled perfect fluid with equation of state parameter, $\gamma_1$, defined in the conformal frame by the metric $\tilde{g}_{\mu\nu}^1$, in the presence of a distinct conformally--disformally coupled perfect fluid with equation of state parameter, $\gamma_2$, defined in the disformal frame by the metric $\tilde{g}_{\mu\nu}^2$. As in the single fluid case, we will be considering exponential couplings and scalar field potential, as follows 
\begin{equation}
C_i(\phi)=e^{2\alpha_i\kappa\phi},\; \hspace{1cm} D_1(\phi)=0,\;\hspace{1cm} D_2(\phi)=\frac{e^{2(\alpha_2+\beta)\kappa\phi}}{M^4},\;\hspace{1cm}V(\phi)=V_0^4 e^{-\lambda\kappa\phi}.
\end{equation}
This system can be viewed as a reduced phase space analysis of the previous higher-dimensional system of Section \ref{sec:cdcd_dust_rad}. Indeed, this dynamical system is four-dimensional, i.e., one dimension less than the previous case. One might think that the results could simply be obtained  from the previous case  by setting $\beta_1 \rightarrow -\infty$. We will see in the following that if we do this, we would be missing out some of the stability conditions.
For simplicity, we choose to eliminate $z_2$ from our system of differential equations, and hence, we end up with four ordinary differential equations for $x,\;y,\;z_1$ and $\sigma_2$. The fixed points for this system are tabulated in Table \ref{table5}. This is a generalisation of the fixed points found in the single fluid case with equation of state parameter, $\gamma$, tabulated in Table \ref{table1}.\vspace{10pt}\\
We will now specify this system to conformally coupled dust and conformally--disformally coupled radiation, i.e. $\gamma_1=1$ and $\gamma_2=4/3$. As expected, the fixed points for this particular system are found to be contained in the fixed points discussed in the previous case when we considered radiation and dust to be both conformally and disformally coupled. Indeed, the fixed points for this system are $(1),\;(6)_{(r)},\;(2),\;(a),\;(6)_{(d)},\;(5)_{(r)},\;(3)_{(r)},\;(4)_{(r)},\;(8)_{(r)},\;(7),$ and $(8)_{(d)}$. Since the existence analysis of these fixed points coincides with that presented in Section \ref{sec:cdcd_dust_rad_existence}, we only discuss the stability of these fixed points. All sets of eigenvalues, $e_{1,2,3,4}$, for each fixed point can be found in Appendix \ref{appendix:c2}.    
\begin{figure}[t!]
\begin{center}
  \includegraphics[width=0.4\textwidth]{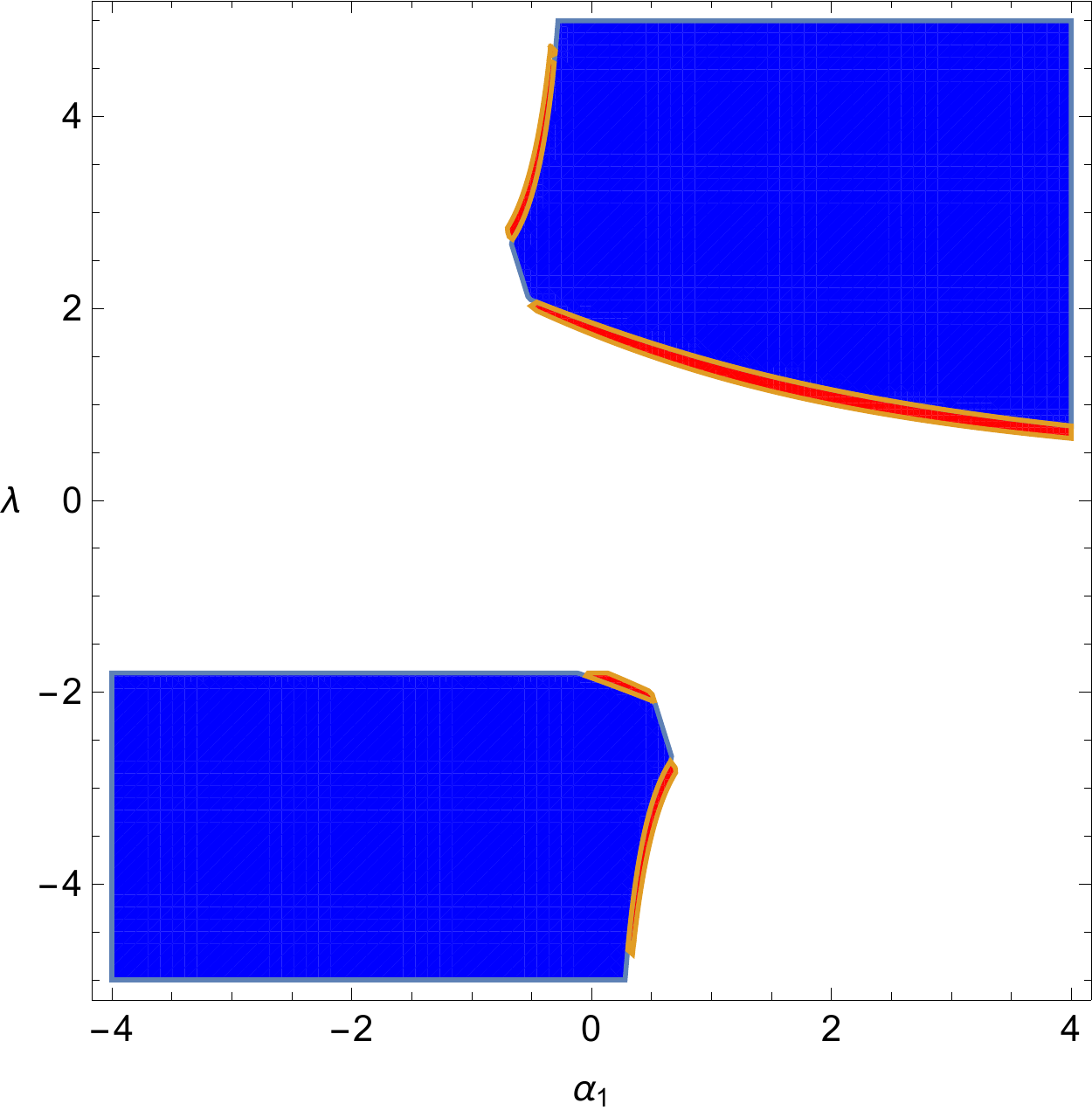}
  \includegraphics[width=0.4\textwidth]{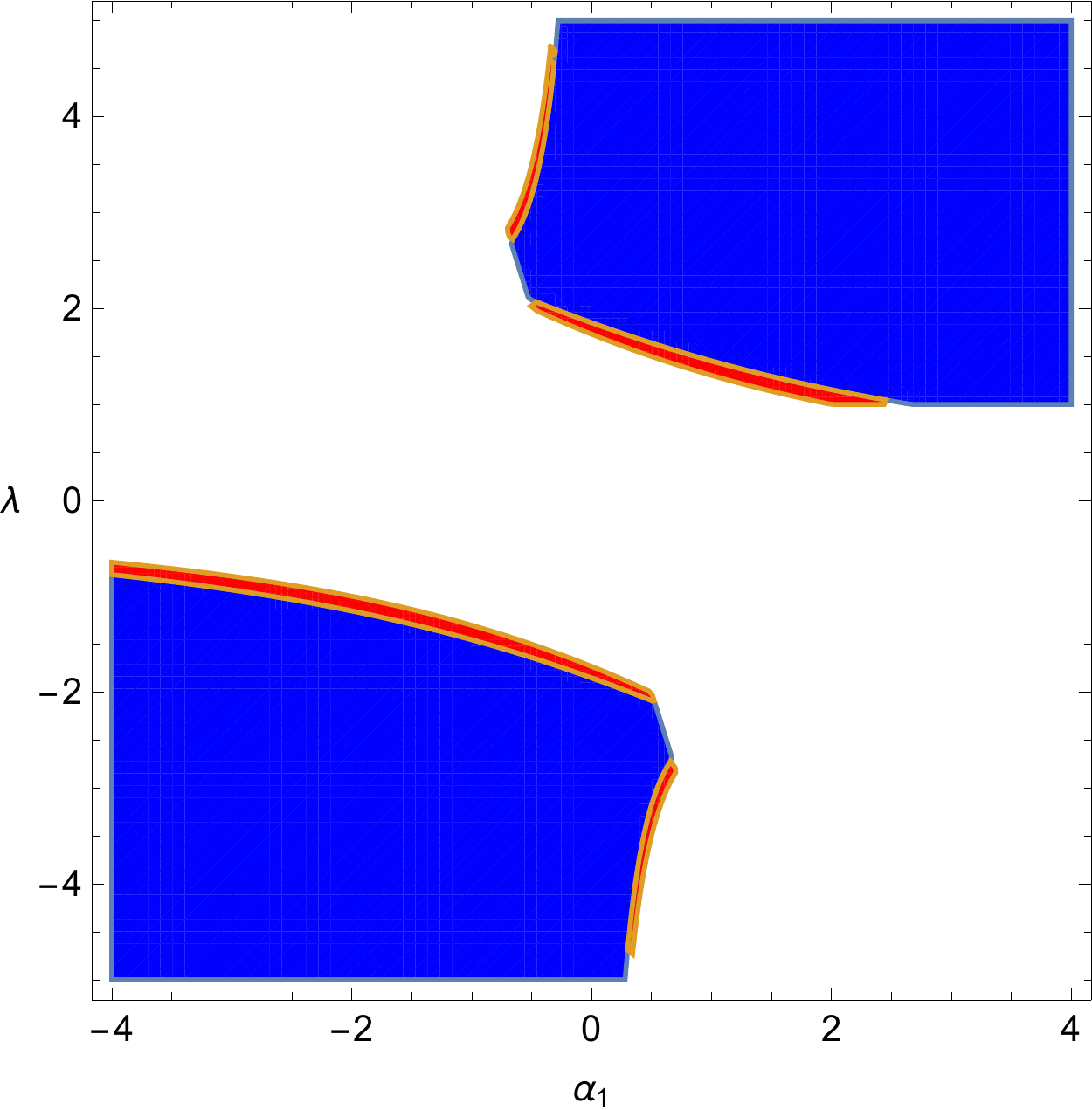}
\end{center}
\begin{center}
\includegraphics[width=0.54\textwidth]{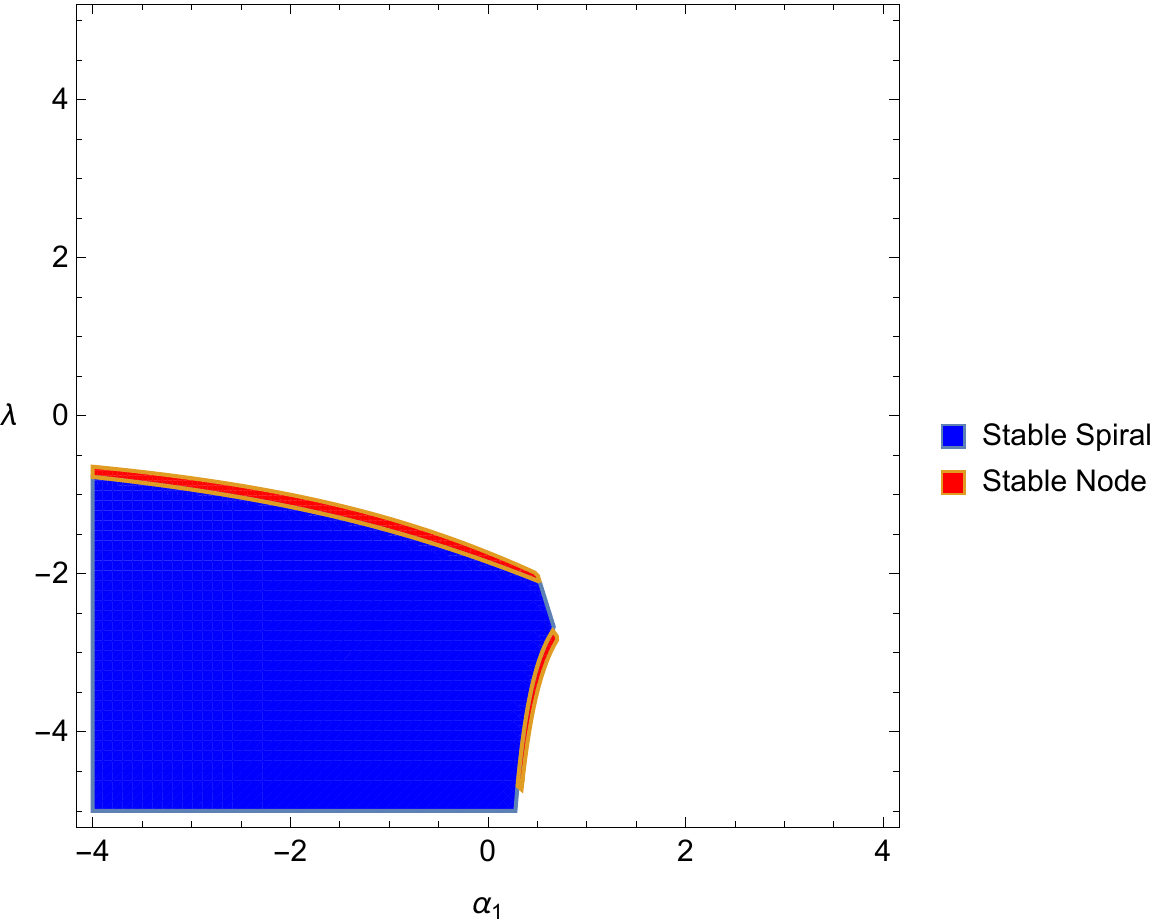}
\caption{The stable regions for the conformal dust and conformal--disformal radiation case, illustrate the parameter values of $\alpha_1$ and $\lambda$ for fixed point $(8)_{(d)}$ when $\beta=-0.9$ (left), $\beta=0.5$ (right), and $\beta=5$ (bottom).}  
\label{fig:beta}
\end{center}
\end{figure}
\begin{figure}[h!]
\centering
\begin{subfigure}[b]{0.495\textwidth}
  \includegraphics[width=1\textwidth]{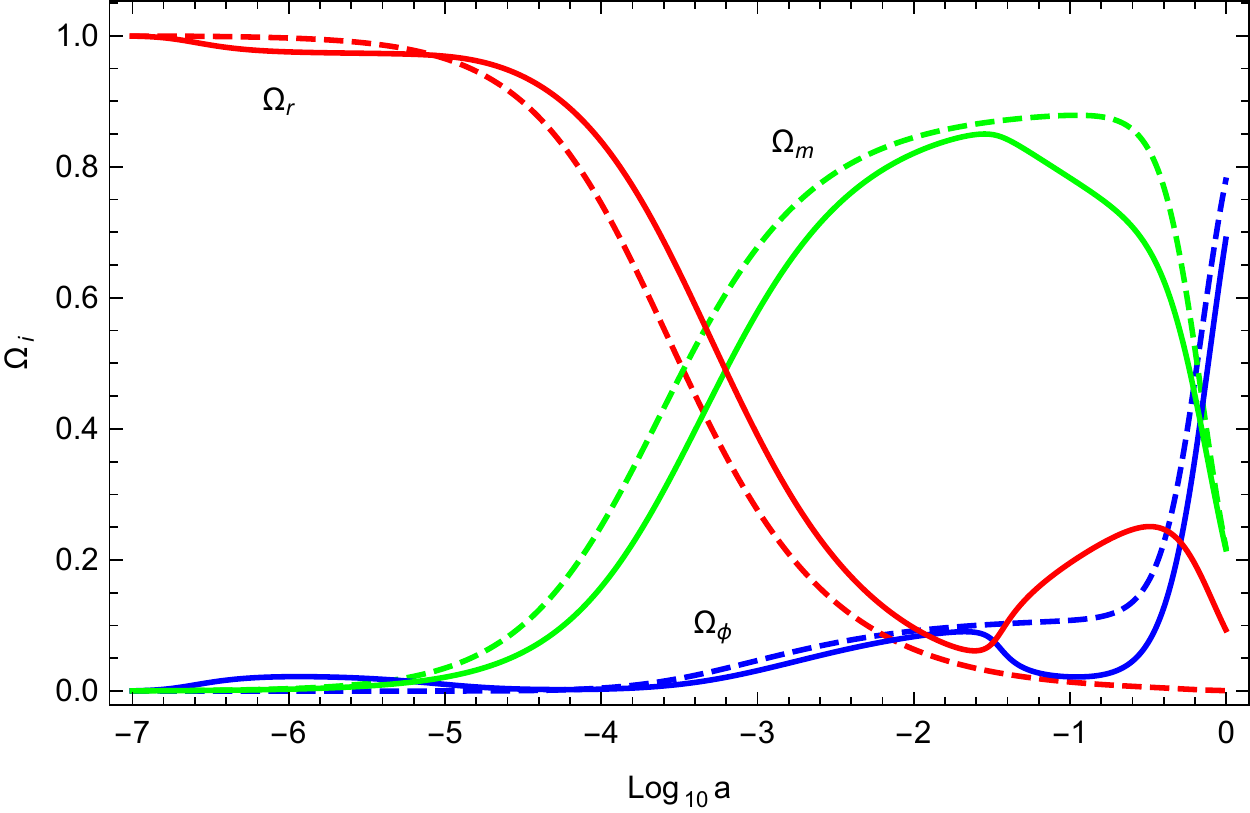} 
\end{subfigure}
\begin{subfigure}[b]{0.495\textwidth}
  \includegraphics[width=1\textwidth]{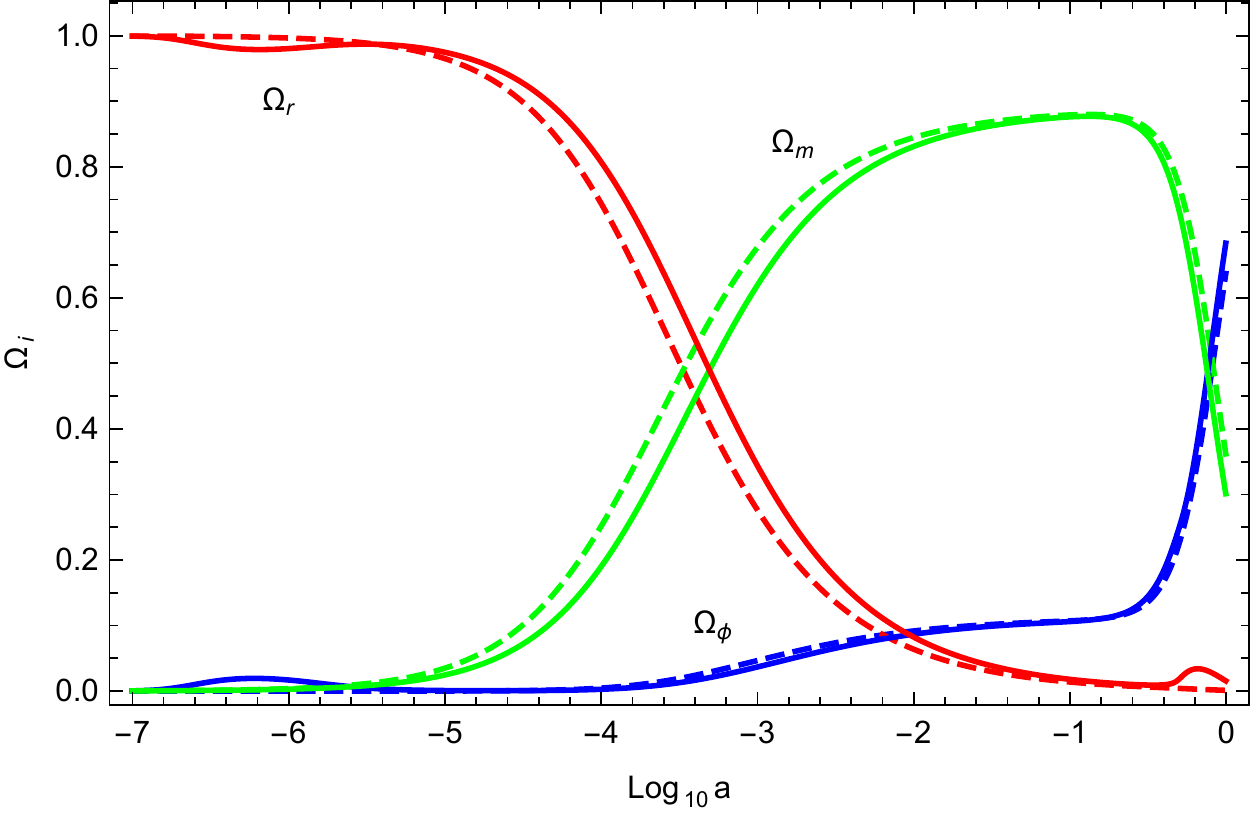} 
\end{subfigure}
\begin{subfigure}[b]{0.495\textwidth}
  \includegraphics[width=1\textwidth]{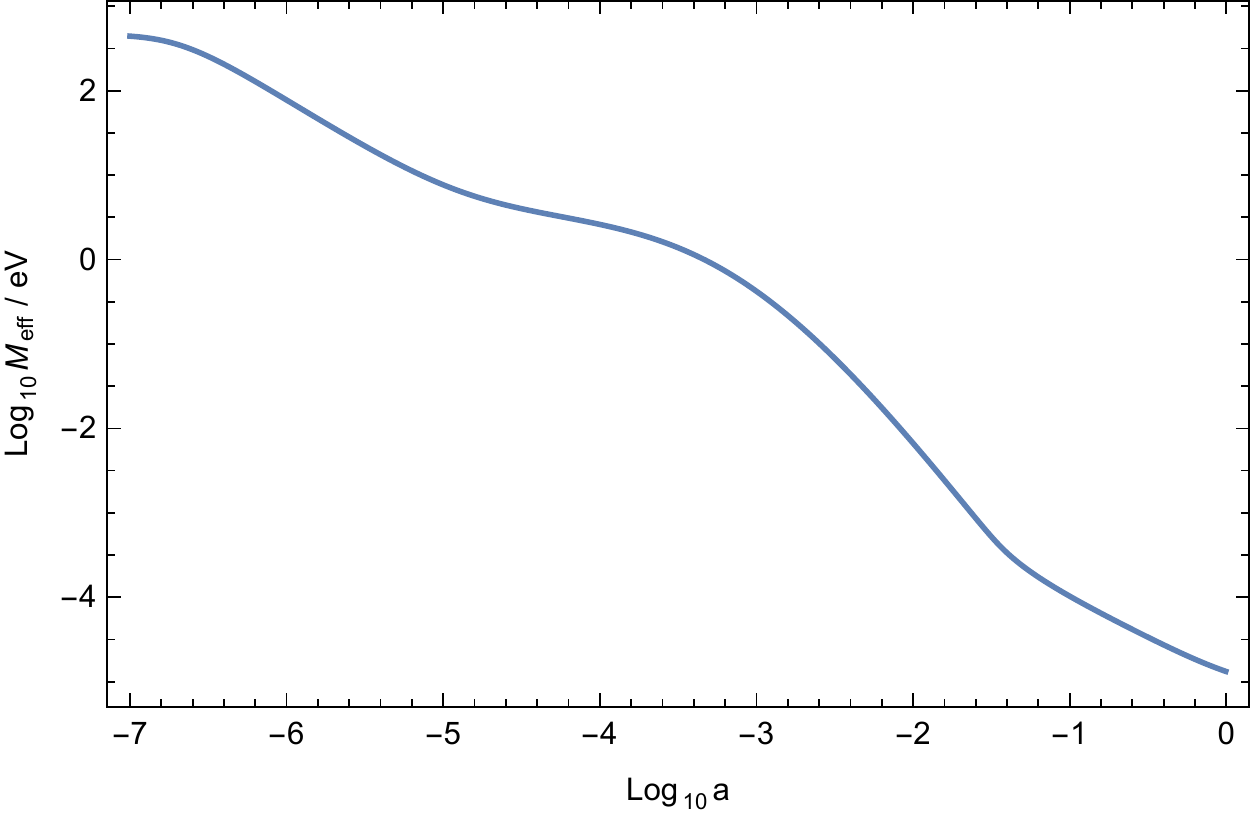} 
  \caption{•}
\end{subfigure}
\begin{subfigure}[b]{0.495\textwidth}
  \includegraphics[width=1\textwidth]{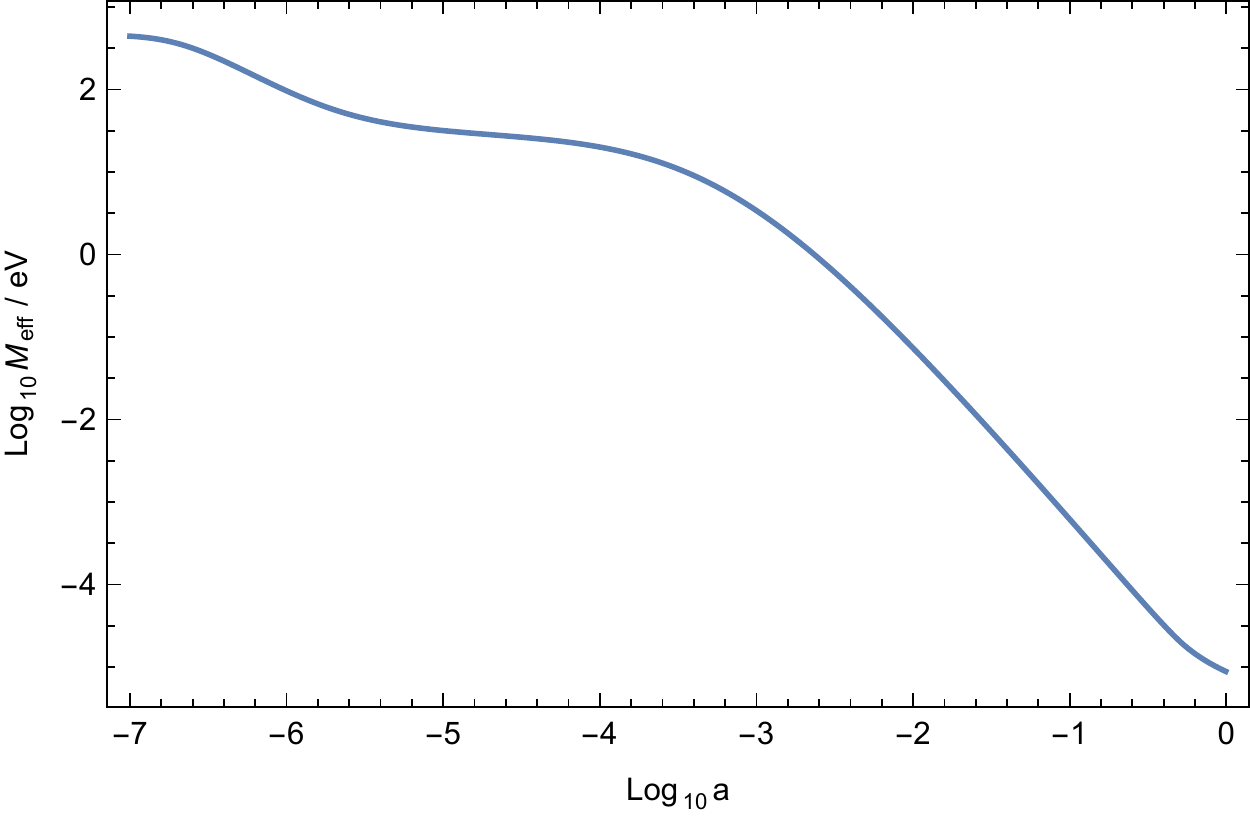} 
  \caption{•}
\end{subfigure}
\caption{The above plots show the evolution of $\Omega_{r,m,\phi}$ and $M_{\text{eff}}$ when the future attractor is  $(7)$, in the conformal dust and conformal--disformal radiation case. In (a) we use the same parameters as those used in Fig \ref{fig:Omega7}, i.e., $\alpha_1=-0.41,\;\beta=5.11,\;\lambda=-0.08$, whereas in (b) we use $\alpha_1=-0.41,\;\beta=4.41,\;\lambda=-0.08$. In each plot we use the same initial conditions and compare the purely conformal case (dashed line) with the conformal dust and conformal--disformal radiation case (solid line) as discussed in Section \ref{sec:ccd_dust_rad}.} 
\label{fig:Omega7r}
\end{figure}
%
\subsubsection{Stability Conditions}
We will now discuss the stability of the fixed points of this reduced system. This analysis will be slightly different from the previous, since we now have four eigenvalues. In the stability analysis that follows, we will only comment on those regions where the fixed point is stable.
\begin{itemize}
\item The scalar field kinetic energy dominated fixed points, (1) and (2), both have $e_1>0$ and hence cannot be stable. The radiation fluid dominated fixed point, $(6)_{(r)}$, can only be a saddle point, since $e_1,\;e_3<0$ and $e_2,\;e_4>0$. 
\item Fixed point $(a)$ is found to be a stable node when
\begin{equation*}
\begin{split}
-\sqrt{\frac{2}{3}}&\leq\alpha_1<-\frac{1}{\sqrt{2}}\;, 
\hspace{1cm} \;\lambda>-4\alpha_1\;,\;\hspace{1cm} \beta<-2\alpha_1,\\
\text{or},\hspace{1cm}\frac{1}{\sqrt{2}}&<\alpha_1\leq\sqrt{\frac{2}{3}}\;,\hspace{1cm}\;\lambda<-4\alpha_1\;,\;\hspace{1cm} \beta>-2\alpha_1\;,
\end{split}
\end{equation*}
and a stable spiral when 
\begin{equation*}
\begin{split}
\alpha_1&<-\sqrt{\frac{2}{3}}\;,\hspace{1cm}\;\beta<-2\alpha_1\;,\;\hspace{1cm}\lambda>-4\alpha_1, \\
\text{or},\hspace{1cm}\alpha_1&>\sqrt{\frac{2}{3}}\;,\;\hspace{1cm}\beta>-2\alpha_1\;,\;\hspace{1cm}\lambda<-4\alpha_1\;.
\end{split}
\end{equation*}
\item Fixed point $(6)_{(d)}$ is a stable node when one of the following set of inequalities is satisfied
\begin{equation*}
\begin{split}
-\frac{1}{\sqrt{2}}&<\alpha_1<0\;,\hspace{1cm}\lambda>\frac{-3-2\alpha_1^2}{2\alpha_1}\;,\hspace{1cm}\;\beta<\frac{-3-2\alpha_1^2}{4\alpha_1}\;,\\
\text{or},\hspace{1cm}0&<\alpha_1<\frac{1}{\sqrt{2}}\;,\hspace{1cm}\;\lambda<\frac{-3-2\alpha_1^2}{2\alpha_1}\;,\hspace{1cm}\;\beta>\frac{-3-2\alpha_1^2}{4\alpha_1}\;.
\end{split}
\end{equation*}
\item The mixed fixed point, $(5)_{(r)}$, can only be a saddle point in the existence range of $\beta^2>2$. The other disformal fixed point, $(3)_{(r)}$, is found to be a stable node when 
\begin{equation*}
\beta>\sqrt{2}\;,\;\hspace{1cm}\alpha_1<-\beta+\sqrt{\frac{2\beta^2-3}{2}}\;,\;\hspace{1cm}\lambda>2\beta\;.
\end{equation*}
The last disformal fixed point, $(4)_{(r)}$, can also be a stable node if the following inequalities are satisfied
\begin{equation*}
\beta<-\sqrt{2}\;,\;\hspace{1cm}\alpha_1>-\beta-\sqrt{\frac{2\beta^2-3}{2}}\;,\;\hspace{1cm}\lambda<2\beta\;.
\end{equation*}
\item The radiation scaling fixed point, $(8)_{(r)}$, is a stable node when
\begin{equation*}
\begin{split}
-\frac{8}{\sqrt{15}}&\leq\lambda<-2\;,\;\hspace{1cm}\beta>\frac{\lambda}{2}\;,\;\hspace{1cm}\alpha_1>-\frac{\lambda}{4},\\
\text{or},\hspace{1cm}2&<\lambda\leq\frac{8}{\sqrt{15}}\;,\;\hspace{1cm}\beta<\frac{\lambda}{2}\;,\;\hspace{1cm}\alpha_1<-\frac{\lambda}{4}\;,
\end{split}
\end{equation*}
and it can also be a stable spiral when the following inequalities are satisfied
\begin{equation*}
\begin{split}
\lambda&<-\frac{8}{\sqrt{15}}\;,\;\hspace{1cm}\beta>\frac{\lambda}{2}\;,\;\hspace{1cm}\alpha_1>-\frac{\lambda}{4},\\
\text{or},\hspace{1cm}\lambda&>\frac{8}{\sqrt{15}}\;,\;\hspace{1cm}\beta<\frac{\lambda}{2}\;,\;\hspace{1cm}\alpha_1<-\frac{\lambda}{4}\;.
\end{split}
\end{equation*}
\item The scalar field dominated fixed point, (7), is a stable node in the following regions
\begin{equation*}
\begin{split}
-2&<\lambda<0\;,\hspace{1cm}\;\beta>\frac{\lambda}{2}\;,\;\hspace{1cm}\alpha_1>\frac{3-\lambda^2}{\lambda}, \\
\text{or},\hspace{1cm} 0&<\lambda<2\;,\;\hspace{1cm}\beta<\frac{\lambda}{2}\;,\;\hspace{1cm}\alpha_1<\frac{3-\lambda^2}{\lambda}\;.
\end{split}
\end{equation*}
\item The conformal dust scaling fixed point, $(8)_{(d)}$, can either be a stable node or a stable spiral, as depicted in Fig. \ref{fig:beta}. Although this fixed point also appears in the previous two fluid and single fluid cases, the stable regions differ from those presented in Fig \ref{fig:3} and Fig \ref{fig:full_sys}. 
\end{itemize}  
We illustrate two examples in Fig \ref{fig:Omega7r}, showing the evolution of $\Omega_i$ and the effective mass, $M_{\text{eff}}$. In Fig \ref{fig:Omega7r} (a) we use the same parameters and initial conditions as those used in the example shown in Fig \ref{fig:Omega7}. It is evident that the radiation disformal coupling gives rise to a larger contribution to the radiation energy density at late times. Indeed, when the radiation disformal coupling exponent parameter is reduced, as depicted in Fig \ref{fig:Omega7r} (b), this enhanced contribution is diluted.
%
\subsection{Two Fluids--Two conformal-disformal dust components}
\label{sec:cdcd_dust_dust}
%
In this section, we study the full solution of the two fluid conformal--disformal system presented in Section \ref{sec:general_system} for the particular case of two dust components. Similar to the system discussed in Section \ref{sec:cdcd_dust_rad}, this system is also five-dimensional, in which we choose our dynamical variables to be $x,\;y,\;z_1,\;\sigma_1$ and $\sigma_2$. Furthermore, the couplings and scalar field potential are identical to those in (\ref{couplings_potential}). As expected, we recover the single fluid dust case fixed points for both components, although we obtain a \textit{conformal dust dominated} fixed point, $(c)$, in which neither of the fluids is subdominant. We list this fixed point in Table \ref{table10}. Fixed point $(c)$ is characterised by $x=y=\sigma_i=0$ and 
\begin{equation}
\sum_{i=1}^2 z_i^2\lambda_C^i(4-3\gamma_i)=0\;.
\end{equation} 
Furthermore, the radiation dominated fixed point $(6)_{(r)}$ is also obtained in this way, although in that case the radiation fluid dominates the solution. This dust dominated fixed point has already been studied in 
Ref. \cite{Brookfield:2007au,Baldi:2012kt,Amendola:2014kwa}. Indeed, the fixed points for this system are found to be the following: $(1),\;(c),\;(2),\;(6)^1_{(d)},\;(6)^2_{(d)},\;(3)^1_{(d)},\;(4)^1_{(d)},\;(3)^2_{(d)},$ $\;(4)^2_{(d)},\;(7),\;(8)^1_{(d)}$, and $(8)^2_{(d)}$. We use a superscript with the single fluid dust fixed point labels to indicate the dominant fluid, i.e. for superscript $i$ we have $z_i \neq 0$. We only comment on the existence of $(c)$, since for the other fixed points, this analysis follows directly from Section \ref{sec:single_fluid_existence}. This conformal dust dominated fixed point exists when either of the following holds
\begin{equation*}
\alpha_2>0,\;\alpha_1\leq0,\;\;\text{or},\;\;  \alpha_2=0,\;\alpha_1\neq0,\;\;\text{or},\;\;\alpha_2<0, \;\alpha_1\geq0\;.  
\end{equation*}
{\setlength\extrarowheight{9pt}
\setlength{\tabcolsep}{5.5pt}
\begin{table}[t!]
\begin{center}
\begin{tabular}{ c c c c c c c c c c c c c} 
 \hline
\hline
 Name   &  $x$ & $y$ & $z_1$ & $z_2$ & $\sigma_1$ & $\sigma_2$ & $\Omega_\phi$ & $w_\phi$ & $Z_1$ & $Z_2$ & $w_{\text{eff}}$ \\ 
\hline
$(c)$ & 0 & 0 & $\sqrt{\frac{\alpha_2}{\alpha_2-\alpha_1}}$ & $\sqrt{\frac{\alpha_1}{\alpha_1-\alpha_2}}$ & 0 & 0 & 0 & - & 1 & 1 & 0   \\ 
\hline
\end{tabular}
\end{center}
\caption{\label{table10} The relevant quantities of the conformal dust dominated fixed point obtained when considering two conformally--disformally coupled dust components.}
\end{table}}
\begin{figure}[h!]
\begin{center}
  \includegraphics[width=0.4\textwidth]{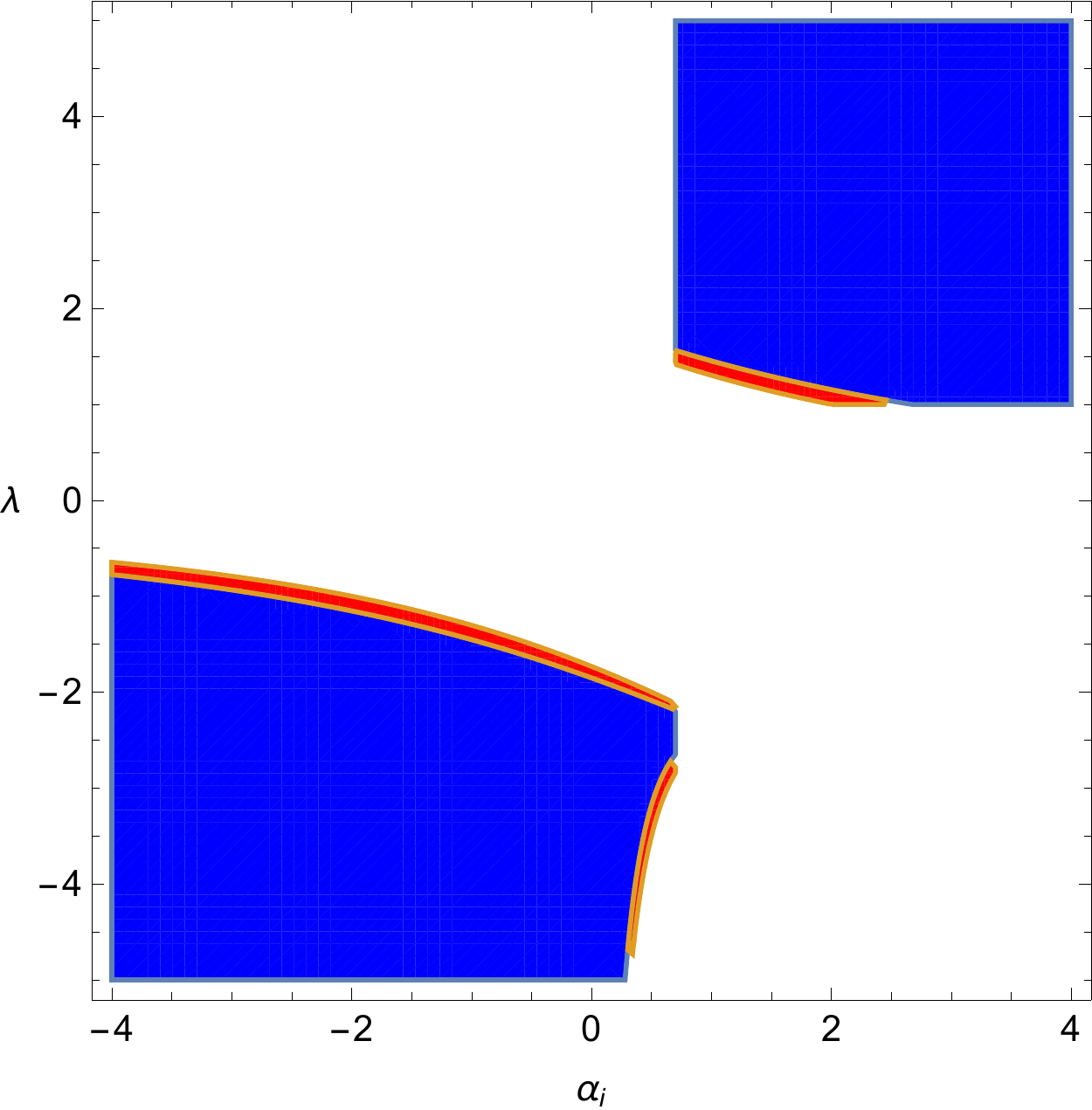}
  \includegraphics[width=0.4\textwidth]{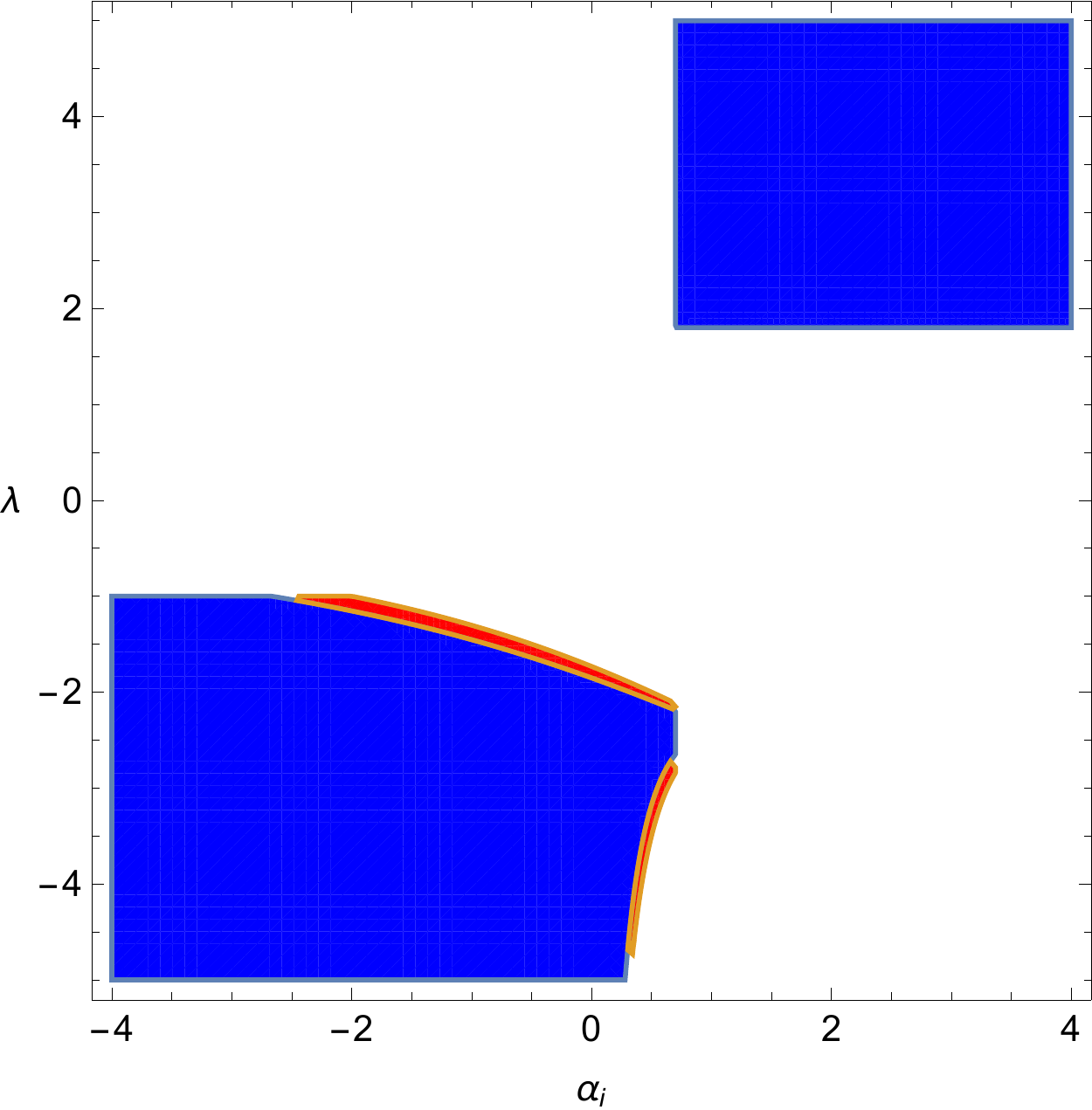}
\end{center}
\begin{center}
\includegraphics[width=0.54\textwidth]{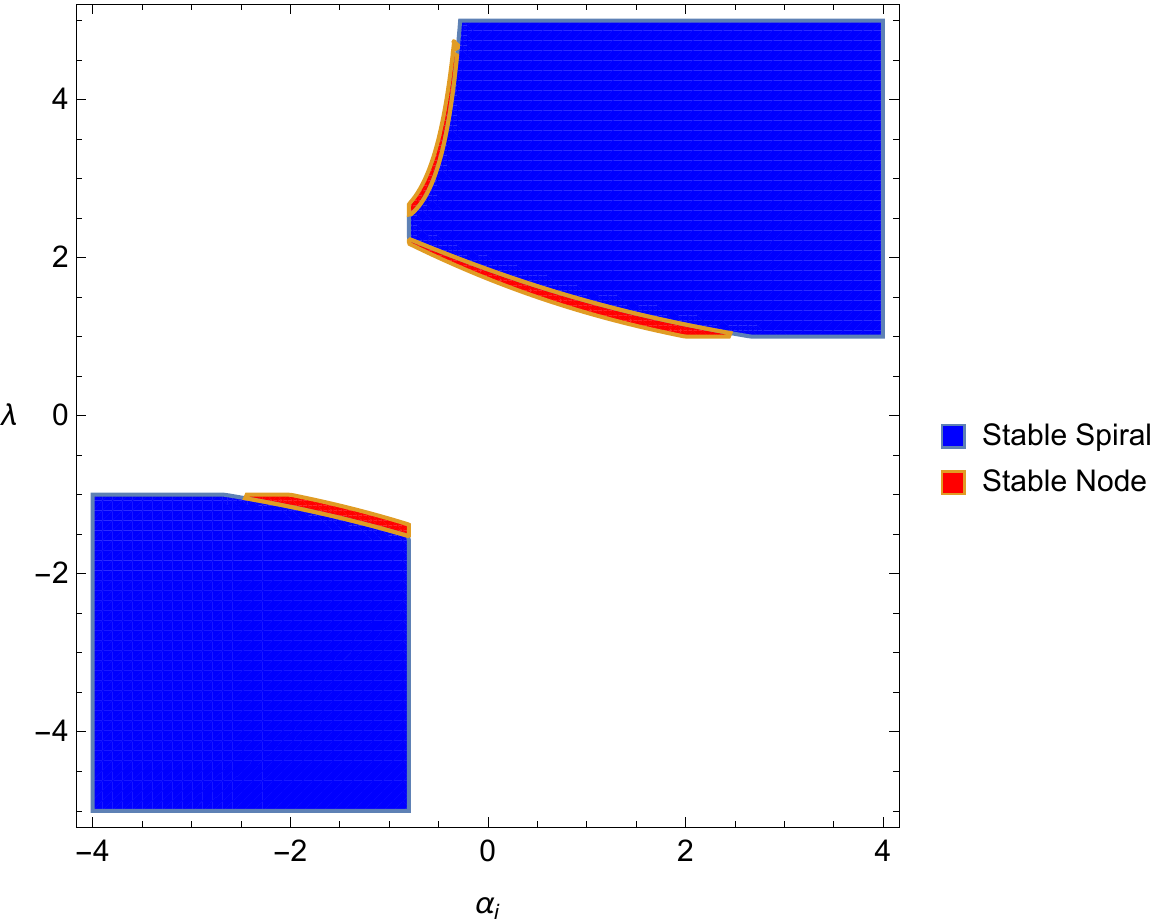}
\caption{The stable regions for a two conformal--disformal dust case, show an illustration of the parameter values of $\alpha_i$ and $\lambda$ for fixed points $(8)_{(d)}^{i=1,2}$ when $(\beta_1,\beta_2,\alpha_{j\neq i})=(0.1 , 0.5, 0.7)$ (left), $(\beta_1,\beta_2,\alpha_{j\neq i})=(-0.5 , 0.9, 0.7)$ (right), and $(\beta_1,\beta_2,\alpha_{j\neq i})=(-0.5 , 0.5, -0.8)$ (bottom).}
\label{fig:dust_dust1}
\end{center}
\end{figure}
%
\subsubsection{Stability Conditions}
\begin{figure}
\centering
\begin{subfigure}[b]{0.47\textwidth}
  \includegraphics[width=1\textwidth]{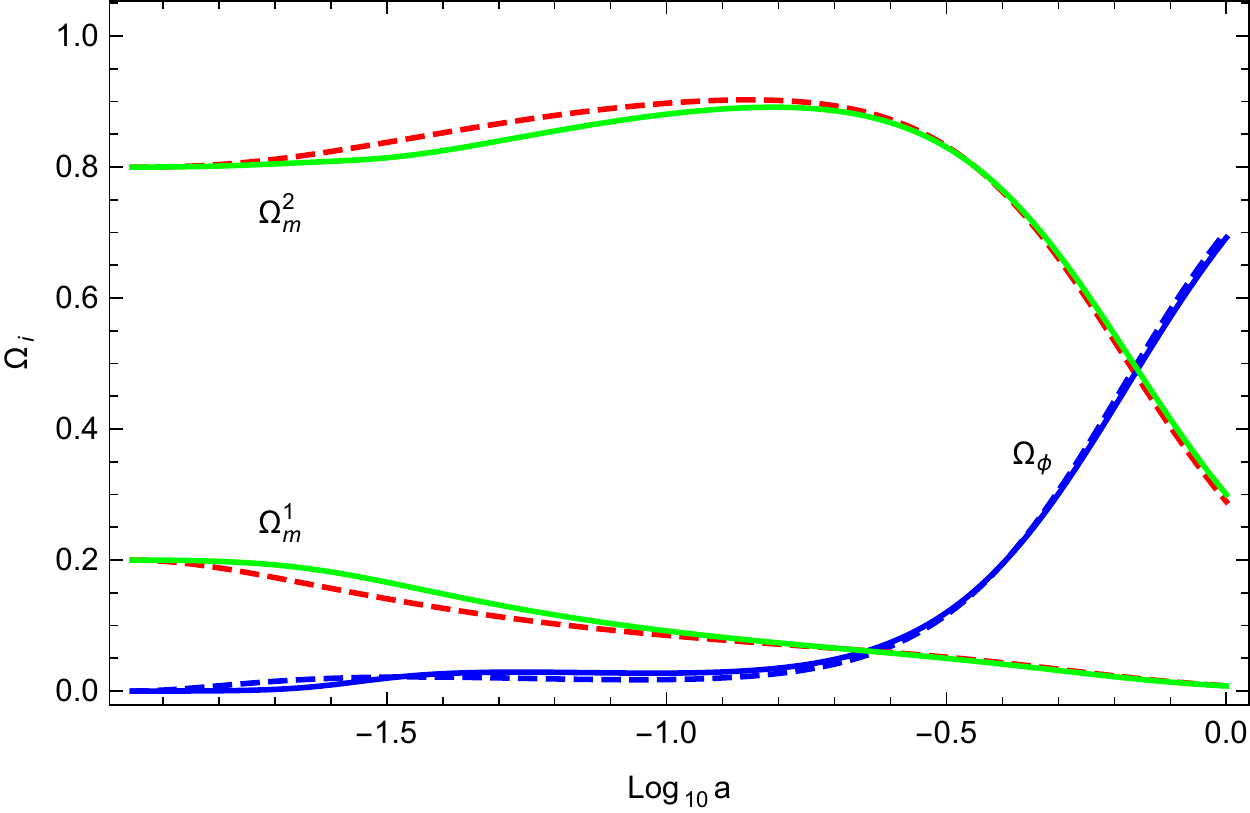}
   \caption{}
 \end{subfigure}
 \begin{subfigure}[b]{0.47\textwidth}
  \includegraphics[width=1\textwidth]{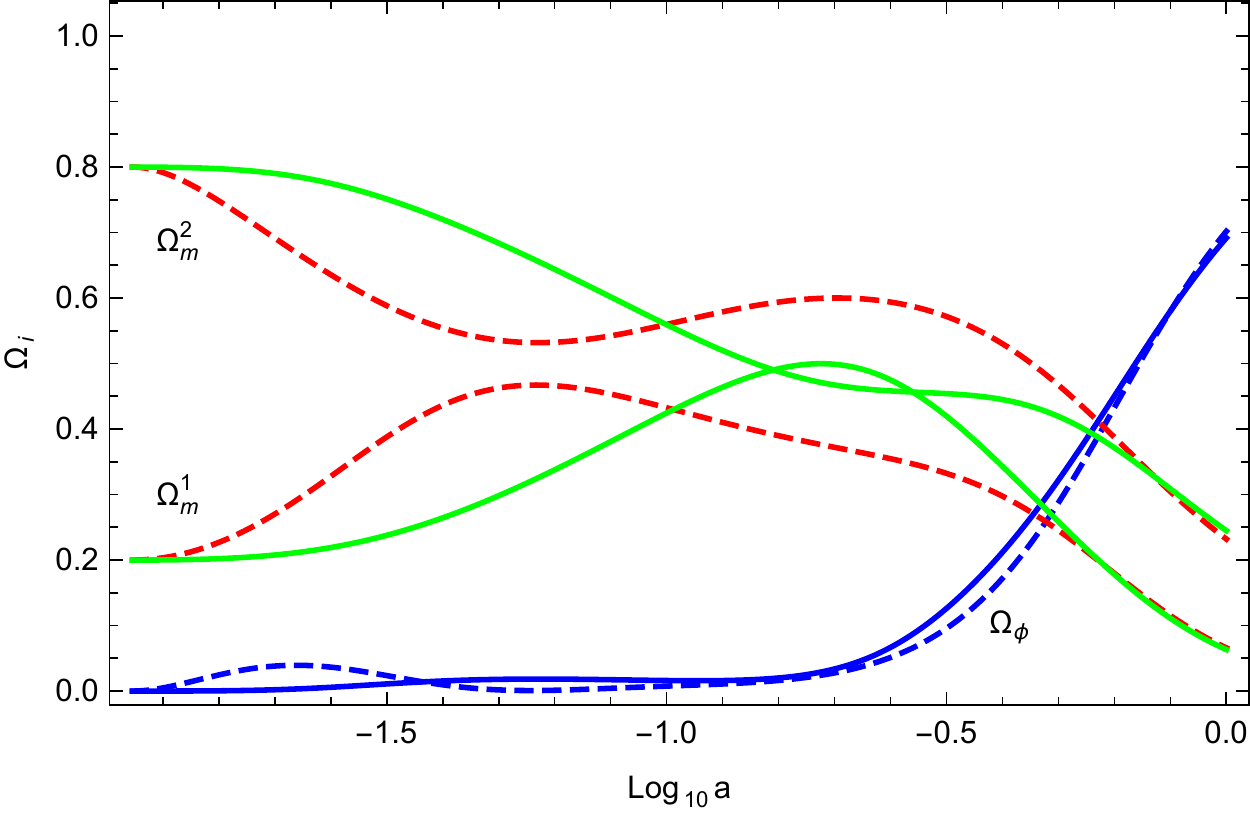}
   \caption{}
 \end{subfigure} 
 \begin{subfigure}[b]{0.47\textwidth}
  \includegraphics[width=1\textwidth]{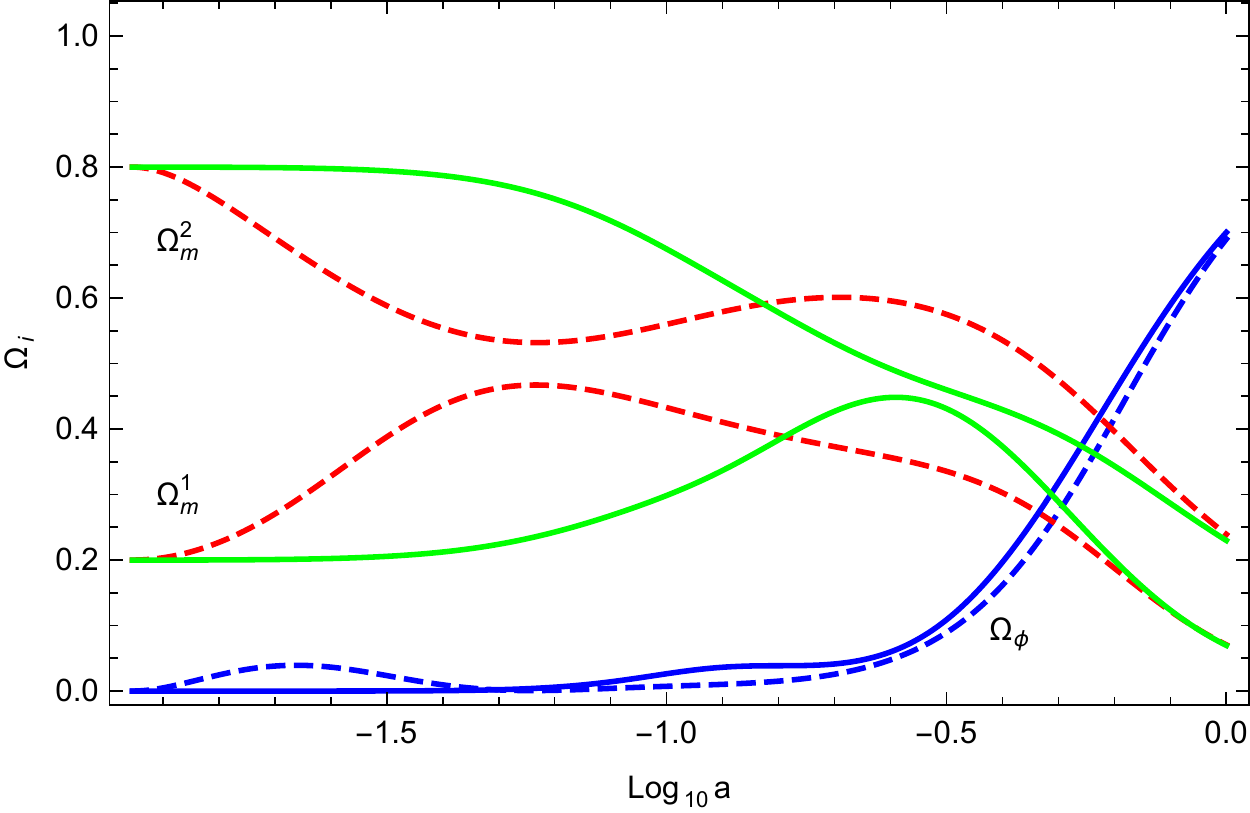}
  \caption{}
 \end{subfigure} 
  \begin{subfigure}[b]{0.47\textwidth}
  \includegraphics[width=1\textwidth]{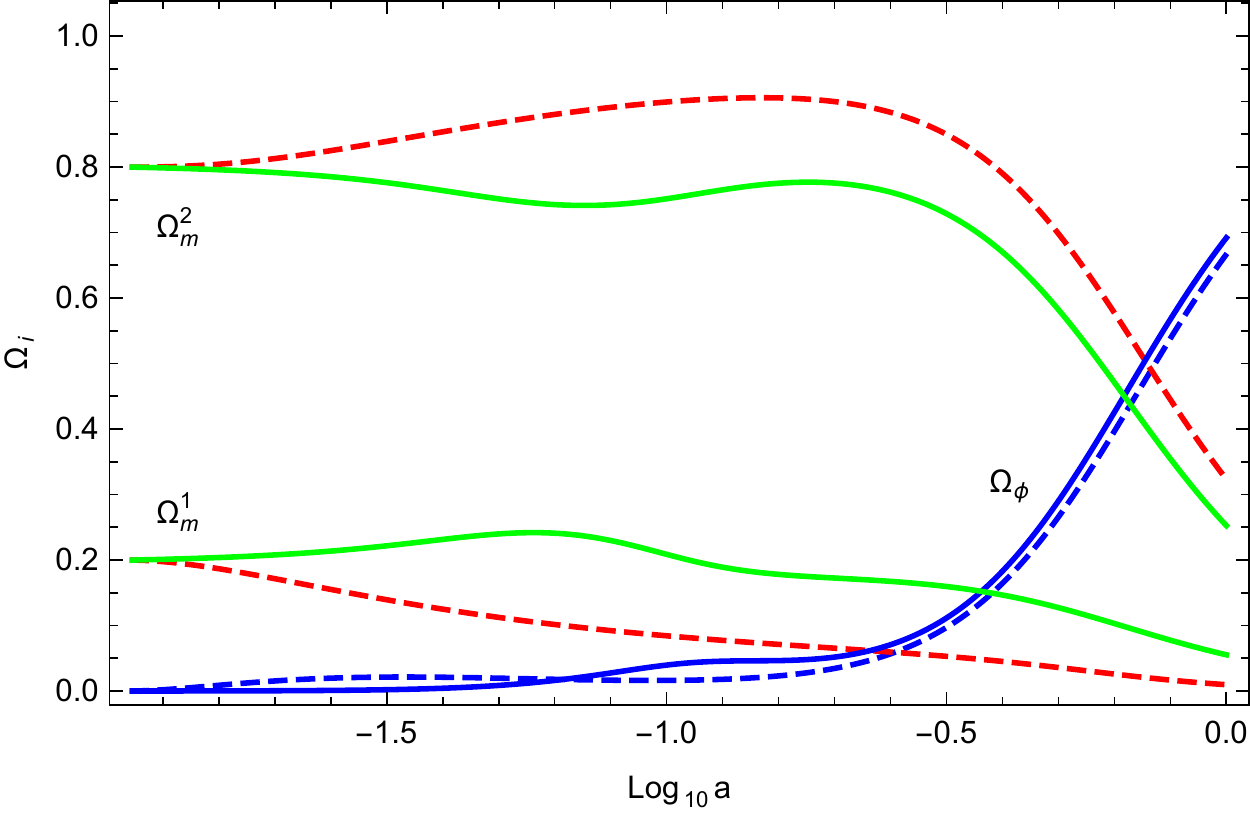}
  \caption{}
 \end{subfigure}  
  \caption{In (a), (b), (c), and (d) we compare disformally coupled cases (solid line) with conformally coupled scenarios (dashed line). Indeed, in (a) cases I and Ic are compared, in (b) cases II and IIc are compared, in (c) we compare III with IIc, and in (d) we compare IV with Ic.} 
\label{fig:cases}
\end{figure}

%
As was done in the previous cases, we also give the stability analysis of these fixed points, in which we only discuss the regions where the fixed point is found to be stable. All eigenvalues are listed in Appendix \ref{appendix:c3}. For each fixed point we have five eigenvalues, $e_{1,2,3,4,5}$, which, in general depend on all five model parameters, $\alpha_{1,2},\;\beta_{1,2},\;\lambda$. This is in contrast with the dust-radiation case in which, although we have the same number of eigenvalues, the eigenvalues in that case did not depend on $\alpha_2$. This is due to the fact that radiation is conformally invariant. 
\begin{itemize}
\item For this case, the two kinetic dominated fixed points can be stable. Indeed, (1) is stable when the following holds
\begin{equation*}
\alpha_1>\sqrt{\frac{3}{2}}\;,\hspace{0.5cm}\;\alpha_2>\sqrt{\frac{3}{2}}\;,\hspace{0.5cm}\;\beta_1>-\sqrt{\frac{3}{2}}\;,\hspace{0.5cm}\beta_2>-\sqrt{\frac{3}{2}}\;,\; \hspace{0.5cm}\lambda<-\sqrt{6}\;,
\end{equation*}
and (2) is a stable node whenever the following inequalities are satisfied 
\begin{equation*}
\alpha_1<-\sqrt{\frac{3}{2}}\;,\hspace{0.5cm}\;\alpha_2<-\sqrt{\frac{3}{2}}\;,\hspace{0.5cm}\;\beta_1<\sqrt{\frac{3}{2}}\;,\hspace{0.5cm} \beta_2<\sqrt{\frac{3}{2}}\;,\hspace{0.5cm}\lambda>\sqrt{6}\;.
\end{equation*}
\item The conformal dust dominated fixed point $(c)$ is found to be a saddle when satisfying the existence condition. 
\begin{figure}[h!]
\centering
\begin{subfigure}[b]{0.44\textwidth}
  \includegraphics[width=1\textwidth]{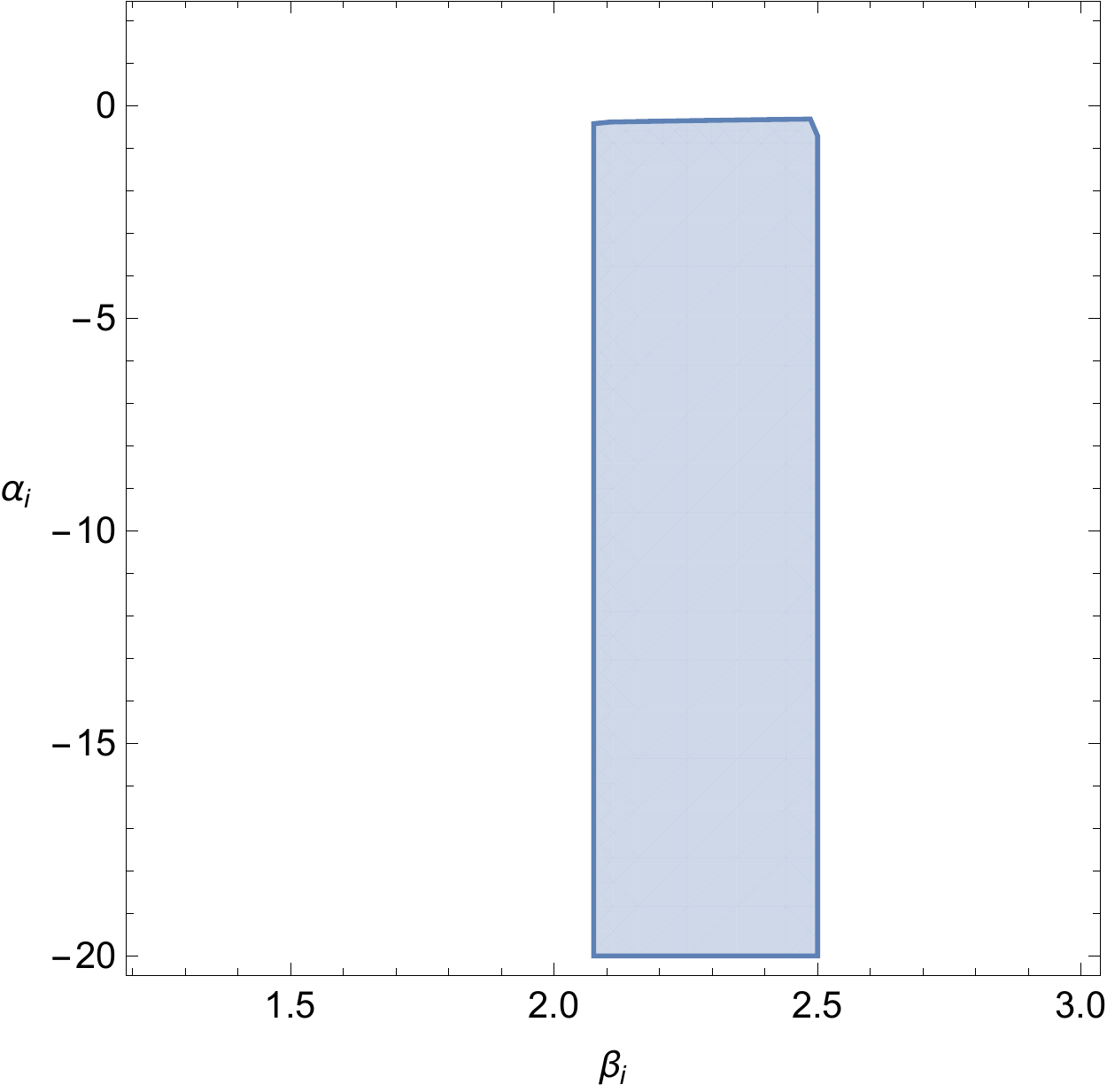}
   \caption{}
 \end{subfigure}
 \begin{subfigure}[b]{0.44\textwidth}
  \includegraphics[width=1\textwidth]{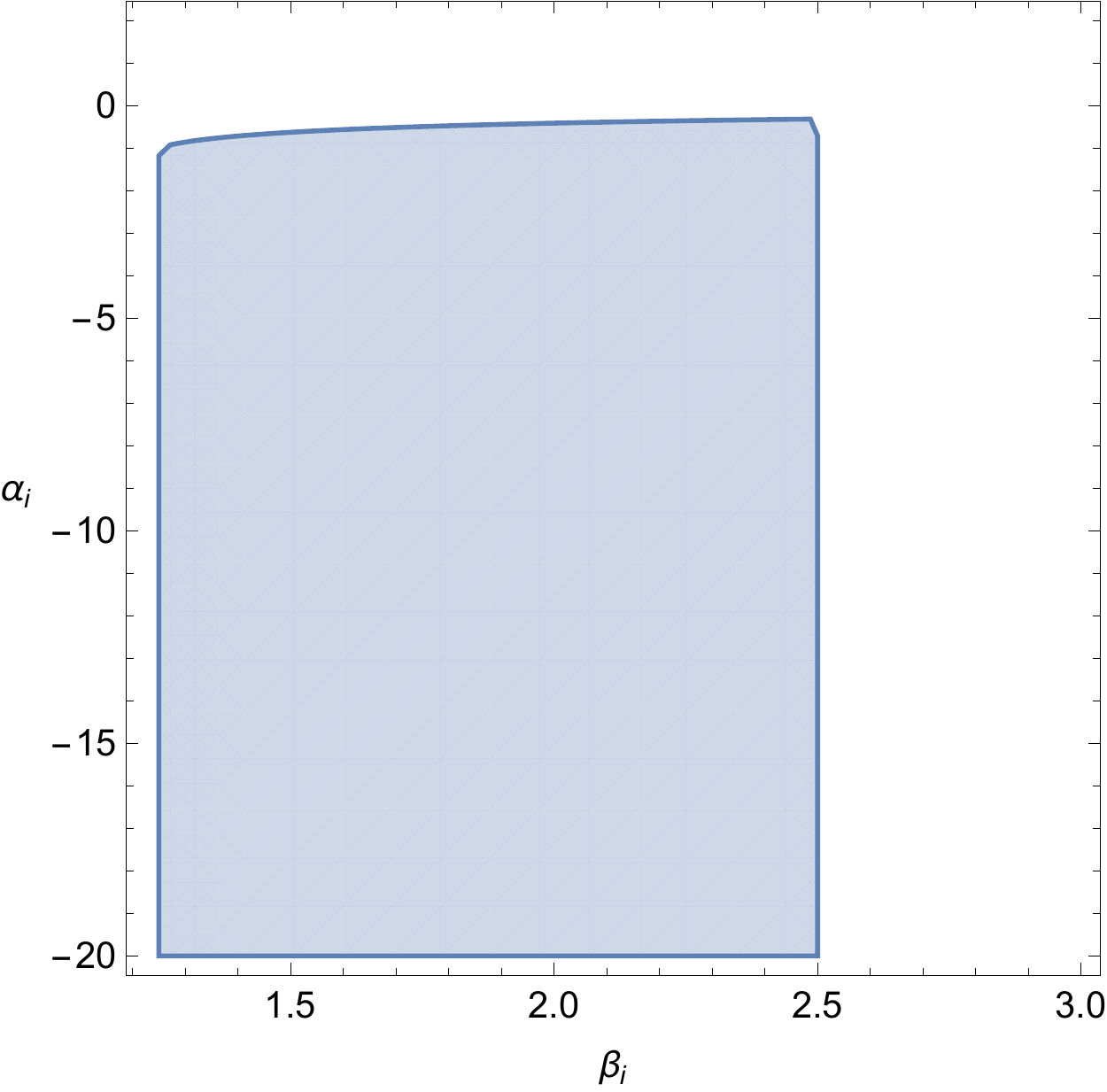}
   \caption{}
 \end{subfigure} 
 \begin{subfigure}[b]{0.44\textwidth}
  \includegraphics[width=1\textwidth]{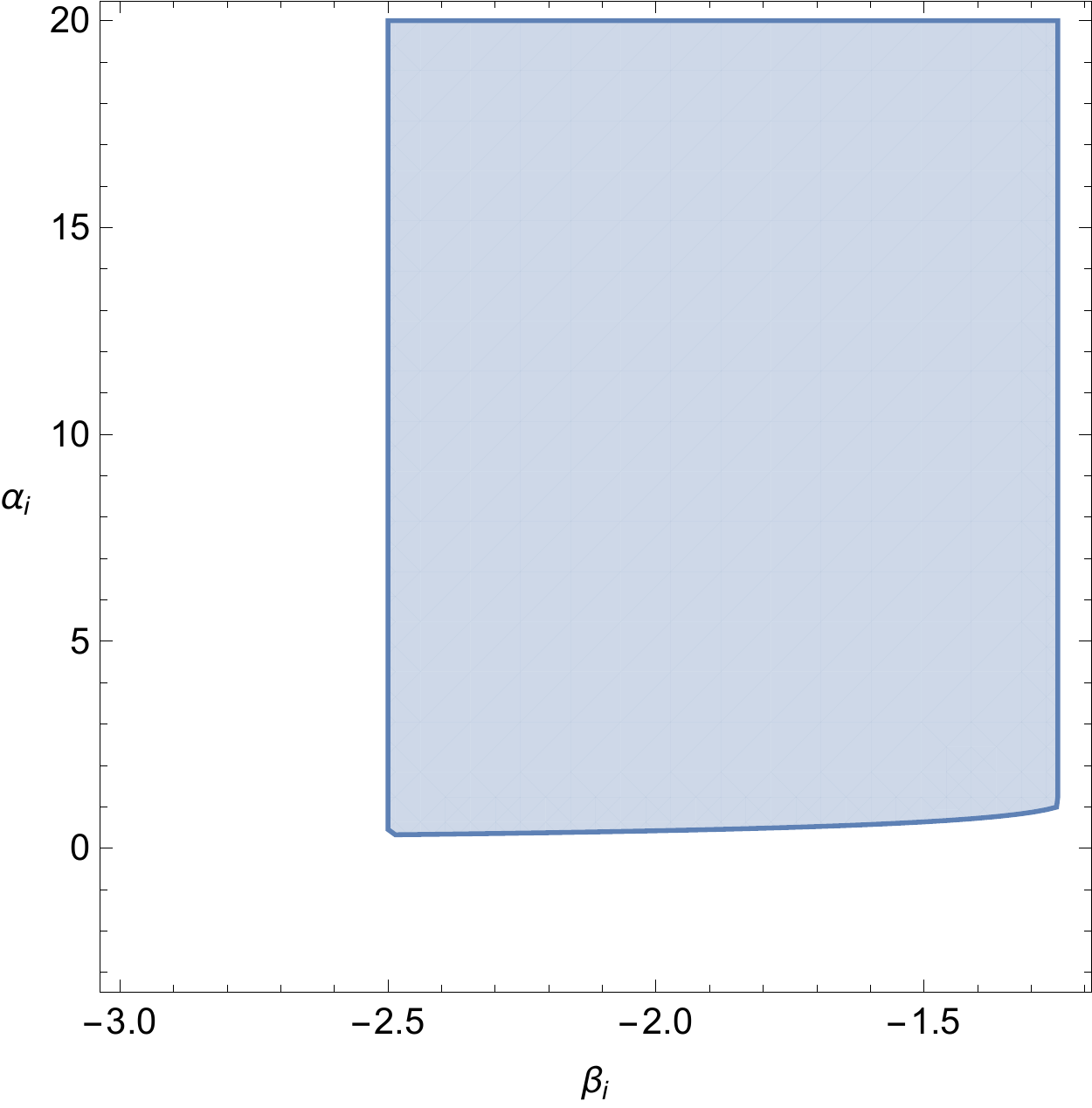}
  \caption{}
 \end{subfigure} 
 \begin{subfigure}[b]{0.44\textwidth}
  \includegraphics[width=1\textwidth]{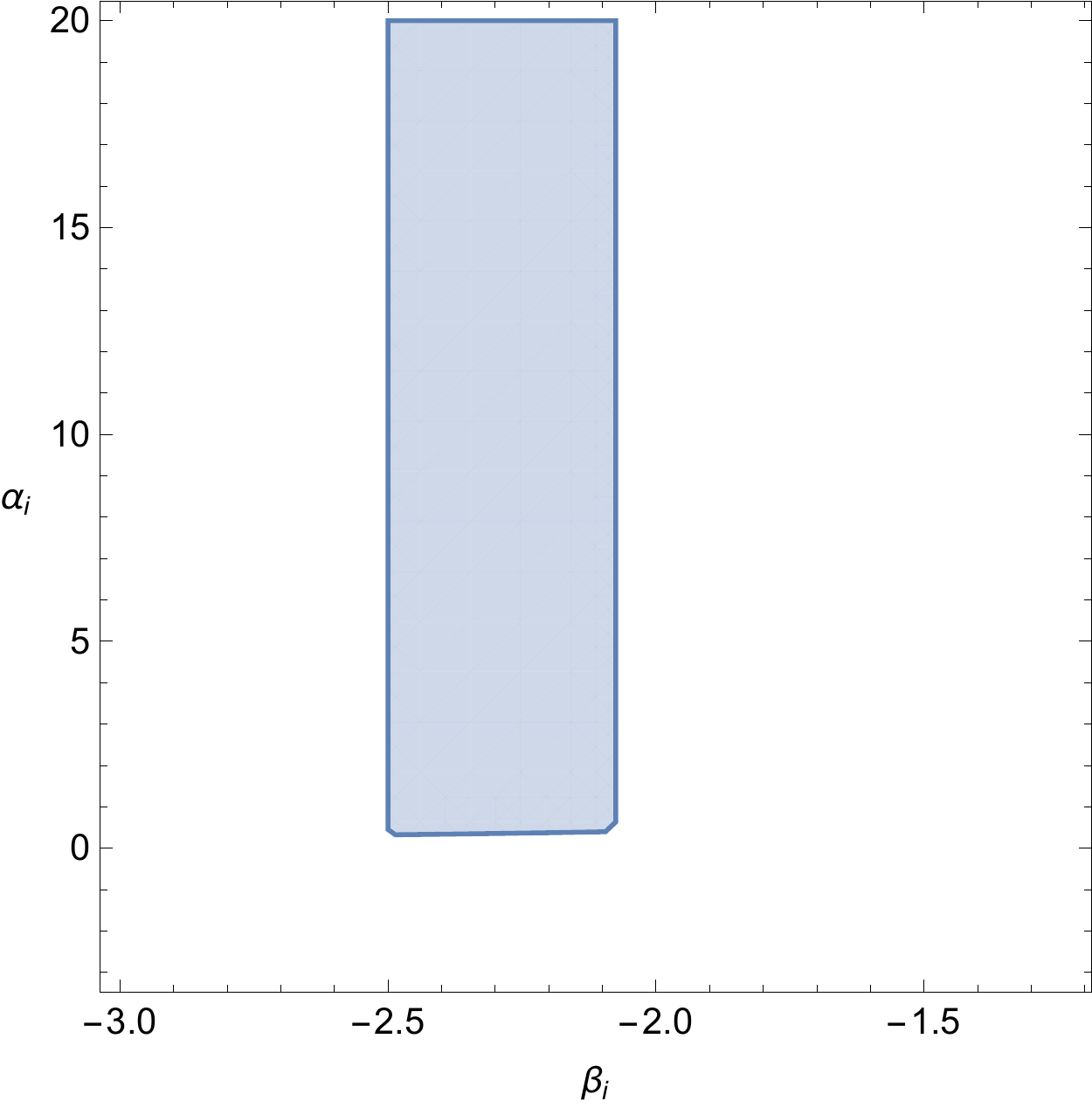}
  \caption{}
 \end{subfigure} 
  \caption{The stable node regions for a two conformal--disformal dust case, show an illustration of the parameter values of $\alpha_i$ and $\beta_i$ for fixed points $(3)_{(d)}^{i=1,2}$ in (a), (b); and $(4)_{(d)}^{i=1,2}$ in (c), (d). The other free parameters are chosen to be as follows; (a): $(\beta_{j\neq i},\alpha_{j\neq i},\lambda)=(0.5 , -0.4, 5)$, (b): $(\beta_{j\neq i},\alpha_{j\neq i},\lambda)=(0.5 , -1, 5)$, (c): $(\beta_{j\neq i},\alpha_{j\neq i},\lambda)=(0.5 , 1, -5)$, and (d): $(\beta_{j\neq i},\alpha_{j\neq i},\lambda)=(0.5 , 0.4, -5)$.} 
\label{fig:dust_dust_dis}
\end{figure}
\item The conformal kinetic fixed point, $(6)^1_{(d)}$, is a stable node when the parameters satisfy the following inequalities,
\begin{align*}
-\sqrt{\frac{3}{2}}&<\alpha _1<0,\;\;\alpha _2<\alpha _1,\;\;\beta _1<\frac{-2 \alpha _1^2-3}{4 \alpha
   _1},\;\;\beta
   _2<\frac{-2 \alpha _1^2-3}{4 \alpha _1},\;\;\lambda >\frac{-2 \alpha _1^2-3}{2 \alpha _1},\\
\text{or},\;\;\,0&<\alpha _1<\sqrt{\frac{3}{2}},\;\;\alpha _2>\alpha _1,\;\;\beta _1>\frac{-2 \alpha _1^2-3}{4 \alpha
   _1},\;\;\beta
   _2>\frac{-2 \alpha _1^2-3}{4 \alpha _1},\;\;\lambda <\frac{-2 \alpha _1^2-3}{2 \alpha _1},
\end{align*}
and the other conformal kinetic dust fixed point, $(6)^2_{(d)}$, is a stable node when either one of the following conditions is satisfied
\begin{equation*}
\begin{split}
\;\;\;\;-\sqrt{\frac{3}{2}}&<\alpha _2<0,\;\;\alpha _1<\alpha _2,\;\;\beta _1<\frac{-2 \alpha _2^2-3}{4 \alpha
   _2},\;\;\beta
   _2<\frac{-2 \alpha _2^2-3}{4 \alpha _2},\;\;\lambda >\frac{-2 \alpha _2^2-3}{2 \alpha _2},\\
\text{or},\;\;\,0&<\alpha _2<\sqrt{\frac{3}{2}},\;\;\alpha _1>\alpha _2,\;\;\beta _1>\frac{-2 \alpha _2^2-3}{4 \alpha
   _2},\;\;\beta
   _2>\frac{-2 \alpha _2^2-3}{4 \alpha _2},\;\;\lambda <\frac{-2 \alpha _2^2-3}{2 \alpha _2}.
\end{split}
\end{equation*}
The regions of stability for the conformal scaling fixed points, $(8)^{1,2}_{(d)}$, are shown in Fig \ref{fig:dust_dust1}. The regions of stability differ from the previous cases, since the eigenvalues now depend on all five model parameters, leading to more degrees of freedom in the $(\alpha_i,\lambda)$ parameter space.
{\setlength\extrarowheight{9pt}
\begin{table}[t!]
\begin{center}
\begin{tabular}{ c c c c c c} 
 \hline
\hline
 Case &  $\lambda$ & $\alpha_1$ & $\beta_1$ & $\alpha_2$ & $\beta_2$  \\ 
\hline
I & $\sqrt{\frac{3}{2}}$ & -1.5 & -- & 0 & -5    \\ 
Ic & $\sqrt{\frac{3}{2}}$ & -1.5 & -- & 0.035 & -- \\
II & $\sqrt{\frac{3}{2}}$ & -1.5 & -- & 1 & -5 \\
IIc & $\sqrt{\frac{3}{2}}$ & -1.5 & -- & 1 & -- \\
III & $\sqrt{\frac{3}{2}}$ & -1.5 & 0.5 & 1 & -5 \\
IV & $\sqrt{\frac{3}{2}}$ & 0 & 0.5 & 0 & -5 \\
 \hline
\hline
\end{tabular}
\end{center}
\caption{\label{table:cases} Listed are, respectively, the cases considered in Fig \ref{fig:cases} together with the respective parameter values. Cases I, II, III, and IV are all disformally coupled cases, whereas cases Ic and IIc are conformally coupled cases.}
\end{table}}
\item The regions of stability for the four disformal fixed points $(3)^{1,2}_{(d)}$, $(4)^{1,2}_{(d)}$ are illustrated in Fig \ref{fig:dust_dust_dis}.
\item The scalar field dominated fixed point (7), is a stable node when the parameters satisfy either one of the following conditions
\begin{equation*}
\begin{split}
-\sqrt{6}&<\lambda <0\;,\hspace{0.5cm}\alpha _1>\frac{3-\lambda ^2}{\lambda }\;,\hspace{0.5cm}\alpha
   _2>\frac{3-\lambda ^2}{\lambda }\;,\hspace{0.5cm}\beta _1>\frac{\lambda }{2}\;,\hspace{0.5cm}\beta _2>\frac{\lambda }{2}\;,\\
\text{or},\;\;\,0&<\lambda <\sqrt{6}\;,\hspace{0.5cm}\alpha _1<\frac{3-\lambda ^2}{\lambda }\;,\hspace{0.5cm}\alpha
   _2<\frac{3-\lambda ^2}{\lambda }\;,\hspace{0.5cm}\beta _1<\frac{\lambda }{2}\;,\hspace{0.5cm}\beta _2<\frac{\lambda }{2}\;.
\end{split}
\end{equation*}
\end{itemize}
We illustrate some examples in Fig \ref{fig:cases}, in which we consider different coupling cases as tabulated in Table \ref{table:cases}. The purely conformal cases are denoted by Ic, and IIc. We use these conformal cases to compare with the disformal cases I, II, III, and IV. In case I, we consider the first dust component to be conformally coupled and the second component to be purely disformally coupled. In case II, the first dust component is conformally coupled, whereas the second component is conformally--disformally coupled. In case III the two dust components are both conformally--disformally coupled, whereas in case IV the two components are purely disformally coupled. We remark that when we neglect the disformal coupling we consider $M_i\rightarrow\infty$ in Eq. (\ref{couplings_potential}), hence $\beta_i$ is arbitrary in these circumstances.  
%
\section{Cosmological consequences}
\label{sec:consequences}
In the following section we will summarize our findings and present some cosmological consequences.

\subsection{General remarks}

The only fixed points that admit accelerated solutions are points (7) and $(8)_{(d)}$. When the attractor of the system is the scalar field dominated fixed point (7), the matter and radiation sectors vanish completely as soon as this point is reached. Hence, in order to account for the present non--zero matter and radiation densities, the initial conditions should be fine tuned in such a way that the scalar field dominated attractor is not reached at the present time. On the other hand, the conformal dust scaling fixed point $(8)_{(d)}$, is a solution for which the matter and scalar field energy density parameters $\Omega_{m,\phi}$, stabilize to a constant finite value and remain indefinitely constant. The values of these energy density parameters are fixed when the conformal coupling strength parameter, $\alpha$, and the scalar field potential exponent, $\lambda$, are specified, and are independent from the choice of initial conditions. As a consequence, the coincidence of the current values of the energy density parameters is solved when the attractor of the system is point $(8)_{(d)}$. However, trajectories with this global attractor are known to lack a matter dominated epoch (see for example Ref. \cite{Amendola:1999er}) which could lead to negative consequences on the growth of perturbations.

All trajectories starting deep in the radiation era depart from the neighbourhood of the radiation dominated fixed point $(6)_{(r)}$, even if initially the field's kinetic energy dominated, i.e. near the kination fixed points (1) and (2), as the system would quickly evolve towards the saddle point $(6)_{(r)}$. After leaving the radiation dominated fixed point, the trajectory could pass near the radiation disformal fixed point, mixed fixed point, and the disformal dust radiation fixed point. The existence of these transient saddle points depends on the parameter choice of a given model. Indeed, in Fig. \ref{fig:Omega7} all three saddle points are present in the field--radiation--dominated era ($\phi$RDE), while in Fig. \ref{fig:Omega7r} only the points $(5)_{(r)}$ and $(3)_{(r)}$ exist, with $(5)_{(r)}$ being the transient fixed point in the $\phi$RDE. In this $\phi$RDE, energy is transferred from radiation to the coupled scalar field as the trajectory passes near these transient fixed points. A further energy transfer to the coupled matter sector is also possible if the trajectory evolves temporarily towards point $(b)$. As a consequence, radiation--matter equivalence happens at a different e-fold number when compared with the purely conformal system which is missing the mentioned fixed points. 

In the matter dominated era, conformally coupled models are known to enter a field--matter--dominated era, or $\phi$MDE \cite{Amendola:1999er}, characterised by an energy transfer from matter to the coupled scalar field. This happens as the trajectory passes near the conformal dust kinetic saddle point $(6)_{(d)}$, before it is attracted towards the scalar field dominated fixed point (7). In the uncoupled case, fixed point $(6)_{(d)}$ is fixed at $x=y=0$, and hence does not lead to the energy transfer observed in the conformally coupled model. When matter and radiation are disformally coupled, the $\phi$MDE could be modified due to an evolution of the trajectory towards one of the disformal fixed points. This is clearly seen in Fig. \ref{fig:Omega7} in which the system, after leaving the point $(6)_{(d)}$, passes near the disformal saddle point $(3)_{(d)}$. As a result, there is a transfer of energy from the coupled scalar field to matter; the reverse process of the conformally coupled model. An increase in the radiation energy density could also appear in the $\phi$MDE. This is illustrated in Fig. \ref{fig:Omega7r}, in which radiation is disformally coupled while matter is only conformally coupled to the scalar field. The energy transfer from matter and the scalar field to radiation occurs, because of an evolution of the trajectory towards the disformal radiation saddle point $(3)_{(r)}$, before the trajectory is attracted towards fixed point (7). 

The conformal dust radiation fixed point could also be a transient fixed point, although this requires a large conformal coupling $\left(\alpha^2>1/2\right)$. This might be in conflict with local observations in the purely conformal scenario, however such constraints could be relaxed when considering multiple couplings, such as the additional disformal coupling \cite{vandeBruck:2016hpz}.

The conformal dust dominated fixed point, although it is not able to give an accelerated expansion of the Universe, it could be a transient fixed point for the evolution of two coupled dust fluids. In conformally coupled models, the trajectory first evolves towards the conformal dust kinetic fixed point, resulting in an increase of the scalar field energy density parameter, and then the trajectory evolves towards saddle point $(c)$, as it is attracted towards the global attractor point (7). The transition towards the conformal dust dominated fixed point is clearly seen in case IIc depicted in Fig. \ref{fig:cases}, characterised by a decrease  in the scalar field energy density before the system evolves towards the dark energy dominated era. On the other hand, in the presence of a disformal coupling, the disformal fixed point of the disformally coupled dust component $(4)_{(d)}$, delays the evolution towards the conformal dust kinetic and the conformal dust dominated fixed points. As shown in Fig. \ref{fig:cases}, this disformal fixed point can even force the system to evolve directly towards point $(c)$, and hence no transfer of energy to the coupled scalar field takes place, which would then shift the field--matter equivalence e-fold number.  

\subsection{Variation of the fine--structure constant induced by disformal couplings}
\label{sec:varying_alpha}
It has previously been shown \cite{vandeBruck:2015rma} that when radiation and matter are purely disformally coupled with the scalar field, such that the disformal couplings are of different strengths, this gives rise to a variation in the evolution of the fine--structure constant, $\alpha$. Indeed, if we consider the action (\ref{action}) with two species, matter and radiation, i.e.
\begin{equation}
\begin{split}
\mathcal{S} =& \int d^4 x \sqrt{-g} \left[ \frac{M_{\rm Pl}^2}{2} R - \frac{1}{2} g^{\mu\nu}\partial_\mu \phi \partial_\nu \phi - V(\phi) \right] + \mathcal{S}_{\text{matter}} \left[\tilde g_{\mu\nu}^{(m)}\right]\\
& - \frac{1}{4}\int d^4x \sqrt{-\tilde{g}^{(r)}} \tilde{g}^{\mu\nu}_{(r)} \tilde{g}^{\alpha\beta}_{(r)} F_{\mu\alpha}F_{\nu\beta}\;,
\end{split}
\end{equation}
such that the matter metric, $\tilde{g}^{(m)}_{\mu\nu}$, and the radiation metric, $\tilde{g}^{(r)}_{\mu\nu}$, are disformally related with the gravitational sector metric $g_{\mu\nu}$ by Eq. (\ref{disformal_relation}), we can identify the fine--structure constant to be 
\begin{equation}
\alpha \propto Z_{\alpha}\;,
\end{equation}    
where
\begin{equation}
Z_\alpha^2=1+\frac{B}{A}\tilde{g}^{\mu\nu}_{(m)}\phi_{,\mu}\phi_{,\nu}\;,
\end{equation}
with
\begin{equation}
A=\frac{C_r}{C_m},\;\;\;\;B=D_r-\frac{C_r D_m}{C_m}\;,
\end{equation}
such that $C_{r,m}$ and $D_{r,m}$ are the radiation and matter conformal and disformal couplings, respectively. 
In terms of the dynamical system variables introduced in equations (\ref{dynamical_variables1})-(\ref{dynamical_variables2}), the fine--structure constant simplifies to 
\begin{equation}\label{alpha}
\alpha\propto\frac{Z_r}{Z_m}\;,
\end{equation}
where the variable $Z_i$ is defined as given in Eq. (\ref{Zi}), and we use subscripts $r,\;m$ for radiation and matter, respectively.
Furthermore, the evolution of the fine--structure constant can be written as 
\begin{equation}\label{delta_alpha}
\frac{\Delta\alpha}{\alpha}=\left(\frac{Z_r}{Z_m}\right)\left(\frac{Z_m^0}{Z_r^0}\right)-1\;,
\end{equation}
where $Z_m^0=Z_m(z=0)$ and $Z_r^0=Z_r(z=0)$. 

From Eq. (\ref{alpha}), it is evident that for a disformal fixed point which is characterised by a metric singularity in either the radiation metric or the matter metric, the fine--structure constant cannot be defined. This observation supports the arguments presented in Ref. \cite{Sakstein:2014aca}, in which such fixed points were considered as unviable fixed points. However, one could consider a trajectory such that the system discussed in Section \ref{sec:cdcd_dust_rad} reaches $(5)_{(r)}$ at BBN and evolve towards the attractor $(8)_{(d)}$, in which case Eq. (\ref{delta_alpha}) reduces to $\Delta\alpha/\alpha=\sqrt{1+4/\left(2-3\beta_r^2\right)}-1$. By considering this evolution and by using the BBN constraint of $|\Delta\alpha/\alpha|<6\times 10^{-2}$ \cite{Cyburt:2004yc,Chiba:2011bz}, we get that $|\beta_r|>3.48$.       
%
\section{Conclusions}
\label{sec:conclusions}

In this article we performed a thorough analysis of generalized couplings of a scalar field with a single or two matter fluids. More specifically, we first generalized previous literature results by studying a scalar field conformally and disformally coupled to a perfect fluid with an arbitrary equation of state. We investigated the generic existence conditions and evaluated the stability conditions for the particular cases of dust and radiation.

We verified that depending on the value of the logarithmic slope of the disformal coupling, $\lambda_D^i = -2 (\alpha_i+\beta_i)$, the fixed points may have a region of parameter space that render it stable. In general, the introduction of disformal couplings allows any given fixed point to have a relevant cosmological role at some given epoch of the history of the Universe. For example, the kinetic fixed point for dust turns out to be stable when the disformal coupling is introduced. 

We extended this analysis to a system with two perfect fluids. Here the analysis is far more complicated and despite not being able to present results for generic fluids, we were able to understand in detail the cases of a conformal--disformal dust and conformal--disformal radiation and, two conformal--disformal dust components, which are the most relevant cases for cosmological applications. In the case of a conformal dust and conformal--disformal radiation of Section \ref{sec:ccd_dust_rad}, we gave a curious example when the radiation component may become important at late times. We concluded that disformal couplings on their own are repellors and therefore the matter metric singularity is safely avoided.

We also looked at some cosmological consequences arising from the obtained fixed points. The introduction of the disformal fixed points lead to an intermediate phase between the radiation era and equivalence, denoted by $\phi$RDE, and to the modification of an intermediate phase between the matter era and accelerated era, denoted by $\phi$MDE, where the latter intermediate phase is also present in purely conformal models. The variation of the fine--structure constant induced by considering disformally coupled matter and radiation has also been addressed in this work. 

It would now be important to evaluate the background and perturbations' evolution impact on the observables constrained by current and future probes. It would also be useful to extend the results of our work to more general actions such as k-essence or multiple fields scenarios.
\vspace{1cm}

\begin{acknowledgments}
The work of CvdB is supported by the Lancaster- Manchester-Sheffield Consortium for Fundamental Physics under STFC Grant No. ST/L000520/1. N.J.N thanks the School of Mathematics and Statistics of the University of Sheffield for hospitality. N.J.N and J.P.M. are supported by the Funda\c{c}\~{a}o para a Ci\^{e}ncia e Tecnologia
(FCT) through the grant UID/FIS/04434/2013. 
\end{acknowledgments}

%
%
\begin{subappendices}

\section*{Appendix A: Beyond exponential form of the couplings and potential}
\renewcommand{\theequation}{A.\arabic{equation}}
\renewcommand{\thesection}{A}
\refstepcounter{section}
\label{appendix:0}
In this paper the functions $V(\phi$), $C_i(\phi)$ and $D_i(\phi)$ were of exponential form, for which $\lambda_V$, $\lambda_C^i$ and $\lambda_D^i$ are constant. While these are well motivated cases, there are other types of models, in which the field sits in the minimum of an effective potential defined by $3H^2\lambda_V y^2 + \kappa(Q_1+Q_2) = 0$,
at finite values of $\phi = \phi_*$. Alternatively the functions can be, for example, of power--law form, in which case $\lambda_V$, $\lambda_C^i$ and $\lambda_D^i$  are varying. 
In these situations, it is useful to introduce another equation related to $x$ which closes the system. The full system now reads 
\begin{eqnarray} 
x^\prime &=& -\left(3 + \frac{H'}{H}\right) x + \sqrt{\frac{3}{2}} \left( \lambda_V y^2 +  \frac{\kappa Q_1}{3H^2} +     \frac{\kappa Q_2}{3 H^2} \right) \;,\label{start1} \\
y^\prime &=& - \sqrt{\frac{3}{2}} \left( \lambda_V x + \sqrt{\frac{2}{3}} \frac{H'}{H} \right) y\;, \\
z_i^\prime &=& -\frac{3}{2} \left( 1+w_i + \frac{2}{3} \frac{H'}{H} + \frac{1}{3} \sqrt{\frac{2}{3}} \frac{\kappa Q_i}{H^2} \frac{x}{z_i^2} \right) z_i\;, \\
\sigma_i^\prime &=& \left( \sqrt{6} (\lambda_C^i - \lambda_D^i) x + 2 \frac{H'}{H} \right) \sigma_i\;, \\
\phi' &=& \frac{\sqrt{6}}{\kappa}\,x \label{additional_eq}\;.
\end{eqnarray}
It then becomes possible to classify all the asymptotic behaviours of the cosmological model in relation to the functional features of the potential and disformal parameters. In particular, we find a non--trivial fixed point given in Table \ref{table0}. Although this requires that $w_1=-1$, this fixed point is distinct from a bare cosmological constant due to the non--vanishing couplings.
{\setlength\extrarowheight{9pt}
\begin{table}
\begin{center}
\begin{tabular}{  c c c c c c} 
 \hline
\hline
  $x$ & $y$ & $\sigma_1$ & $\sigma_2$ & $z_1$ & $z_2$ \\ 
\hline
 0 & $\sqrt{\frac{2\lambda^1_C}{2\lambda^1_C-\lambda_V}}$\;& $\forall \sigma_1$ & $\forall \sigma_2$   & $\sqrt{\frac{\lambda_V}{\lambda_V-2\lambda^1_C}}$ & 0  \\
 \hline
\hline
\end{tabular}
\end{center}
\caption{\label{table0} A non--trivial fixed point of the system (\ref{start1})--(\ref{additional_eq}) when $x=0$ and the scalar field freezes.}
\end{table}}

\section*{Appendix B: Stability}
\renewcommand{\theequation}{B.\arabic{equation}}
\renewcommand{\thesection}{B}
\refstepcounter{section}
\label{appendix:a}
In this section we briefly point out the method used in order to arrive to the eigenvalues with which we then determine the stability of the fixed points. A broader discussion on this subject can be found in Ref. \cite{Sakstein:2014aca}. We will consider a general system of $n$ first-order ordinary differential equations for $n$ variables $X_i$ as a function of some coordinate $t$. Since all systems considered are autonomous, we will consider this general system to be of the same type, by which we can write our system as follows
\begin{equation}
\frac{\text{d}X_i}{\text{d}t}=f_i(\{X_j\})\;.
\end{equation}
We define the fixed points, $\{X_j^c\}$, of our system to be the solutions of the $n$ algebraic equations when we set $f_i=0\;\forall i$. By considering a small perturbation, $\delta X_i$, around a fixed point, $X_i^c$, i.e. considering a point $X_i=X_i^c+\delta X_i$, one obtains, up to first order
\begin{equation}
\frac{\text{d}\delta X_i}{\text{d} t}=\mathcal{M}_{ij}\delta X_j,\;\text{where},\;\mathcal{M}_{ij}=\frac{\partial f_i}{\partial X_j}\;.
\end{equation}
The eigenvalues that are used in order to study the stability of the fixed points correspond to the eigenvalues of the $n\times n$ matrix $\mathcal{M}_{ij}$ evaluated at the fixed point $X_i^c$. For example, for the single fluid system with equation of state parameter, $\gamma$, described by the system (\ref{single1})-(\ref{single3}) together with Eq. (\ref{couplings}), the matrix elements of the $3\times3$ matrix, $\mathcal{M}_{ij}$, are the following
\begin{align*}
\begin{split}
\mathcal{M}_{11}=&-3+\frac{3}{2}\left(1-x^2-y^2\right)\left(\gamma-18(\gamma-1)\sigma x^2\right)+3x^2\left(3-\gamma+6(\gamma-1)\sigma x^2\right)\\
&+\sqrt{\frac{3}{2}}\frac{-2x+12\sigma x(1-y^2)}{(1+3\sigma(1-3x^2-y^2))^2}\biggl((3\gamma-18(\gamma-1)\sigma x^2-4)\alpha\biggr.\\
&\biggl.+3\sigma\left(\sqrt{6}x(\gamma-6(\gamma-1)\sigma x^2)-\lambda y^2+2(\alpha-\beta)x^2\right)\biggr)\\
&+\sqrt{\frac{3}{2}}\frac{1-x^2-y^2}{1+3\sigma(1-3x^2-y^2)}\biggl(-36(\gamma-1)\alpha x\sigma\biggr.\\
&\biggl.+3\sigma\left(\sqrt{6}(\gamma-18(\gamma-1)\sigma x^2)+4(\alpha-\beta)x\right)\biggr)
\end{split}\\
\begin{split}
\mathcal{M}_{12}=&\;3xy\left(6(\gamma-1)\sigma x^2-\gamma\right)+\sqrt{6}\lambda y\\
&+\sqrt{\frac{3}{2}}\frac{-2y+12\sigma yx^2}{\left(1+3\sigma(1-3x^2-y^2)\right)^2}\biggl(\left(3\gamma-18(\gamma-1)\sigma x^2-4\right)\alpha\biggr.\\
&+\biggl.3\sigma\left(\sqrt{6}x\left(\gamma-6(\gamma-1)\sigma x^2\right)-\lambda y^2+2(\alpha-\beta)x^2\right)\biggr)
+3\sqrt{6}\frac{\left(x^2+y^2-1\right)\lambda\sigma y}{1+3\sigma\left(1-3x^2-y^2\right)}
\end{split}\\
\begin{split}
\mathcal{M}_{13}=&\;9(\gamma-1)\left(x^2+y^2-1\right)x^3+\sqrt{\frac{3}{2}}\frac{1-x^2-y^2}{1+3\sigma\left(1-3x^2-y^2\right)}\biggl(-18(\gamma-1)\alpha x^2\biggr.\\
&\biggl.-36\sqrt{6}(\gamma-1)\sigma x^3+3\left(\sqrt{6}\gamma x-\lambda y^2+2(\alpha-\beta)x^2\right)\biggr)\\
&-\sqrt{\frac{3}{2}}\frac{3\left(1-3x^2-y^2\right)\left(1-x^2-y^2\right)}{\left(1+3\sigma\left(1-3x^2-y^2\right)\right)^2}\biggl(\left(3\gamma-18(\gamma-1)\sigma x^2-4\right)\alpha\biggr.\\
&+\biggl.3\sigma\left(\sqrt{6}x\left(\gamma-6(\gamma-1)\sigma x^2\right)-\lambda y^2+2(\alpha-\beta)x^2\right)\biggr)
\end{split}\\
\mathcal{M}_{21}=&-\sqrt{\frac{3}{2}}\lambda y-18(\gamma-1)\sigma x y\left(1-x^2-y^2\right)+3xy\left(2-\gamma+6(\gamma-1)\sigma x^2\right)\\
\mathcal{M}_{22}=&-\sqrt{\frac{3}{2}}\lambda x+\frac{3}{2}\left(\gamma-6(\gamma-1)\sigma x^2\right)(1-3y^2)+\frac{3}{2}\left(2-\gamma+6(\gamma-1)\sigma x^2\right)x^2\\
\mathcal{M}_{23}=&\;9(\gamma-1)\left(x^2+y^2-1\right)x^2y\\
\mathcal{M}_{31}=&\;2\sqrt{6}\beta\sigma+36(\gamma-1)\sigma^2 x\left(1-x^2-y^2\right)+6\sigma x\left(\gamma-2-6(\gamma-1)\sigma x^2\right)\\
\mathcal{M}_{32}=&\;6\sigma y\left(\gamma-6(\gamma-1)\sigma x^2\right)\\
\mathcal{M}_{33}=&\;2\sqrt{6}\beta x-6x^2-3\left(1-x^2-y^2\right)\left(\gamma-12(\gamma-1)\sigma x^2\right)
\end{align*}
\end{subappendices}

\begin{subappendices}
\section*{Appendix C: Eigenvalues--Single Fluid Case}
\renewcommand{\theequation}{C.\arabic{equation}}
\renewcommand{\thesection}{C}
\refstepcounter{section}
\label{appendix:b}
\subsubsection*{\textnormal{\underline{Arbitrary EOS}}}
(1) 
\begin{equation*}
e_1=-2(3+\sqrt{6}\beta),\;e_2=3(2-\gamma)-\sqrt{6}\alpha(4-3\gamma),\;e_3=3+\sqrt{3/2}\lambda
\end{equation*}
(2)
\begin{equation*}
e_1=-2(3-\sqrt{6}\beta),\;e_2=3(2-\gamma)+\sqrt{6}\alpha(4-3\gamma),\;e_3=3-\sqrt{3/2}\lambda
\end{equation*}
(3)
\begin{equation*}
\begin{split}
e_1=&\frac{3\left(2\beta^2-3\right)}{2 \left(\beta  \left(2 \beta
   +u_2\right)-3\right)^4}\biggl\{u_1\left[3-2\beta\left(u_2+2\beta\right)\right]-\left(2\beta^2-3\right)\biggr.\\
   &\biggl.\times\biggl[9+18\gamma+2\beta\left[2\alpha\left(8\beta^2-9\right)(3\gamma-4)+\beta\left(21-48\gamma+8\beta^2(4\gamma-3)\right)\right]\biggr.\biggr.\\
   &\biggl.\biggl.\;+u_2\left[2\alpha\left(8\beta^2-3\right)(3\gamma-4)+\beta\left(3-24\gamma+8\beta^2(4\gamma-3)\right) \right]   \biggr]\biggr\}\;,
\end{split}
\end{equation*}
\begin{equation*}
\begin{split}   
e_2=&\frac{3\left(2\beta^2-3\right)}{2 \left(\beta  \left(2 \beta
   +u_2\right)-3\right)^4}\biggl\{u_1\left[-3+2\beta\left(u_2+2\beta\right)\right]-\left(2\beta^2-3\right)\biggr.\\
   &\biggl.\times\biggl[9+18\gamma+2\beta\left[2\alpha\left(8\beta^2-9\right)(3\gamma-4)+\beta\left(21-48\gamma+8\beta^2(4\gamma-3)\right)\right]\biggr.\biggr.\\
   &\biggl.\biggl.\;+u_2\left[2\alpha\left(8\beta^2-3\right)(3\gamma-4)+\beta\left(3-24\gamma+8\beta^2(4\gamma-3)\right) \right]   \biggr]\biggr\}\;,
\end{split}
\end{equation*}
\begin{equation*}
e_3=-\frac{1}{2} \left(u_2-2 \beta \right) (2 \beta -\lambda )\;,
\end{equation*}
where $u_1$ and $u_2$ are defined as follows
\begin{equation*}
\begin{split}
u_1^2=&\frac{\left(3-2\beta^2\right)^4}{\left(-3+\beta\left(u_2+2\beta\right)\right)^4}\biggl\{81-4482\beta^2+23688\beta^4-32640\beta^6+12800\beta^8\biggr.\\
&\biggl.+8\alpha^2\left[-27+4\beta^2\left(9-8\beta^2\right)^2\right](4-3\gamma)^2-4\left[81-2862\beta^2+11808\beta^4-14208\beta^6+5120\beta^8\right]\gamma\biggr.\\
&\biggl.+4\left[81+64\beta^2\left(3-4\beta^2\right)^2\left(-3+2\beta^2\right)\right]\gamma^2\biggr.\\
&\biggl.+2 u_2\left[-9+18\gamma+4\alpha\beta\left(8\beta^2-9\right)(3\gamma-4)+2\beta^2\left(51-48\gamma+8\beta^2(4\gamma-5)\right) \right]\biggr.\\
&\biggl.\times\left[2\alpha\left(8\beta^2-3\right)(3\gamma-4)+\beta\left(21-24\gamma+8\beta^2(4\gamma-5)\right) \right]\biggr.\\
&\biggl.+16\alpha\beta(3\gamma-4)\left[27(5-7\gamma)+2\beta^2\left(96\beta^2(8-7\gamma)+64\beta^4(4\gamma-5)+9(56\gamma-55)  \right)\right]\biggr\}\;,
\end{split}
\end{equation*}
\begin{equation*}
u_2^2=-6+4\beta^2
\end{equation*}
(4)
\begin{equation*}
\begin{split}
e_1=&\frac{3\left(2\beta^2-3\right)}{2 \left(\beta  \left(u_2-2\beta\right)+3\right)^4}\biggl\{u_3\left[3+2\beta\left(u_2-2\beta\right)\right]-\left(2\beta^2-3\right)\biggr.\\
   &\biggl.\times\biggl[9+18\gamma+2\beta\left[2\alpha\left(8\beta^2-9\right)(3\gamma-4)+\beta\left(21-48\gamma+8\beta^2(4\gamma-3)\right)\right]\biggr.\biggr.\\
   &\biggl.\biggl.\;+u_2\left[-2\alpha\left(8\beta^2-3\right)(3\gamma-4)+\beta\left(-3+24\gamma+8\beta^2(3-4\gamma)\right) \right]   \biggr]\biggr\}\;,
\end{split}
\end{equation*}
\begin{equation*}
\begin{split}   
e_2=&\frac{3\left(2\beta^2-3\right)}{2 \left(\beta  \left(u_2-2\beta\right)+3\right)^4}\biggl\{u_3\left[-3-2\beta\left(u_2-2\beta\right)\right]-\left(2\beta^2-3\right)\biggr.\\
   &\biggl.\times\biggl[9+18\gamma+2\beta\left[2\alpha\left(8\beta^2-9\right)(3\gamma-4)+\beta\left(21-48\gamma+8\beta^2(4\gamma-3)\right)\right]\biggr.\biggr.\\
   &\biggl.\biggl.\;+u_2\left[-2\alpha\left(8\beta^2-3\right)(3\gamma-4)+\beta\left(-3+24\gamma+8\beta^2(3-4\gamma)\right) \right]   \biggr]\biggr\}\;,
\end{split}
\end{equation*}
\begin{equation*}
e_3=\frac{1}{2} \left(u_2+2 \beta \right) (2 \beta -\lambda )\;,
\end{equation*}
where
\begin{equation*}
\begin{split}
u_3^2=&\frac{\left(3-2\beta^2\right)^4}{\left(3+\beta\left(u_2-2\beta\right)\right)^4}\biggl\{81-4482\beta^2+23688\beta^4-32640\beta^6+12800\beta^8\biggr.\\
&\biggl.+8\alpha^2\left[-27+4\beta^2\left(9-8\beta^2\right)^2\right](4-3\gamma)^2-4\left[81-2862\beta^2+11808\beta^4-14208\beta^6+5120\beta^8\right]\gamma\biggr.\\
&\biggl.+4\left[81+64\beta^2\left(3-4\beta^2\right)^2\left(-3+2\beta^2\right)\right]\gamma^2\biggr.\\
&\biggl.-2u_2\left[-9+18\gamma+4\alpha\beta\left(8\beta^2-9\right)(3\gamma-4)+2\beta^2\left(51-48\gamma+8\beta^2(4\gamma-5)\right) \right]\biggr.\\
&\biggl.\times\left[2\alpha\left(8\beta^2-3\right)(3\gamma-4)+\beta\left(21-24\gamma+8\beta^2(4\gamma-5)\right) \right]\biggr.\\
&\biggl.+16\alpha\beta(3\gamma-4)\left[27(5-7\gamma)+2\beta^2\left(96\beta^2(8-7\gamma)+64\beta^4(4\gamma-5)+9(56\gamma-55)  \right)\right]\biggr\}
\end{split}
\end{equation*}
(5)
\begin{equation*}
\begin{split}
e_1=&-3\left(u_4+u_5\right)\biggl\{2(\gamma-1)\gamma\left(8(2\alpha+\beta)^2-24\alpha(2\alpha+\beta)\gamma+3\left(6\alpha^2-1\right)\gamma^2\right)^2(-2\beta+\alpha(3\gamma-4))^5\biggr.\\
&\biggl.\times\biggl[4\alpha^4(4-3\gamma)^4-24\alpha^3\beta(3\gamma-4)^3+3\gamma^2\left(8\beta^2+3(\gamma-4)\gamma\right)-12\alpha^2(4-3\gamma)^2\left(\gamma^2-4\beta^2\right)\biggr.\biggr.\\
&\biggl.\biggl.\;\;\;\;+4\alpha\beta(3\gamma-4)\left(3\gamma^2-8\beta^2\right) \biggr]\biggr\}^{-1} \;,
\end{split}
\end{equation*}
\begin{equation*}
\begin{split}
e_2=&\;3\left(u_4-u_5\right)\biggl\{2(\gamma-1)\gamma\left(8(2\alpha+\beta)^2-24\alpha(2\alpha+\beta)\gamma+3\left(6\alpha^2-1\right)\gamma^2\right)^2(-2\beta+\alpha(3\gamma-4))^5\biggr.\\
&\biggl.\times\biggl[4\alpha^4(4-3\gamma)^4-24\alpha^3\beta(3\gamma-4)^3+3\gamma^2\left(8\beta^2+3(\gamma-4)\gamma\right)-12\alpha^2(4-3\gamma)^2\left(\gamma^2-4\beta^2\right)\biggr.\biggr.\\
&\biggl.\biggl.\;\;\;\;+4\alpha\beta(3\gamma-4)\left(3\gamma^2-8\beta^2\right) \biggr]\biggr\}^{-1} \;,
\end{split}
\end{equation*}
\begin{equation*}
e_3=\frac{3 \gamma  (\lambda -2 \beta )}{\alpha  (6 \gamma -8)-4 \beta }\;,
\end{equation*}
where $u_4$ and $u_5$ are as follows
\begin{equation*}
\begin{split}
u_4^2=&-\left(2\beta+\alpha(4-3\gamma)\right)^8(\gamma-1)^2\gamma^2\left(8(2\alpha+\beta)^2-24\alpha(2\alpha+\beta)\gamma+3\left(6\alpha^2-1\right)\gamma^2\right)^4\\
&\times\biggl[4\alpha^4(4-3\gamma)^4-24\alpha^3\beta(3\gamma-4)^3+3\gamma^2\left(8\beta^2+3(\gamma-4)\gamma\right)-12\alpha^2(4-3\gamma)^2\left(\gamma^2-4\beta^2\right)\biggr.\\
&\biggl.+4\alpha\beta(3\gamma-4)\left(3\gamma^2-8\beta^2\right)\biggr]\biggl\{4\alpha^6(4-3\gamma)^6(4\gamma-1)+8\alpha^5\beta(\gamma-5)(3\gamma-4)^5(4\gamma-1)\biggr.\\
&\biggl. +3(\gamma-2)\gamma^2\left[18\gamma^4+8\beta^4(15\gamma-14)-3\beta^2\gamma(8+\gamma(17\gamma-6))\right]\biggr.\\
&\biggl.-4\alpha^4(4-3\gamma)^4\left[3\gamma^2(2\gamma-5)+\beta^2(40+\gamma(65\gamma-176))\right]\biggr.\\
&\biggl.-\alpha^2(4-3\gamma)^2\left[9\gamma^3\left(4\gamma^2+\gamma-4\right)-12\beta^2\gamma^2(102+\gamma(41\gamma-102))+16\beta^4(20+\gamma(67\gamma-96)) \right]\biggr.\\
&\biggl. +4\alpha^3\beta(3\gamma-4)^3\left[-3\gamma^2\left(8\gamma^2-26\gamma+37\right)+2\beta^2(40+\gamma(99\gamma-184)) \right]\biggr.\\
&\biggl.+2\alpha\beta(3\gamma-4)\bigl[-6\beta^2\gamma^2(124+5\gamma(-32+13\gamma))+16\beta^4(4+\gamma(17\gamma-20))\bigr.\biggr.\\
&\bigl.\biggl.+9\gamma^3(-8+\gamma(6+\gamma(4\gamma-5))) \bigr] \biggr\}\;,
\end{split}
\end{equation*}
\begin{equation*}
\begin{split}
u_5=&(2\beta+\alpha(4-3\gamma))^4(\gamma-1)\gamma\left[8(2\alpha+\beta)^2-24\alpha(2\alpha+\beta)\gamma+3(6\alpha^2-1)\gamma^2 \right]^2\\
&\times\biggl\{4\alpha^4\beta(4-3\gamma)^4(\gamma-8)+4\alpha^5(3\gamma-4)^5+3\beta(\gamma-2)\gamma^2\left(8\beta^2+3(\gamma-4)\gamma\right)\biggr.\\
&\biggl.-12\alpha^3(3\gamma-4)^3\left(2\beta^2(\gamma-4)+\gamma^2\right)+\alpha(3\gamma-4)\left[-32\beta^4(\gamma-2)+12\beta^2\gamma^3+9(\gamma-4)\gamma^3\right]\biggr.\\
&\biggl.+4\alpha^2\beta(4-3\gamma)^2\left(-3(\gamma-3)\gamma^2+4\beta^2(3\gamma-8)\right) \biggr\}
\end{split}
\end{equation*}
(6)
\begin{equation*}
e_1=\frac{3 (\gamma -2)^2-2 \alpha ^2 (4-3 \gamma )^2}{2 (\gamma -2)},\;e_2=\frac{2 \alpha  (3 \gamma -4) (\alpha  (3 \gamma -4)-2 \beta )}{\gamma -2}-3 \gamma,
\end{equation*}
\begin{equation*}
e_3=\frac{-2 \alpha ^2 (4-3 \gamma )^2+2 \alpha  (3 \gamma -4) \lambda +3 (\gamma -2) \gamma }{2
   (\gamma -2)}
\end{equation*}
(7)
\begin{equation*}
e_1=\lambda  (2 \beta -\lambda ),\;e_2=\frac{1}{2} \left(\lambda ^2-6\right),\;e_3=\lambda(4\alpha+\lambda)-3\gamma(1+\alpha\lambda)
\end{equation*}
(8)
\begin{equation*}
e_1=\frac{3\gamma(2\beta-\lambda)}{\alpha(4-3\gamma)+\lambda}\;,
\end{equation*}
\begin{equation*}
e_2=-\frac{-3 (\alpha  (6 \gamma -8)+(\gamma -2) \lambda ) \left[\alpha(4-3\gamma)+\lambda
   \right]^4-\sqrt{3}u_6}{4 (\alpha  (4-3 \gamma )+\lambda )^5}\;,
\end{equation*}
\begin{equation*}
e_3=-\frac{\sqrt{3}u_6-3 \left[\alpha(4-3\gamma)+\lambda \right]^4 (\alpha  (6 \gamma
   -8)+(\gamma -2) \lambda )}{4 (\alpha  (4-3 \gamma )+\lambda )^5}\;,
\end{equation*}
where
\begin{equation*}
\begin{split}
u_6^2=&(\alpha  (4-3 \gamma )+\lambda )^8 \biggl[16 \alpha ^3 (3 \gamma -4)^3 \lambda +4 \alpha ^2 (4-3
   \gamma )^2 \left(12 \gamma -8 \lambda ^2+3\right)\biggr.\\
   &\biggl.-4 \alpha  (3 \gamma -4) \lambda  \left(6
   \gamma ^2-3 \gamma -4 \lambda ^2+6\right)-3 (\gamma -2) \left(24 \gamma ^2-9 \gamma  \lambda
   ^2+2 \lambda ^2\right)\biggr]
\end{split}
\end{equation*}
One can easily obtain the eigenvalues for a perfect fluid with a specified equation of state parameter $\gamma$. We do not list the eigenvalues for dust and radiation here, since both cases are a particular case of the generalised eigenvalues presented above. 

\end{subappendices}
\begin{subappendices}
\section*{Appendix D: Two Fluid Case}
\renewcommand{\theequation}{D.\arabic{equation}}
\renewcommand{\thesection}{D}
\refstepcounter{section}
\setcounter{table}{0}
\renewcommand{\thetable}{D\arabic{table}}
\label{appendix:c}
\section*{Appendix D1: Eigenvalues--Conformal-disformal dust and conformal-disformal radiation}
\renewcommand{\theequation}{D1.\arabic{equation}}
\renewcommand{\thesection}{D1}
\refstepcounter{section}
\setcounter{table}{0}
\renewcommand{\thetable}{D1\arabic{table}}
\label{appendix:c1}

$(a)$
\begin{equation*}
\begin{split}
&e_1=-\frac{\alpha_1^5+\sqrt{\alpha_1^8\left(2-3\alpha_1^2\right)}}{2\alpha_1^5},\;e_2=-\frac{1}{2}+\frac{\sqrt{\alpha_1^8\left(2-3\alpha_1^2\right)}}{2\alpha_1^5},\;e_3=-4-\frac{2\beta_1}{\alpha_1},\;\\
&e_4=-4-\frac{2\beta_2}{\alpha_1},\;e_5=2+\frac{\lambda}{2\alpha_1}
\end{split}
\end{equation*}
$(b)$
\begin{equation*}
\begin{split}
&e_1=-4+\frac{4\beta_2}{\beta_1},\;e_2=2-\frac{\lambda}{\beta_1},\\
&e_3=\frac{-12+u_9^{2/3}+16\alpha_1^2+20\alpha_1\beta_1+13\beta_1^2+u_9^{1/3}(4\alpha_1+\beta_1)}{3u_9^{1/3}\beta_1},\\
&e_4=\frac{i\left(i+\sqrt{3}\right)u_9^{2/3}+2u_9^{1/3}(4\alpha_1+\beta_1)+\left(-1-i\sqrt{3}\right)\left(-12+16\alpha_1^2+20\alpha_1\beta_1+13\beta_1^2\right)}{6u_9^{1/3}\beta_1},\\
&e_5=\frac{\left(-1-i\sqrt{3}\right)u_9^{2/3}+2u_9^{1/3}(4\alpha_1+\beta_1)+i\left(i+\sqrt{3}\right)\left(-12+16\alpha_1^2+20\alpha_1\beta_1+13\beta_1^2\right)}{6u_9^{1/3}\beta_1},
\end{split}
\end{equation*}
where
\begin{equation*}
\begin{split}
&u_7^2=-\left(1+4\alpha_1^2+5\alpha_1\beta_1+\beta_1^2\right)^2(-16+9\beta_1^2),\\
&u_8=18+8\alpha_1^2+10\alpha_1\beta_1-7\beta_1^2,\\
&u_9=6\sqrt{3}u_7+8u_8\alpha_1+5u_8\beta_1
\end{split}
\end{equation*}
\section*{Appendix D2: Eigenvalues--Conformal dust and conformal-disformal radiation}
\renewcommand{\theequation}{D2.\arabic{equation}}
\renewcommand{\thesection}{D2}
\refstepcounter{section}
\setcounter{table}{0}
\renewcommand{\thetable}{D2\arabic{table}}
\label{appendix:c2}
(1)
\begin{equation*}
e_1=2,\;e_2=\frac{1}{2}\left(3-\sqrt{6}\alpha_1\right),\;e_3=-2(3+\sqrt{6}\beta),\;e_4=3+\sqrt{\frac{3}{2}}\lambda
\end{equation*}
$(6)_{(r)}$
\begin{equation*}
e_1=-4,\;e_2=2,\;e_3=-1,\;e_4=\frac{1}{2}
\end{equation*}
(2)
\begin{equation*}
e_1=2,\;e_2=\frac{1}{2}\left(3+\sqrt{6}\alpha_1\right),\;e_3=-2(3-\sqrt{6}\beta),\;e_4=3-\sqrt{\frac{3}{2}}\lambda
\end{equation*}
$(a)$
\begin{equation*}
e_1=-\frac{\alpha_1^5+\sqrt{\alpha_1^8\left(2-3\alpha_1^2\right)}}{2\alpha_1^5},\;e_2=-\frac{1}{2}+\frac{\sqrt{\alpha_1^8\left(2-3\alpha_1^2\right)}}{2\alpha_1^5},\;e_3=-4-\frac{2\beta}{\alpha_1},\;e_4=2+\frac{\lambda}{2\alpha_1}
\end{equation*}
$(6)_{(d)}$
\begin{equation*}
e_1=-\frac{3}{2}+\alpha_1^2,\;e_2=-1+2\alpha_1^2,\;e_3=-3-2\alpha_1(\alpha_1+2\beta),\;e_4=\frac{3}{2}+\alpha_1(\alpha_1+\lambda)
\end{equation*}
$(5)_{(r)}$
\begin{equation*}
\begin{split}
&e_1=\frac{1}{2}+\frac{\alpha_1}{\beta},\\
&e_2=-\frac{27 \beta ^{11}-72 \beta ^9+60 \beta ^7-16 \beta ^5+\sqrt{\beta ^8 \left(2-3 \beta
   ^2\right)^4 \left(81 \beta ^6-312 \beta ^4+368 \beta ^2-128\right)}}{2 \beta ^5 \left(2-3
   \beta ^2\right)^2 \left(3 \beta ^2-4\right)},\\
&e_3=\frac{-27 \beta ^{11}+72 \beta ^9-60 \beta ^7+16 \beta ^5+\sqrt{\beta ^8 \left(2-3 \beta
   ^2\right)^4 \left(81 \beta ^6-312 \beta ^4+368 \beta ^2-128\right)}}{2 \beta ^5 \left(2-3
   \beta ^2\right)^2 \left(3 \beta ^2-4\right)},\\
&e_4=2-\frac{\lambda}{\beta}
\end{split}
\end{equation*}
$(3)_{(r)}$
\begin{equation*}
\begin{split}
&e_1=-\frac{3}{2}-\frac{1}{2}(\alpha_1+2\beta)\left(-2\beta+u_2\right),\\
&e_2=-\frac{14\beta^2+7\beta u_2-33+\sqrt{225-66\beta^2+8\beta^4-30\beta u_2+4\beta^3 u_2}}{8\beta^2+4\beta u_2-6},\\
&e_3=\frac{-14\beta^2-7\beta u_2+33+\sqrt{225-66\beta^2+8\beta^4-30\beta u_2+4\beta^3 u_2}}{8\beta^2+4\beta u_2-6},\\
&e_4=-\frac{1}{2}(-2\beta+u_2)(2\beta-\lambda)
\end{split}
\end{equation*}
$(4)_{(r)}$
\begin{equation*}
\begin{split}
&e_1=-\frac{3}{2}+\frac{1}{2}(\alpha_1+2\beta)\left(2\beta+u_2\right),\\
&e_2=-\frac{-14\beta^2+7\beta u_2+33+\sqrt{225-66\beta^2+8\beta^4+30\beta u_2-4\beta^3 u_2}}{-8\beta^2+4\beta u_2+6},\\
&e_3=\frac{14\beta^2-7\beta u_2-33+\sqrt{225-66\beta^2+8\beta^4+30\beta u_2-4\beta^3 u_2}}{-8\beta^2+4\beta u_2+6},\\
&e_4=\frac{1}{2}(2\beta+u_2)(2\beta-\lambda)
\end{split}
\end{equation*}
$(8)_{(r)}$
\begin{equation*}
e_1=-4+\frac{8\beta}{\lambda},\;e_2=\frac{1}{2}+\frac{2\alpha_1}{\lambda},\;e_3=-\frac{\lambda^5+\sqrt{\lambda^8\left(64-15\lambda^2\right)}}{2\lambda^5},\;e_4=-\frac{1}{2}+\frac{\sqrt{\lambda^8\left(64-15\lambda^2\right)}}{2\lambda^5}
\end{equation*}
(7)
\begin{equation*}
e_1=(2\beta-\lambda)\lambda,\;e_2=\frac{1}{2}\left(\lambda^2-6\right),\;e_3=\lambda^2-4,\;e_4=\frac{1}{2}(-3+\lambda(\alpha_1+\lambda))
\end{equation*}
$(8)_{(d)}$
\begin{equation*}
\begin{split}
&e_1=\frac{6\beta-3\lambda}{\alpha_1+\lambda},\;e_2=-\frac{4\alpha_1+\lambda}{\alpha_1+\lambda},\\
&e_3=-\frac{6\alpha_1^2+9\alpha_1\lambda+3\lambda^2+\sqrt{3}\sqrt{-(\alpha_1+\lambda)^2\left(-72+16\alpha_1^3\lambda+21\lambda^2+4\alpha_1\lambda\left(4\lambda^2-9\right)+4\alpha_1^2\left(8\lambda^2-15\right)\right)}}{4(\alpha_1+\lambda)^2},\\
&e_4=\frac{-6\alpha_1^2-9\alpha_1\lambda-3\lambda^2+\sqrt{3}\sqrt{-(\alpha_1+\lambda)^2\left(-72+16\alpha_1^3\lambda+21\lambda^2+4\alpha_1\lambda\left(4\lambda^2-9\right)+4\alpha_1^2\left(8\lambda^2-15\right)\right)}}{4(\alpha_1+\lambda)^2}\\
\end{split}
\end{equation*}
\section*{Appendix D3: Eigenvalues--Two conformal-disformal dust components}
\renewcommand{\theequation}{D3.\arabic{equation}}
\renewcommand{\thesection}{D3}
\refstepcounter{section}
\setcounter{table}{0}
\renewcommand{\thetable}{D3\arabic{table}}
\label{appendix:c3}
(1)
\begin{equation*}
e_1=\frac{1}{2}\left(3-\sqrt{6}\alpha_1\right),\;e_2=3-\sqrt{6}\alpha_2,\;e_3=-2\left(3+\sqrt{6}\beta_1\right),\;e_4=-2\left(3+\sqrt{6}\beta_2\right),\;e_5=3+\sqrt{\frac{3}{2}}\lambda
\end{equation*}
$(c)$
\begin{equation*}
\begin{split}
&e_1=-3,\;e_2=-3,\;e_3=\frac{3}{2},\;e_4=\frac{1}{4}\left(-3+\frac{\sqrt{3}}{\alpha_2}\sqrt{\frac{\alpha_2}{\alpha_2-\alpha_1}}\sqrt{-\alpha_2(\alpha_1-\alpha_2)(3+16\alpha_1\alpha_2)}\right),\\
&e_5=-\frac{1}{4}\left(3+\frac{\sqrt{3}}{\alpha_2}\sqrt{\frac{\alpha_2}{\alpha_2-\alpha_1}}\sqrt{-\alpha_2(\alpha_1-\alpha_2)(3+16\alpha_1\alpha_1)}\right)
\end{split}
\end{equation*}
(2)
\begin{equation*}
e_1=\frac{1}{2}\left(3+\sqrt{6}\alpha_1\right),\;e_2=3+\sqrt{6}\alpha_2,\;e_3=-2\left(3-\sqrt{6}\beta_1\right),\;e_4=-2\left(3-\sqrt{6}\beta_2\right),\;e_5=3-\sqrt{\frac{3}{2}}\lambda
\end{equation*}
$(6)_{(d)}^1$
\begin{equation*}
\begin{split}
&e_1=-\frac{3}{2}+\alpha_1^2,\;e_2=2\alpha_1(\alpha_1-\alpha_2),\;e_3=-3-2\alpha_1(\alpha_1+2\beta_1),\;e_4=-3-2\alpha_1(\alpha_1+2\beta_2),\;\\
&e_5=\frac{3}{2}+\alpha_1(\alpha_1+\lambda)
\end{split}
\end{equation*}
$(6)_{(d)}^2$
\begin{equation*}
\begin{split}
&e_1=\alpha_2(\alpha_2-\alpha_1),\;e_2=-\frac{3}{2}+\alpha_2^2,\;e_3=-3-2\alpha_2(\alpha_2+2\beta_1),\;e_4=-3-2\alpha_2(\alpha_2+2\beta_2),\;\\
&e_5=\frac{3}{2}+\alpha_2(\alpha_2+\lambda)
\end{split}
\end{equation*}
$(3)_{(d)}^1$
\begin{equation*}
\begin{split}
&e_1=\frac{3 \left(2 \beta _1^2-3\right) \left(2 \alpha _2 \beta _1+\sqrt{2} \alpha _2
   u_{10}+3\right)}{\left(2 \beta _1^2+\sqrt{2} \beta _1 u_{10}-3\right){}^2},\;
e_2=\frac{3 \left(2 \beta _1^2-3\right) \left(\left(2 \alpha _1-\beta _1\right) \left(2 \beta
   _1+\sqrt{2} u_{10}\right)+9\right)-3 u_{11}}{2 \left(2 \beta _1^2+\sqrt{2} \beta _1
   u_{10}-3\right){}^2},\\
&e_3=\frac{3 \left(\left(2 \beta _1^2-3\right) \left(\left(2 \alpha _1-\beta _1\right) \left(2 \beta
   _1+\sqrt{2} u_{10}\right)+9\right)+u_{11}\right)}{2 \left(2 \beta _1^2+\sqrt{2} \beta _1
   u_{10}-3\right){}^2},\;
e_4=2 \left(\beta _1-\beta _2\right) \left(\sqrt{2} u_{10}-2 \beta _1\right),\\
&e_5=-\frac{1}{2} \left(2 \beta _1-\lambda \right) \left(\sqrt{2} u_{10}-2 \beta _1\right)\;,
\end{split}
\end{equation*}
where
\begin{equation*}
u_{10}^2=-3+2\beta_1^2,
\end{equation*}
\begin{equation*}
\begin{split}
u_{11}^2=&\frac{1}{\left(2 \beta _1^2+\sqrt{2} \beta _1 u_{10}-3\right){}^2}\left(2 \beta _1^2-3\right){}^3 \biggl(16 \alpha _1 \left(16 \beta _1^4-12 \beta _1^2-9\right)
   \beta _1+8 \alpha _1^2 \left(32 \beta _1^4-48 \beta _1^2+9\right)+64 \beta _1^6\biggr.\\
   &\biggl.-54 \beta _1^2+4
   \sqrt{2} u_{10} \left(4 \beta _1 \left(2 \alpha _1+\beta _1\right)+3\right) \left(\alpha _1
   \left(4 \beta _1^2-3\right)+2 \beta _1^3\right)-27\biggr)
\end{split}
\end{equation*}
$(4)_{(d)}^1$
\begin{equation*}
\begin{split}
&e_1=\left(\alpha _2+2 \beta _1\right) \left(2 \beta _1+\sqrt{2} u_{10}\right)-3,\;e_2=-\frac{3 \left(\left(2 \beta _1^2-3\right) \left(\left(\beta _1-2 \alpha _1\right)
   \left(-\left(\sqrt{2} u_{10}-2 \beta _1\right)\right)-9\right)+u_{12}\right)}{2 \left(-2 \beta
   _1^2+\sqrt{2} \beta _1 u_{10}+3\right){}^2},\\
&e_3=\frac{3 u_{12}-3 \left(2 \beta _1^2-3\right) \left(\left(\beta _1-2 \alpha _1\right)
   \left(-\left(\sqrt{2} u_{10}-2 \beta _1\right)\right)-9\right)}{2 \left(-2 \beta _1^2+\sqrt{2}
   \beta _1 u_{10}+3\right){}^2},\;e_4=-2 \left(\beta _1-\beta _2\right) \left(2 \beta _1+\sqrt{2} u_{10}\right),\\
&e_5=\frac{1}{2} \left(2 \beta _1-\lambda \right) \left(2 \beta _1+\sqrt{2} u_{10}\right)\;,
\end{split}
\end{equation*}
where
\begin{equation*}
\begin{split}
u_{12}^2=&-\frac{1}{\left(-2 \beta _1^2+\sqrt{2} \beta _1 u_{10}+3\right){}^2}\left(2 \beta _1^2-3\right){}^3 \biggl(16 \alpha _1 \left(-16 \beta _1^4+12 \beta _1^2+9\right)
   \beta _1-8 \alpha _1^2 \left(32 \beta _1^4-48 \beta _1^2+9\right)\biggr.\\
   &\biggl.-64 \beta _1^6
   +54 \beta _1^2+4
   \sqrt{2} u_{10} \left(4 \beta _1 \left(2 \alpha _1+\beta _1\right)+3\right) \left(\alpha _1
   \left(4 \beta _1^2-3\right)+2 \beta _1^3\right)+27\biggr)
\end{split}
\end{equation*}
$(3)_{(d)}^2$
\begin{equation*}
\begin{split}
&e_1=2 \left(\beta _2-\beta _1\right) \left(u_{13}-2 \beta _2\right),\;e_2=-\frac{1}{2} \left(\alpha _1+2 \beta _2\right) \left(u_{13}-2 \beta _2\right)-\frac{3}{2},\\
&e_3=\frac{3 \left(2 \beta _2^2-3\right) \left(4 \alpha _2 \beta _2-2 \beta _2^2+u_{13} \left(2 \alpha
   _2-\beta _2\right)+9\right)-3 u_{14}}{2 \left(2 \beta _2^2+\beta _2 u_{13}-3\right){}^2},\;\\
&e_4=\frac{3 \left(\left(2 \beta _2^2-3\right) \left(4 \alpha _2 \beta _2-2 \beta _2^2+u_{13} \left(2
   \alpha _2-\beta _2\right)+9\right)+u_{14}\right)}{2 \left(2 \beta _2^2+\beta _2
   u_{13}-3\right){}^2},\\
&e_5=-\frac{1}{2} \left(2 \beta _2-\lambda \right) \left(u_{13}-2 \beta _2\right)\;,
\end{split}
\end{equation*}
where
\begin{equation*}
u_{13}^2=-6+4\beta_2^2,
\end{equation*}
\begin{equation*}
\begin{split}
u_{14}^2=&\frac{1}{\left(2 \beta _2^2+\beta _2 u_{13}-3\right){}^2}\left(2 \beta _2^2-3\right){}^3 \left(16 \alpha _2 \biggl(16 \beta _2^4-12 \beta _2^2-9\right)
   \beta _2+8 \alpha _2^2 \left(32 \beta _2^4-48 \beta _2^2+9\right)+64 \beta _2^6\biggr.\\
   &\biggl.-54 \beta _2^2+4
   u_{13} \left(4 \beta _2 \left(2 \alpha _2+\beta _2\right)+3\right) \left(\alpha _2 \left(4
   \beta _2^2-3\right)+2 \beta _2^3\right)-27\biggr)
\end{split}
\end{equation*}
$(4)_{(d)}^2$
\begin{equation*}
\begin{split}
&e_1=2 \left(\beta _1-\beta _2\right) \left(2 \beta _2+u_{13}\right),\;e_2=\frac{1}{2} \left(\left(\alpha _1+2 \beta _2\right) \left(2 \beta _2+u_{13}\right)-3\right),\\
&e_3=\frac{3 \left(2 \beta _2^2-3\right) \left(4 \alpha _2 \beta _2-2 \beta _2^2+u_{13} \left(\beta
   _2-2 \alpha _2\right)+9\right)-3 u_{15}}{2 \left(-2 \beta _2^2+\beta _2 u_{13}+3\right){}^2},\\
&e_4=\frac{3 \left(\left(2 \beta _2^2-3\right) \left(4 \alpha _2 \beta _2-2 \beta _2^2+u_{13}
   \left(\beta _2-2 \alpha _2\right)+9\right)+u_{15}\right)}{2 \left(-2 \beta _2^2+\beta _2
   u_{13}+3\right){}^2},\\
&e_5=\frac{1}{2} \left(2 \beta _2-\lambda \right) \left(2 \beta _2+u_{13}\right)\;,
\end{split}
\end{equation*}
where
\begin{equation*}
\begin{split}
u_{15}^2=&\frac{1}{\left(-2 \beta _2^2+\beta _2 u_{13}+3\right){}^2}\left(2 \beta _2^2-3\right){}^3 \biggl(16 \alpha _2 \left(16 \beta _2^4-12 \beta _2^2-9\right)
   \beta _2+8 \alpha _2^2 \left(32 \beta _2^4-48 \beta _2^2+9\right)+64 \beta _2^6\biggr.\\
   &\biggl.-54 \beta _2^2-4
   u_{13} \left(4 \beta _2 \left(2 \alpha _2+\beta _2\right)+3\right) \left(\alpha _2 \left(4
   \beta _2^2-3\right)+2 \beta _2^3\right)-27\biggr)
\end{split}
\end{equation*}
(7)
\begin{equation*}
e_1=(2\beta_1-\lambda)\lambda,\;e_2=(2\beta_2-\lambda)\lambda,\;e_3=\frac{1}{2}\left(\lambda^2-6\right),\;e_4=\frac{1}{2}\left(\lambda(\alpha_1+\lambda)-3\right),\;e_5=\lambda(\alpha_2+\lambda)-3
\end{equation*}
$(8)_{(d)}^1$
\begin{equation*}
\begin{split}
&e_1=-\frac{3(\alpha_1-\alpha_2)}{\alpha_1+\lambda},\;e_2=\frac{3(2\beta_1-\lambda)}{\alpha_1+\lambda},\;e_3=\frac{3(2\beta_2-\lambda)}{\alpha_1+\lambda},\\
&e_4=-\frac{3(\alpha_1+\lambda)^3(2\alpha_1+\lambda)+\sqrt{3}\sqrt{-(\alpha_1+\lambda)^6\left(-72+16\alpha_1^3\lambda+21\lambda^2+4\alpha_1\lambda(4\lambda^2-9)+4\alpha_1^2(8\lambda^2-15)\right)}}{4(\alpha_1+\lambda)^4},\\
&e_5=\frac{-3(\alpha_1+\lambda)^3(2\alpha_1+\lambda)+\sqrt{3}\sqrt{-(\alpha_1+\lambda)^6\left(-72+16\alpha_1^3\lambda+21\lambda^2+4\alpha_1\lambda(4\lambda^2-9)+4\alpha_1^2(8\lambda^2-15)\right)}}{4(\alpha_1+\lambda)^4}
\end{split}
\end{equation*}
$(8)_{(d)}^2$
\begin{equation*}
\begin{split}
&e_1=\frac{3(\alpha_1-\alpha_2)}{2(\alpha_2+\lambda)},\;e_2=\frac{3(2\beta_1-\lambda)}{\alpha_2+\lambda},\;e_3=\frac{3(2\beta_2-\lambda)}{\alpha_2+\lambda},\\
&e_4=-\frac{3(\alpha_2+\lambda)^3(2\alpha_2+\lambda)+\sqrt{3}\sqrt{-(\alpha_2+\lambda)^6\left(-72+16\alpha_2^3\lambda+21\lambda^2+4\alpha_2\lambda(4\lambda^2-9)+4\alpha_2^2(8\lambda^2-15)\right)}}{4(\alpha_2+\lambda)^4},\\
&e_5=\frac{-3(\alpha_2+\lambda)^3(2\alpha_2+\lambda)+\sqrt{3}\sqrt{-(\alpha_2+\lambda)^6\left(-72+16\alpha_2^3\lambda+21\lambda^2+4\alpha_2\lambda(4\lambda^2-9)+4\alpha_2^2(8\lambda^2-15)\right)}}{4(\alpha_2+\lambda)^4}
\end{split}
\end{equation*}
\subsection*{Fixed Points--Conformally coupled perfect fluid with equation of state parameter $\gamma_1$, and conformally-disformally coupled perfect fluid with equation of state parameter $\gamma_2$}
\renewcommand{\thetable}{D\arabic{table}}
{\setlength{\tabcolsep}{0.5pt}
\setlength\extrarowheight{9pt}
\begin{landscape}
\begin{table}
\begin{center}
\begin{tabular}{ c c c c c} 
 \hline
\hline
  $x$ & $y$ & $z_1$ & $\sigma_2$  \\ 
\hline 
 -1 & 0 & 0 & 0   \\ 
 1 & 0 & 0 & 0  \\ 
 $\frac{\sqrt{2} \beta -\sqrt{2 \beta ^2-3}}{\sqrt{3}}$ & 0 & 0 & $\frac{1}{18} \left(2 \beta  \left(2\beta+\sqrt{4 \beta ^2-6} \right)-3\right)$   \\ 
 $\frac{\sqrt{2} \beta + \sqrt{2 \beta ^2-3} }{\sqrt{3}}$ & 0 & 0 & $\frac{1}{18} \left(2 \beta \left(2\beta - \sqrt{4 \beta ^2-6}\right)-3\right)$    \\
 $\frac{\sqrt{\frac{3}{2}}\gamma_1}{2\beta+\alpha_1(4-3\gamma_1) }$ & 0 & $\sqrt{\frac{-8(2\alpha_1+\beta)^2+8(2\alpha_1+\beta)(3\alpha_1+\beta)\gamma_1-3\left(1+6\alpha_1^2+4\alpha_1\beta\right)\gamma_1^2}{2(2\beta+\alpha_1(4-3\gamma_1))^2(\gamma_1-1)}}$ & $\frac{(2\beta+\alpha_1(4-3\gamma_1))^2}{9\gamma_1^2}$   \\ 
 $\frac{\sqrt{\frac{2}{3}}\alpha_1(4-3\gamma_1)}{\gamma_1-2}$ & 0 & $\sqrt{\frac{3(\gamma_1-2)^2-2\alpha_1^2(4-3\gamma_1)^2}{3(\gamma_1-2)^2}}$ & 0\\
 $\frac{\sqrt{6}(\gamma_2-\gamma_1)}{\alpha_1(6\gamma_1-8)+\alpha_2(8-6\gamma_2)}$ & 0 & $\sqrt{\frac{2\alpha_2^2(4-3\gamma_2)^2+3(\gamma_1-\gamma_2)(\gamma_2-2)-2\alpha_1\alpha_2(3\gamma_1-4)(3\gamma_2-4)}{2(\alpha_1(3\gamma_1-4)+\alpha_2(4-3\gamma_2))^2}}$ & 0 \\
 $\frac{\sqrt{\frac{3}{2}}\gamma_2}{2\beta+\alpha_2(4-3\gamma_2)}$ & 0 & 0 & $\frac{(2\beta+\alpha_2(4-3\gamma_2))^2\left(4\alpha_2\beta(4-3\gamma_2)+2\alpha_2^2(4-3\gamma_2)^2-3(\gamma_2-2)\gamma_2\right)}{9(\gamma_2-1)\gamma_2\left(8(2\alpha_2+\beta)^2-24\alpha_2(2\alpha_2+\beta)\gamma_2+3(6\alpha_2^2-1)\gamma_2^2\right)}$ \\
 $\frac{\sqrt{\frac{2}{3}}\alpha_2(4-3\gamma_2)}{\gamma_2-2}$ & 0 & 0 & 0   \\ 
 $\frac{\sqrt{\frac{3}{2}}\gamma_1}{\alpha_1(4-3\gamma_1)+\lambda}$ & $\sqrt{\frac{2\alpha_1^2(4-3\gamma_1)^2-3(\gamma_1-2)\gamma_1+\alpha_1(8-6\gamma_1)\lambda}{2(\alpha_1(4-3\gamma_1)+\lambda)^2}}$ & $\sqrt{\frac{\lambda(4\alpha_1+\lambda)-3\gamma_1(1+\alpha_1\lambda)}{(\alpha_1(4-3\gamma_1)+\lambda)^2}}$ & 0 &  \\ 
 $\frac{\lambda }{\sqrt{6}}$ & $\sqrt{1-\frac{\lambda ^2}{6}}$ & 0 & 0 \\  
 $\frac{\sqrt{\frac{3}{2}}\gamma_2}{(4-3\gamma_2)\alpha_2+\lambda}$ & $\sqrt{\frac{2\alpha_2^2(4-3\gamma_2)^2-3(\gamma_2-2)\gamma_2+\alpha_2(8-6\gamma_2)\lambda }{2(\alpha_2(4-3\gamma_2)+\lambda)^2}}$ & 0 & 0 \\ 
\hline
\hline
\end{tabular}
\end{center}
\caption{\label{table5} The fixed points of the system (\ref{start})--(\ref{end}) for conformally coupled fluid with equation of state parameter, $\gamma_1$, together with a conformally--disformally coupled fluid with equation of state parameter, $\gamma_2$, as described in Section \ref{sec:ccd_dust_rad}.}
\end{table}

\end{landscape}}
\end{subappendices}

\bibliographystyle{JHEP}
\bibliography{fullbib}

\providecommand{\href}[2]{#2}\begingroup\raggedright\begin{thebibliography}{10}

\bibitem{Copeland:2006wr}
E.~J. Copeland, M.~Sami, and S.~Tsujikawa, {\it {Dynamics of dark energy}},
  {\em Int. J. Mod. Phys.} {\bf D15} (2006) 1753--1936,
  [\href{http://arxiv.org/abs/hep-th/0603057}{{\tt hep-th/0603057}}].

\bibitem{Sotiriou:2008rp}
T.~P. Sotiriou and V.~Faraoni, {\it {f(R) Theories Of Gravity}},  {\em Rev.
  Mod. Phys.} {\bf 82} (2010) 451--497,
  [\href{http://arxiv.org/abs/0805.1726}{{\tt arXiv:0805.1726}}].

\bibitem{Clifton:2011jh}
T.~Clifton, P.~G. Ferreira, A.~Padilla, and C.~Skordis, {\it {Modified Gravity
  and Cosmology}},  {\em Phys. Rept.} {\bf 513} (2012) 1--189,
  [\href{http://arxiv.org/abs/1106.2476}{{\tt arXiv:1106.2476}}].

\bibitem{Wetterich:1987fm}
C.~Wetterich, {\it {Cosmology and the Fate of Dilatation Symmetry}},  {\em
  Nucl. Phys.} {\bf B302} (1988) 668--696.

\bibitem{Peebles:1987ek}
P.~J.~E. Peebles and B.~Ratra, {\it {Cosmology with a Time Variable
  Cosmological Constant}},  {\em Astrophys. J.} {\bf 325} (1988) L17.

\bibitem{Wetterich:1994bg}
C.~Wetterich, {\it {The Cosmon model for an asymptotically vanishing time
  dependent cosmological 'constant'}},  {\em Astron. Astrophys.} {\bf 301}
  (1995) 321--328, [\href{http://arxiv.org/abs/hep-th/9408025}{{\tt
  hep-th/9408025}}].

\bibitem{Caldwell:1997ii}
R.~R. Caldwell, R.~Dave, and P.~J. Steinhardt, {\it {Cosmological imprint of an
  energy component with general equation of state}},  {\em Phys. Rev. Lett.}
  {\bf 80} (1998) 1582--1585,
  [\href{http://arxiv.org/abs/astro-ph/9708069}{{\tt astro-ph/9708069}}].

\bibitem{Barreiro:1999zs}
T.~Barreiro, E.~J. Copeland, and N.~J. Nunes, {\it {Quintessence arising from
  exponential potentials}},  {\em Phys. Rev.} {\bf D61} (2000) 127301,
  [\href{http://arxiv.org/abs/astro-ph/9910214}{{\tt astro-ph/9910214}}].

\bibitem{Amendola:1999er}
L.~Amendola, {\it {Coupled quintessence}},  {\em Phys. Rev.} {\bf D62} (2000)
  043511, [\href{http://arxiv.org/abs/astro-ph/9908023}{{\tt
  astro-ph/9908023}}].

\bibitem{Holden:1999hm}
D.~J. Holden and D.~Wands, {\it {Selfsimilar cosmological solutions with a
  nonminimally coupled scalar field}},  {\em Phys. Rev.} {\bf D61} (2000)
  043506, [\href{http://arxiv.org/abs/gr-qc/9908026}{{\tt gr-qc/9908026}}].

\bibitem{Copeland:2003cv}
E.~J. Copeland, N.~J. Nunes, and M.~Pospelov, {\it {Models of quintessence
  coupled to the electromagnetic field and the cosmological evolution of
  alpha}},  {\em Phys. Rev.} {\bf D69} (2004) 023501,
  [\href{http://arxiv.org/abs/hep-ph/0307299}{{\tt hep-ph/0307299}}].

\bibitem{Carroll:1998zi}
S.~M. Carroll, {\it {Quintessence and the rest of the world}},  {\em Phys. Rev.
  Lett.} {\bf 81} (1998) 3067--3070,
  [\href{http://arxiv.org/abs/astro-ph/9806099}{{\tt astro-ph/9806099}}].

\bibitem{Bekenstein}
J.~D. Bekenstein, {\it {The Relation between physical and gravitational
  geometry}},  {\em Phys. Rev.} {\bf D48} (1993) 3641--3647,
  [\href{http://arxiv.org/abs/gr-qc/9211017}{{\tt gr-qc/9211017}}].

\bibitem{Zuma2}
M.~Zumalac\'{a}rregui, T.~S. Koivisto, and D.~F. Mota, {\it {DBI Galileons in
  the Einstein Frame: Local Gravity and Cosmology}},  {\em Phys. Rev.} {\bf
  D87} (2013) 083010, [\href{http://arxiv.org/abs/1210.8016}{{\tt
  arXiv:1210.8016}}].

\bibitem{Zuma5}
M.~Zumalac\'{a}rregui, T.~S. Koivisto, D.~F. Mota, and P.~Ruiz-Lapuente, {\it
  {Disformal Scalar Fields and the Dark Sector of the Universe}},  {\em JCAP}
  {\bf 1005} (2010) 038, [\href{http://arxiv.org/abs/1004.2684}{{\tt
  arXiv:1004.2684}}].

\bibitem{Zumalacarregui:2012us}
M.~Zumalacarregui, T.~S. Koivisto, and D.~F. Mota, {\it {DBI Galileons in the
  Einstein Frame: Local Gravity and Cosmology}},  {\em Phys. Rev.} {\bf D87}
  (2013) 083010, [\href{http://arxiv.org/abs/1210.8016}{{\tt
  arXiv:1210.8016}}].

\bibitem{Jack}
C.~van~de Bruck and J.~Morrice, {\it {Disformal couplings and the dark sector
  of the universe}},  {\em JCAP} {\bf 1504} (2015), no.~04 036,
  [\href{http://arxiv.org/abs/1501.03073}{{\tt arXiv:1501.03073}}].

\bibitem{Sakstein:2014aca}
J.~Sakstein, {\it {Towards Viable Cosmological Models of Disformal Theories of
  Gravity}},  {\em Phys. Rev.} {\bf D91} (2015), no.~2 024036,
  [\href{http://arxiv.org/abs/1409.7296}{{\tt arXiv:1409.7296}}].

\bibitem{Sakstein:2014isa}
J.~Sakstein, {\it {Disformal Theories of Gravity: From the Solar System to
  Cosmology}},  {\em JCAP} {\bf 1412} (2014) 012,
  [\href{http://arxiv.org/abs/1409.1734}{{\tt arXiv:1409.1734}}].

\bibitem{Bettoni:2013diz}
D.~Bettoni and S.~Liberati, {\it {Disformal invariance of second order
  scalar-tensor theories: Framing the Horndeski action}},  {\em Phys. Rev.}
  {\bf D88} (2013) 084020, [\href{http://arxiv.org/abs/1306.6724}{{\tt
  arXiv:1306.6724}}].

\bibitem{Zuma1}
M.~Zumalac\'{a}rregui and J.~Garc\'{i}a-Bellido, {\it {Transforming gravity:
  from derivative couplings to matter to second-order scalar-tensor theories
  beyond the Horndeski Lagrangian}},  {\em Phys. Rev.} {\bf D89} (2014) 064046,
  [\href{http://arxiv.org/abs/1308.4685}{{\tt arXiv:1308.4685}}].

\bibitem{Koivisto}
T.~Koivisto, D.~Wills, and I.~Zavala, {\it {Dark D-brane Cosmology}},  {\em
  JCAP} {\bf 1406} (2014) 036, [\href{http://arxiv.org/abs/1312.2597}{{\tt
  arXiv:1312.2597}}].

\bibitem{Carsten}
C.~van~de Bruck, J.~Morrice, and S.~Vu, {\it {Constraints on Nonconformal
  Couplings from the Properties of the Cosmic Microwave Background Radiation}},
   {\em Phys. Rev. Lett.} {\bf 111} (2013) 161302,
  [\href{http://arxiv.org/abs/1303.1773}{{\tt arXiv:1303.1773}}].

\bibitem{Steinhardt:1999nw}
P.~J. Steinhardt, L.-M. Wang, and I.~Zlatev, {\it {Cosmological tracking
  solutions}},  {\em Phys. Rev.} {\bf D59} (1999) 123504,
  [\href{http://arxiv.org/abs/astro-ph/9812313}{{\tt astro-ph/9812313}}].

\bibitem{Ng:2001hs}
S.~C.~C. Ng, N.~J. Nunes, and F.~Rosati, {\it {Applications of scalar attractor
  solutions to cosmology}},  {\em Phys. Rev.} {\bf D64} (2001) 083510,
  [\href{http://arxiv.org/abs/astro-ph/0107321}{{\tt astro-ph/0107321}}].

\bibitem{Barrow:2013uza}
J.~D. Barrow and A.~A.~H. Graham, {\it {General Dynamics of Varying-Alpha
  Universes}},  {\em Phys. Rev.} {\bf D88} (2013) 103513,
  [\href{http://arxiv.org/abs/1307.6816}{{\tt arXiv:1307.6816}}].

\bibitem{Nunes:2000yc}
A.~Nunes and J.~P. Mimoso, {\it {On the potentials yielding cosmological
  scaling solutions}},  {\em Phys. Lett.} {\bf B488} (2000) 423--427,
  [\href{http://arxiv.org/abs/gr-qc/0008003}{{\tt gr-qc/0008003}}].

\bibitem{Amendola:2014kwa}
L.~Amendola, T.~Barreiro, and N.~J. Nunes, {\it {Multifield coupled
  quintessence}},  {\em Phys. Rev.} {\bf D90} (2014), no.~8 083508,
  [\href{http://arxiv.org/abs/1407.2156}{{\tt arXiv:1407.2156}}].

\bibitem{Gumjudpai:2005ry}
B.~Gumjudpai, T.~Naskar, M.~Sami, and S.~Tsujikawa, {\it {Coupled dark energy:
  Towards a general description of the dynamics}},  {\em JCAP} {\bf 0506}
  (2005) 007, [\href{http://arxiv.org/abs/hep-th/0502191}{{\tt
  hep-th/0502191}}].

\bibitem{Copeland:1997et}
E.~J. Copeland, A.~R. Liddle, and D.~Wands, {\it {Exponential potentials and
  cosmological scaling solutions}},  {\em Phys. Rev.} {\bf D57} (1998)
  4686--4690, [\href{http://arxiv.org/abs/gr-qc/9711068}{{\tt gr-qc/9711068}}].

\bibitem{Brookfield:2007au}
A.~W. Brookfield, C.~van~de Bruck, and L.~M.~H. Hall, {\it {New interactions in
  the dark sector mediated by dark energy}},  {\em Phys. Rev.} {\bf D77} (2008)
  043006, [\href{http://arxiv.org/abs/0709.2297}{{\tt arXiv:0709.2297}}].

\bibitem{P2}
{\bf Planck} Collaboration, P.~A.~R. Ade et~al., {\it {Planck 2015 results.
  XIII. Cosmological parameters}},  \href{http://arxiv.org/abs/1502.01589}{{\tt
  arXiv:1502.01589}}.

\bibitem{Ferreira:1997hj}
P.~G. Ferreira and M.~Joyce, {\it {Cosmology with a primordial scaling field}},
   {\em Phys. Rev.} {\bf D58} (1998) 023503,
  [\href{http://arxiv.org/abs/astro-ph/9711102}{{\tt astro-ph/9711102}}].

\bibitem{Cyburt:2004yc}
R.~H. Cyburt, B.~D. Fields, K.~A. Olive, and E.~Skillman, {\it {New BBN limits
  on physics beyond the standard model from $^4He$}},  {\em Astropart. Phys.}
  {\bf 23} (2005) 313--323, [\href{http://arxiv.org/abs/astro-ph/0408033}{{\tt
  astro-ph/0408033}}].

\bibitem{PhysRevD.64.103508}
R.~Bean, S.~H. Hansen, and A.~Melchiorri, {\it Early-universe constraints on
  dark energy},  {\em Phys. Rev. D} {\bf 64} (Oct, 2001) 103508.

\bibitem{Baldi:2012kt}
M.~Baldi, {\it {Multiple Dark Matter as a self-regulating mechanism for dark
  sector interactions}},  {\em Annalen Phys.} {\bf 524} (2012) 602--617,
  [\href{http://arxiv.org/abs/1204.0514}{{\tt arXiv:1204.0514}}].

\bibitem{vandeBruck:2016hpz}
C.~van~de Bruck, J.~Mifsud, and J.~Morrice, {\it {Testing coupled dark energy
  models with their cosmological background evolution}},
  \href{http://arxiv.org/abs/1609.09855}{{\tt arXiv:1609.09855}}.

\bibitem{vandeBruck:2015rma}
C.~van~de Bruck, J.~Mifsud, and N.~J. Nunes, {\it {The variation of the
  fine-structure constant from disformal couplings}},  {\em JCAP} {\bf 1512}
  (2015), no.~12 018, [\href{http://arxiv.org/abs/1510.00200}{{\tt
  arXiv:1510.00200}}].

\bibitem{Chiba:2011bz}
T.~Chiba, {\it {The Constancy of the Constants of Nature: Updates}},  {\em
  Prog. Theor. Phys.} {\bf 126} (2011) 993--1019,
  [\href{http://arxiv.org/abs/1111.0092}{{\tt arXiv:1111.0092}}].

\end{thebibliography}\endgroup

\end{document}